\documentclass[twocolumn, switch]{article} 

\usepackage{preprint}
\usepackage{algorithm}
\usepackage{algorithmic}
\usepackage{amsmath}
\usepackage{makecell}
\usepackage{multirow}
\usepackage{tabularx}
\usepackage{geometry} 
\usepackage{booktabs} 
\usepackage{array}    
\usepackage{caption}  
\usepackage{setspace}
\usepackage{amsmath, amsthm, amssymb, amsfonts}
\usepackage{subcaption}

\usepackage{natbib}

\usepackage{xcolor}		
\usepackage[colorlinks = true,
            linkcolor = purple,
            urlcolor  = blue,
            citecolor = cyan,
            anchorcolor = black]{hyperref}	
\usepackage{booktabs} 		
\usepackage{nicefrac}		
\usepackage{microtype}		
\usepackage{lineno}		
\usepackage{float}			
\usepackage{enumitem}

\usepackage{lipsum}		
\usepackage{pdflscape}
\usepackage{newfloat}
\DeclareFloatingEnvironment[name={Supplementary Figure}]{suppfigure}
\usepackage{sidecap}
\sidecaptionvpos{figure}{c}

\usepackage{titlesec}
\titlespacing\section{0pt}{12pt plus 3pt minus 3pt}{1pt plus 1pt minus 1pt}
\titlespacing\subsection{0pt}{10pt plus 3pt minus 3pt}{1pt plus 1pt minus 1pt}
\titlespacing\subsubsection{0pt}{8pt plus 3pt minus 3pt}{1pt plus 1pt minus 1pt}

\title{Neither Consent nor Property: A Policy Lab for Data Law}

\usepackage{eso-pic}
\usepackage{tikz}
\usepackage{xcolor}
\PassOptionsToPackage{colorlinks=true,linkcolor=gray,urlcolor=gray}{hyperref}

\newcommand{\AddMyWatermarks}{%
  \begin{tikzpicture}[remember picture, overlay]
    \node[color=gray!90, scale=1] at ([xshift=0in,yshift=-5in]current page.center) {%
      This is a preprint manuscript. Version: \today.%
    };
  \end{tikzpicture}%
}

\AddToShipoutPictureBG*{\AddMyWatermarks}

\usepackage{titling}
\usepackage{orcidlink}
\usepackage{footmisc}
\usepackage[most]{tcolorbox}
\newtcbtheorem[auto counter, number within = section]{prompt}{Prompt}{
	colbacktitle = black!60!white, colframe = black!60!white,
	colback = black!5!white,
	fonttitle=\bfseries,
}{t}

\newcommand{\Author}[3]{
  \textbf{#1}\textsuperscript{#2},\ \orcidlink{#3} %
}

\author{
  \Author{Haoyi Zhang}{1}{0009-0008-2582-4467} \and
  \Author{Tianyi Zhu}{1}{0009-0001-0821-8018}
}

\date{%
  \textsuperscript{1}Law School, Peking University\\
  [1em]
 \footnotesize The authors have contributed equally.\\ \url{<zhanghaoyi,tianyizhu>@law.pku.edu.cn}\\
}

\begin{document}

\twocolumn[ 
  \begin{@twocolumnfalse} 

\maketitle
\thispagestyle{empty}

\begin{abstract}
Regulators currently govern the AI data economy based on intuition rather than evidence, struggling to choose between inconsistent regimes of informed consent, immunity, and liability. To fill this policy vacuum, this paper develops a novel computational policy laboratory: a spatially explicit Agent-Based Model (ABM) of the data market. To solve the problem of missing data, we introduce a two-stage methodological pipeline. First, we translate decision rules from multi-year fieldwork (2022-2025) into agent constraints. This ensures the model reflects actual bargaining frictions rather than theoretical abstractions. Second, we deploy Large Language Models (LLMs) as ``subjects'' in a Discrete Choice Experiment (DCE). This novel approach recovers precise preference primitives, such as willingness-to-pay elasticities, which are empirically unobservable in the wild. Calibrated by these inputs, our model places rival legal institutions side-by-side to simulate their welfare effects. The results challenge the dominant regulatory paradigm. We find that property-rule mechanisms, such as informed consent, fail to maximize welfare. Counterintuitively, social welfare peaks when liability for substantive harm is shifted to the downstream buyer. This aligns with the ``least cost avoider'' principle, because downstream users control post-acquisition safeguards, they are best positioned to mitigate risk efficiently. By ``de-romanticizing'' seller-centric frameworks, this paper provides an economic justification for emerging doctrines of downstream reachability.

\end{abstract}
\vspace{0.1cm}
\keywords{Data market \and Data law \and Comparative institutional analysis \and Computational social science \and Agent-based modeling}
\JEL {C63 \and D47 \and D61 \and K11 \and K12 \and K24}
\vspace{0.35cm}
  \end{@twocolumnfalse} 
] 

\onecolumn
\begingroup
\setstretch{0.8}
\setcounter{tocdepth}{2}
\thispagestyle{empty}
\tableofcontents
\endgroup
\clearpage
\pagenumbering{arabic}
\setcounter{page}{1}
\twocolumn

\section{Introduction}
Regulators are governing the AI economy under conditions of information asymmetry. The market for data, the most essential input for AI, is characterized by structural opacity that effective governance has become nearly impossible. This fundamental lack of visibility regarding who trades, on what terms, and with what risk has resulted in a fragmented policy landscape, split between inconsistent consent gates and liability rules \citep{solove2013consent, viljoen2021consent, citron2022consent, acquisti2015consent}. This paper proposes a novel methodological framework to move legal analysis from pure theoretical speculation to empirical observation. We develop an Agent-Based Model (ABM) calibrated via Large Language Model (LLM) proxy experiments, to stimulate the data market's dynamics and test the compararive efficiency of rival legal institutions. 

The epistemic barrier to regulating data markets is structural, not accidental. Traditional empirical tools falter because the underlying transactions are legally sensitive, bilaterally negotiated, and systematically obscured. As the U.S. Federal Trade Commission has documented, the data brokerage system operates through pipelines with ``limited transparency into who trades what.'' \footnote{Federal Trade Commission, \textit{Data Brokers: A Call for Transparency and Accountability} (May 2014), at iv (noting that the industry is ``complex, with multiple layers of data brokers providing data to each other'' and that consumers are generally unaware of these data flows).} This problem is even more acute in China, after a \$1.2 billion fine against Didi Global, where a ``compliance chill'' has pushed high-stakes transactions further into the shadows.\footnote{Cyberspace Administration of China, \textit{Decision on the Administrative Penalty of Didi Global Inc.} (July 21, 2022). The penalty of RMB 8.026 billion ended a year-long cybersecurity review that removed Didi from app stores, signaling a shift toward strict enforcement of the Data Security Law. \href{https://www.reuters.com/technology/china-fines-didi-global-12-bln-violating-data-security-laws-2022-07-21/}{\texttt{https://www.reuters.com/technology/china\-fines\-didi\-global\-12\-bln\-violating\-data\-security\-laws\-2022-07-21/}}.} Without a public panel of prices or quantities, researchers and policymakers lack the micro-evidence required to assess how legal rules actually shape market behavior.

To construct a model of a data market where observational data is unavailable, we adopt a two-stage research design that grounds simulation in empirical reality. First, to uncover the hidden logic of market actors, we lead a multi-sited fieldwork conducted between 2022 and 2025. Rather than relying on abstract theoretical payoffs, this qualitative inquiry identifies the specific micro-foundations of transaction. We observe how parties assess value under conditions of limited verifiability and identify the binding constraints on the bargaining process. These findings allow us to translate observed behaviors into formal state variables and action strategies-specifically, the buyer's reliance on institutional signals to navigate quality uncertainty, and the seller's dynamic sensitivity to enforcement salience.

Second, to calibrate these rules with empirical precision, we address the lack of transaction microdata through a novel application of Generative AI. Because ``hand-tuning'' parameters invites researchers bias and yield economically nonsensical results, we generate the missing preference data through a Discrete Choice Experiement (DCE) using Large Language Models (LLMs) as proxy respondents. By subjecting LLM agents to thousands of choice tasks, we recover internally consistent preference primitives. Specifically, the elasticities of willingness-to-pay (WTP) and willingness-to-accept (WTA) with respect to institutional variables. This method offers a transparency rarely attainable in opaque data markets. The entire protocol is auditable, the reasoning process of the agents is logged, and the resulting parameters capture complex interaction effects \citep{wang2024surveyllm, rathje2024gpt, ziems2024can, park2023llm}, such as the interplay between liability risk and enforcement intensity that are practically infeasible to estimate through human expert panels. 

We integrate these primitives into a spatially explicit ABM designed to function as a policy laboratory. Unlike static equilibrium models, an ABM is necessary here because the data market is defined by dynamic feedback loops \citep{abar2017ABM, groeneveld2017ABM, railsback2019ABM}. Our simulation maps the Chinese economy onto a lattice of 14,526 hexagonal cells, populated by hospitals as sellers and AI firms as buyers. The spacial distribution mirrors empirical industrial clusters. In this geographical sandbox, agents engage in decentralized bilateral bargaining where prices are emerge endogenously from the clash between buyer valuation and seller risk tolerance. Crucially, the model captures the dynamic co-evolution of market activity and legal oversight. For example, enforcement mechanism is not a fixed parameter, but a response to local transaction volumes. This creates a feedback loop: ``hot spots'' of trade attract stricter audits, forcing agents to continuously update their risk-adjusted reservation prices. This dynamic architecture allows us to move beyond simple throughput metrics. We evaluate rival legal regimes by calculating total welfare as the sum of consumer and producer surplus minus unpriced externalities. Finally, to demonstrate that our simulation reflects reality rather than artifact, we validate this architecture in a dedicated chapter, demonstrating that the model reproduces key stylized facts of the real-world data economy.

Our comparative analysis offers a sharp critique of the \textit{status quo}. First, we find that property-rule mechanisms, such as informed consent and anonymization carve-outs, do not expand trade volume and fail to maximize welfare. By treating certain data as ``outside the law'', these regimes remove the damages backstop necessary to discipline risky exchanges. Consequently, the market fails to internalize the objective harm like re-identification, causing aggregate welfare to fall once social costs are deducted. Second, and contrary to the dominant regulatory instinct, our model demonstrates that social welfare is maximized when liability is shifted to the buyer. While current frameworks like the General Data Protection Regulation (GDPR) are heavily seller-centric, our simulation show that assigning substantive risk to the buyer simultaneously raises welfare and market participation. This finding aligns with the ``least cost avoider'' principle. Downstream users control post-acquisition safeguards and are best positioned to mitigate risk efficiently. Shifting liability to this locus induces efficient investment in care. Thus, our work provides a empirical foundation for emerging doctrines of ``reachability,'' such as the direct liability for Business Associates under Health Insurance Portability and Accountability Act (HIPAA).

In sum, this paper offers both a methodological and a substantive contribution. Methodologically, it pioneers a pipeline for modeling opaque markets, replacing subjective priors with experimentally disciplined primitives. Substantively, it ``de-romanticizes'' seller-centric regulation, providing rigorous evidence that the path to a healthy data economy lies in making downstream buyers legally accountable. By illuminating the hidden trade-offs of data governance, we provide policymakers with the evidence required to move from guesswork to effective institutional design.

\section{Literature review}
This paper contributes to three strands of literature. First, drawing on field research, it brings renewed attention to the role of reputation in shaping data transaction mechanisms. Second, it assesses the policy implications of alternative legal and regulatory frameworks for data transactions through formal modeling and simulation. Third, in terms of methodology, the paper adopts a computational social science approach for counterfactual comparative institutional analysis, complementing---and in key respects departing from---both traditional normative analyses and standard econometric policy evaluations.

\subsection{Data market mechanisms}
Trading datasets---especially institutionally generated, regulated-core datasets---differs from trading conventional goods because the asset is nonrival and cheaply replicable, so market exchange is not primarily about allocating scarce physical units but about engineering excludability and allocating rights. In dataset markets, ``sale'' typically means a licensed transfer of access and usage rights, often with field-of-use limits, redistribution bans, term limits, and update entitlements, rather than a clean transfer of ownership. This basic non-rivalry changes both incentives and mechanism design: private sellers tend to under-share absent credible control over downstream reuse, while social value can be amplified by multi-user reuse, creating a wedge between efficient diffusion and privately sustainable trade \citep{jones2020nonrivalry}. Consequently, pricing cannot be separated from governance: the market must bundle pricing with rules and technologies that make exclusion credible and define what exactly the buyer is paying for \citep{DriessenMonsieurVanDenHeuvel2022,FernandezSubramaniamFranklin2020}.

Because the dataset is an experience good whose value is hard to verify \textit{ex ante} and highly buyer-specific, dataset trading mechanisms devote substantial attention to reducing information asymmetry without giving away the asset. Platforms and contracts implement partial revelation through standardized metadata, provenance disclosures, samples or limited previews, and service-level terms including freshness, correction obligations, and update cadence, which jointly ``productize'' a dataset into a tradable unit \citep{KennedySubramaniamGalhotraFernandez2022,DriessenMonsieurVanDenHeuvel2022}. In practice, this supports the prevalence of familiar digital-service pricing forms---one-off licenses for static archives, subscriptions for continuously refreshed datasets, and tiered plans that screen along coverage, latency, history length, and permitted uses---mechanisms that substitute for the per-unit spot pricing typical of commodity trade \citep{AzcoitiaIordanouLaoutaris2023,ZhuZhangZhangLiuRen2024}. In regulated institutional contexts, these contractual tiers are not merely commercial differentiation but also compliance-compatible implementation choices that preserve custody and auditability, making ``rights and governance'' the effective product \citep{DriessenMonsieurVanDenHeuvel2022}.

A further mechanism-design difference is that dataset markets face a distinct market-integrity problem: the main threat is not theft of a rival unit but leakage via copying, internal sharing, or misuse that destroys the seller’s ability to sustain pricing over time. This pushes dataset trade toward enforcement complements---auditing, watermarking and fingerprinting, monitoring, and legally enforceable penalties---which become core components of the mechanism rather than ancillary safeguards \citep{Fernandez2022StrategicBuyers}. Strategic behavior also appears in repeated interaction, where buyers may attempt to game platform rules or extract value in ways that undermine posted-price logic; robust mechanisms therefore emphasize enforceable licenses and governance structures that deter strategic extraction while keeping markets thick enough to function \citep{KennedySubramaniamGalhotraFernandez2022,Fernandez2022StrategicBuyers}. Taken together, the mainstream dataset-trading literature portrays a market whose defining mechanisms are license-based product differentiation, asymmetric-information mitigation through controlled preview and quality governance, and enforcement-backed excludability---features that have no close analogue in traditional goods markets, where rivalry and physical scarcity do most of the work.

However, in the literature on market mechanisms, reputation---a key factor shaping data quality assessment and exchange---has long been underexamined. Interest in reputation as a determinant of data pricing has only surged in recent years, despite its early recognition in foundational work on data quality. \citet{wang1996quality}’s seminal study identified reputation as part of consumers’ intrinsic assessment of data quality, yet subsequent research largely prioritized intrinsic, data-centric attributes such as accuracy, completeness, and timeliness. In that period, the dominant agenda was to improve the data product itself, rather than to theorize and measure external signals that condition its market value. 

Only recently has a small but growing body of scholarship explicitly incorporated reputation into data pricing frameworks \citep{majumdar2025pricing}. This renewed attention is closely linked to advances in information economics: theories of asymmetric information and signaling \citep{Akerlof1970Lemons,spence1973signaling}, especially as applied to digital platforms \citep{edelman2007auction,cabral2010price,athey2011auction}, underscore the need for trust-building mechanisms when quality is costly to verify. At the same time, the expansion of big-data markets has increased the frequency and scale of data transactions, making credibility and provenance salient determinants of willingness to pay. These developments suggest that reputation is not merely a peripheral attribute, but a central mechanism through which market participants infer quality under uncertainty. Accordingly, further empirical research is needed to develop and validate pricing models that explicitly incorporate reputation, and to assess how such mechanisms operate across different institutional and market settings. Building on fieldwork, modeling, and simulation, this paper revisits reputation as an overlooked but consequential feature of data transaction design and advances the emerging research agenda on reputation-informed data pricing.

\subsection{Legal Frameworks}\label{sec:frameworks}
This section reviews the institutional arrangements governing data transactions. To structure our analysis of these diverse governance mechanisms, we draw upon the framework established by \citet{calabresi1972property}, which distinguishes between entitlements protected by liability rules and those protected by property rules. Accordingly, we classify data institutions into two distinct categories: liability rules, which permit outcomes subject to \textit{ex post} compensation; and property rules, which rely on \textit{ex ante} permission to define the validity and boundaries of the data entitlement.

\subsubsection{Liability Rules}
In the tradition of \citet{calabresi1972property}, liability rules govern the data economy by establishing a pricing mechanism for non-compliance and harm rather than forbidding the activity outright. Under these regimes, the primary focus is on allocating the costs of \textit{ex post} violations, such as data breaches or misuse, to the appropriate party, thereby allowing the market to value the risk.

\textbf{1. Controller-centric liability (Baseline).} 
The default regime in major jurisdictions operates as a controller-centric liability baseline. Under this model, data exchange is permitted, but unlawful processing is deterred and priced through \textit{ex post} compensation claims and administrative enforcement. Primary legal exposure attaches to the ``controller''—the actor that determines the purposes and means of processing. This structure is designed to force the entity with the greatest oversight capacity to internalize the expected costs of privacy harms and regulatory non-compliance. Doctrinally, this maps closely to the GDPR. Data subjects hold a right to compensation for material or non-material damage caused by infringing processing under Article 82, creating a private liability channel. Simultaneously, supervisory authorities may impose administrative fines under Article 83—scaling to worldwide annual turnover—alongside Member State penalties under Article 84. These mechanisms collectively function as a pricing system for legal risk. Recent enforcement practice in China illustrates the same economic logic. The Personal Information Protection Law (PIPL) authorizes fines of up to 5\% of the preceding year’s turnover for serious violations, thereby encouraging regulated entities to factor expected regulatory costs into their pricing and compliance choices.\footnote{Personal Information Protection Law of the People's Republic of China (PIPL), Art. 66. \url{https://www.pkulaw.com/en_law/d653ed619d0961c0bdfb.html}.} The administrative penalty against Didi Global in July 2022, which imposed an \$1.2 billion fine alongside personal penalties for executives, serves as a market signal that large-scale data compliance failures generate material, turnover-sized exposure for data-intensive firms.\footnote{Cyberspace Administration of China, \textit{Decision on the Administrative Penalty of Didi Global Inc.} (July 21, 2022). \href{https://www.reuters.com/technology/china-fines-didi-global-12-bln-violating-data-security-laws-2022-07-21/}{\texttt{https://www.reuters.com/technology/china\-fines\-didi\-global\-12\-bln\-violating\-data\-security\-laws\-2022-07-21/}}.}

The logic of controller-centric liability extends beyond specific data statutes, it represents a family of attribution rules designed to mobilize organizational governance. In the classic tort setting, modern scholarship treats vicarious liability as a doctrine keyed to the enterprise’s creation of a foreseeable risk environment. By organizing work and benefiting from the activity, the controller is best positioned to reduce risks through systemic precautions and to spread residual losses through insurance \citep{Geistfeld2024Reformulating}. This framework combines deterrence-through-control—shifting incentives to the party best able to redesign the system—with the administrability of targeting a solvent defendant rather than diffuse individual wrongdoers. This strategy is also evident in corporate criminal law and financial regulation, where liability is assigned to gatekeepers or principals to induce monitoring. The broad \textit{respondeat superior} rule in corporate crime is defended precisely as a mechanism for inducing firms to invest in internal controls, shifting the law enforcement burden onto the entity that can most plausibly prevent misconduct at scale \citep{Luskin2020DueCare}. Similarly, gatekeeper liability assigns duties to intermediaries who occupy chokepoints capable of screening misconduct more reliably than direct regulation \citep{Kraakman1986Gatekeepers}. In this sense, the data controller is the modern digital gatekeeper: the law assigns liability to this node not merely based on moral culpability, but to convert private organizational control into public prevention capacity.

\textbf{2. Buyer-shared risk.}
Distinct from the baseline is the buyer-shared risk regime. Here, the economic incidence of expected harm is allocated across both buyer and seller through private contract, even where public law supplies a mandatory liability floor. This architecture is visible in U.S. health data practice, where HIPAA conditions the disclosure of protected health information on obtaining written ``satisfactory assurances'' through a Business Associate Agreement (BAA). Simultaneously, regulations recognize the direct liability of business associates (buyers) for specific obligations.\footnote{See 45 C.F.R. \S\S 164.502(e), 164.504(e). For the Office for Civil Rights (OCR) statement on direct liability under the HITECH Act, see HHS/OCR, \textit{Direct Liability of Business Associates} (2019).} This dual structure encourages parties to explicitly price downstream compliance and breach costs into the transaction through negotiated indemnities and monitoring rights. Analytically, buyer-shared risk operates as a rule-in-use that refines the coarse allocations of public statutes. While public law sets minimum duties and sanctions, private ordering reallocates the marginal expected loss to the party best positioned to invest in downstream safeguards. This logic is consistent with doctrinal accounts of controller-processor liability under EU law \citep{VanAlsenoy2016} and empirical evidence that data-service contracting systematically embeds privacy allocations into standard terms \citep{KamarinouMillardHon2015}. Economically, this form of indemnification preserves precaution incentives: by shifting liability to the downstream user—paired with termination rights—the contract ensures that the actor controlling the risk environment faces the marginal cost of their own negligence \citep{GuoHuaJiang2022}.

To operationalize this allocation, modern commercial agreements employ a sophisticated architecture of ``liability laddering.'' Rather than leaving loss attribution to open-ended tort principles, sophisticated parties construct a tiered stack of caps, baskets, and exclusions. This converts uncertain regulatory risk into a priced and partially insurable set of contingent obligations, allowing trade to proceed despite residual uncertainty \citep{Stanton2025RiskReimagined}. This phenomenon explains the increasing prevalence of ``contractual depth,'' where parties draft dense, modular agreements to distribute risk across multiple audiences (counterparties, insurers, regulators) and to separate routine compliance issues from catastrophic tail risks \citep{HwangJennejohn2022ContractualDepth}. Functionally, these devices serve to make data transactions priceable. By allocating slices of downside risk, they support valuation and closing even under imperfect verification. Furthermore, they protect the negotiated risk split against doctrinal ``end-runs,'' preventing disappointed buyers from recasting bargained-for warranty risks as extra-contractual fraud claims \citep{WestLewis2009ExtraContractual}. Empirically, these terms are not merely bespoke craftsmanship; they evolve as semi-standardized boilerplate in repeat-play markets, where legal practitioners iterate clause design to minimize agency costs and manage recurring disputes over liability caps \citep{ChoiGulatiScott2021ContractProductionProcess}.

\textbf{3. Joint liability.}
The third is the joint-liability design. Where buyer-shared risk relies on private ordering to distribute loss, joint liability makes both sides of the transaction legally reachable—whether in civil, administrative, or criminal forums—setting the expected enforcement price on a two-sided basis. The core economic rationale for this regime is informational. In clandestine or socially harmful trades that are privately beneficial to both counterparties, single-sided enforcement creates a perverse incentive: the non-liable party can be effectively ``deputized'' as a shield to help conceal the transaction. Classic law and economics work on victimless crimes and price-control evasion formalizes this dynamic: when only one party is exposed, the counterpart can bargain over concealment, thereby reducing the effective expected sanction. By contrast, two-sided exposure weakens these mutual concealment incentives and raises the probability-adjusted cost of transacting \citep{LottRoberts1989}. In this sense, joint liability functions less as a mechanism for \textit{ex post} compensation and more as a lever for reshaping \textit{ex ante} trade feasibility by placing legal downside on both margins of the bargain.

In practice, joint-liability regimes are most salient where bilateral surplus coexists with negative externalities and where the state faces detection constraints—conditions characteristic of corruption, collusive exchange, and opaque data markets. A large modern literature treats two-sided reachability as a design platform that can be tuned via asymmetric penalties and leniency. In the context of bribery, for instance, shifting exposure across both the giver and taker creates whistleblowing incentives that substitute for costly public detection, altering the endogenous complexity of the regulatory environment \citep{HuOak2023JEMS,HuOak2023JEBO}. More generally, analyses of leniency in illegal transactions demonstrate how enforcement available against both sides—when paired with self-reporting discounts—destabilizes trust, effectively deterring illegal trade relationships by introducing a ``lemons'' problem into the mechanism of collusion \citep{BuccirossiSpagnolo2006}.

\subsubsection{Property rules}
Contrastingly, institutions functioning as property rules grant the entitlement holder a right to exclude, enforceable via \textit{ex ante} barriers. These mechanisms define the absolute boundaries of the market: transactions are legally valid only if the entitlement holder grants permission or if the subject matter falls outside the protected scope.

\textbf{4. Informed consent.}
Informed consent functions as the canonical \textit{ex ante} authorization rule. In the standard law and economics vocabulary, it tracks a property-rule protection. It grants the entitlement-holder a veto, shifting the legal default from ``may proceed unless later priced'' to ``may proceed only if permission is secured'' \citep{calabresi1972property}. Functionally, this rule is deployed where the legal system treats the underlying interest as sovereignty-like—such as bodily integrity or intimate decision-making—and where legitimacy requires that the affected party be able to say ``no'' without the burden of proving harm \textit{ex post}. Beyond its role as a veto, consent acts as an information-forcing device. By making authorization contingent on disclosure, it attempts to reshape the choice architecture upstream rather than relying purely on downstream deterrence. However, modern accounts stress that ``informed'' consent is not merely a signature or a disclosure checklist. It is a structured decision procedure whose design choices determine the actual autonomy delivered.

A critical conceptual move in this literature is the decoupling of the disclosure duty from the understanding requirement. As \citet{MillumBromwich2021} argue, these requirements rest on distinct normative grounds, implying that the institutional validity of consent cannot be assessed solely by comprehension deficits. This tension is visible in research ethics reforms, where regulators have attempted to operationalize valid consent by mandating salient, front-loaded ``key information,'' recognizing that ever-longer disclosure forms paradoxically undermine comprehension and the rule's justificatory logic \citep{BazzanoEtAl2021}. At the margins, the doctrine reveals a recurring tension between autonomy and welfare-protective paternalism. Doctrines such as therapeutic privilege illustrate the system's hesitancy to grant absolute veto power in all contexts, and recent analyses highlight the conceptual instability of expanding such carve-outs to borderline-capacity cases \citep{MenonEtAl2021}. Ultimately, consent doctrine serves as the site for specifying what counts as ``material'' to decision-making—a standard that is constantly renegotiated as novel intermediaries and decision tools enter the market \citep{Cohen2020}.

\textbf{5. Low-risk carve-outs.}
Alongside gatekeeping rules like consent, governance systems rely on a structural boundary mechanism: defining certain categories of data as sufficiently low-risk to fall outside the core entitlement entirely. In the property-rule framework, this functions as a definition of the \textit{res}: it determines the subject matter to which the property right attaches. In the EU, the doctrinal anchor is the distinction between personal and anonymous data under GDPR Recital 26. The carve-out operates by treating the residual probability of identification as acceptably remote, effectively converting anonymization from a mere compliance technique into a gatekeeping technology for scope exclusion \citep{FinckPallas2020,WeitzenboeckEtAl2022}. Recent scholarship emphasizes that this boundary is best understood through the lens of risk management: the legal question is not whether data is metaphysically ``about'' a person, but whether a specific processing context generates rights-relevant risks sufficient to trigger the GDPR’s apparatus \citep{RuppVonGrafenstein2024}. On this view, low-risk carve-outs perform a system-level function that baseline pricing cannot achieve: they create administrable ``zones of liberty'' that enable routine downstream reuse—such as analytics and research—without the friction of continuously recalculating expected liability.

The same logic appears in U.S. sectoral regimes that treat de-identification as a regulatory off-ramp. HIPAA confirms that once information is de-identified through defined pathways (Safe Harbor or Expert Determination), it effectively exits the statute’s authorization constraints. This converts de-identification into a legal switch that removes the transaction from the consent-and-restriction frame entirely \citep{Evans2023HIPAA}. However, the literature highlights a core tension in this design: because re-identification risk is probabilistic and environment-dependent, checklist-style safe harbors inevitably become either over-inclusive (under-protective) or under-inclusive (functionally unusable). This fragility pushes jurisdictions toward more explicitly risk-based, contextual tests and robust technical governance to prevent downstream linkage \citep{FinckPallas2020,WeitzenboeckEtAl2022}. In comparative terms, low-risk carve-outs operate as an \textit{ex ante} complement to enforcement-based pricing: instead of deterring unlawful processing by raising expected sanctions, they redraw the perimeter of the market so that low-risk uses can proceed with minimal legal friction.

\textbf{6. Provider immunization.} 
Finally, a more aggressive design seeks to lower the ``enforcement price'' of exchange by immunizing the provider itself. In the \citet{calabresi1972property} framework, this functions as a shift in the entitlement. Rather than protecting the victim with a liability rule (damages), the law grants the injurer (provider) a property-rule entitlement to act without fear of \textit{ex post} compensation, provided specific conditions are met. In practice, this design functions as a conditional safe harbor. It converts uncertain tort-and-regulatory exposure into a rules-based compliance gate, encouraging investment in sharing infrastructures that would otherwise be chilled by litigation risk \citep{Bramble2013SafeHarbors}. Such regimes are most common where policymakers believe under-sharing produces negative societal externalities, notably in the context of cybersecurity threat intelligence. For example, the U.S. Cybersecurity Information Sharing Act (CISA) provides liability shields to firms that share cyber-threat indicators, prioritizing the rapid diffusion of security data over the potential privacy harms to individuals contained within that data \citep{Lathrop2020InadequaciesCISA,SchwarczWolffWoods2023Privilege}.

The core functional trade-off is that immunization is intentionally asymmetric. It promotes information flow by suppressing one channel of accountability—private liability—and attempts to backfill deterrence through procedural conditions (e.g., data scrubbing requirements). The literature highlights the calibration challenge here: if the safe harbor is too narrow or compliance-heavy, it fails to unlock supply; if too broad, it erodes the baseline incentives to prevent harm and invest in care, effectively subsidizing data flow by externalizing risk onto third parties \citep{Lathrop2020InadequaciesCISA,SchwarczWolffWoods2023Privilege}. Empirical management scholarship confirms that liability protection acts as a powerful lever for shifting firm behavior, supporting the intuition that immunization changes the fundamental incentives of production rather than merely rearranging the private allocation of loss \citep{YangKwonLee2020ImpactInfoSharing}.

\subsection{Prevailing methods}
Current scholarship on data and privacy law remains largely anchored in traditional doctrinal approaches that focus on legal interpretation and regulatory analysis. The legal community places a clear emphasis on the unique methodologies of doctrinal research and the conceptual systems that define its discourse. As a result, most studies in this area rely primarily on qualitative descriptions and speculative reasoning. This is especially evident in data-related legal studies, where approaches that treat established legal frameworks as the core of analysis are predominant. Scholars often unconsciously strive to apply existing legal concepts to new issues, attempting to integrate emerging problems into well-established doctrinal systems \citep{dai2019horse}. A notable manifestation of this tendency is the overemphasis on devising ways to confer legal rights to the distributional consequences of new technologies \citep{wang2012information,wang2013information,zhang2023data}. However, at its core, data functions as a fundamental means of production within the internet-based production model. In this context, efforts to construct legal rights through traditional formal frameworks are increasingly inadequate in addressing the profound challenges data privacy poses to society \citep{dai2019expansion}.

In recent years, there has been a gradual increase in quantitative research within interdisciplinary law and economics, as well as in empirical legal studies journals. However, mainstream quantitative analyses still predominantly rely on econometric techniques derived from classical statistics \citep{Goldsmith2002empirical,Eisenberg2010els}. In the domain of data law, much of this research focuses on evaluating policy measures aimed at establishing data exchanges. These studies often treat the establishment of data exchanges as a proxy for regulation, employing difference-in-differences (DID) designs to estimate the effects on various economic indicators. 
However, China’s policy pilot programs often exhibit strong positive selectivity \citep{wang2025policy}, with data exchanges frequently being established in regions where the digital economy is more developed. This undermines two of the three critical identification assumptions necessary for a DID causal inference: the parallel trends assumption and the no anticipation assumption. Additionally, the spatial spillover effects within the digital economy raise doubts about whether the stable unit treatment value assumption (SUTVA) holds. These issues have led to significant flaws in the model specifications of many studies in this field \citep{yang2021data,Liu2022data,zheng2023data,wang2024data,zhou2024data}.\footnote{These papers are sourced from leading Chinese journals, and the dynamic effect diagrams of event study specifications in these studies often display pronounced growth trends prior to the event.} As a result, the effectiveness of this strand of literature remains relatively limited.

To date, few works have examined the decision-making mechanisms and market behaviors of data sellers and buyers within the framework of computational social science---particularly through the use of agent-based modeling and simulation---to explore how legal rules interact with market dynamics. Some scholars have incorporated theories and methods from evolutionary game theory, social network analysis, and related fields into multi-agent systems to simulate complex social phenomena such as group decision-making \citep{Fernández2023land,Sen2025move}. Although this line of research generally lies outside the domain of legal studies, the models and methodologies developed therein provide valuable methodological references for examining law-related behavioral and institutional dynamics.

In fact, multi-agent systems (MAS), as a core paradigm within distributed artificial intelligence, provide an exceptionally suitable modeling framework for the simulation and analysis of law and economics problems \citep{shoham2009mas,weiss1999multiagent}. An agent refers to a software entity endowed with autonomy, responsiveness, rational reasoning (or bounded rationality), and social interaction capabilities, typically modeled as pursuing the maximization (or satisficing) of its own utility \citep{wooldridge1995intelligent,jennings2000agent}. A multi-agent system consists of multiple such agents that communicate and interact with one another, generating patterns of cooperation (and conflict) while also dynamically engaging with their surrounding environment \citep{stone2000survey,wooldridge2009imas}. Through these interactions, agents both shape and adapt to their environment, leading to processes of mutual adaptation and co-evolution among agents and between agents and their environment \citep{millerpage2007cas,tesfatsion2006ace}.
In recent years, MAS, and more broadly, agent-based modeling, has been increasingly adopted across international academia. Beyond the natural sciences and engineering, it has found wide application in social, economic, and military domains to simulate human behavior, conceptual change, and the emergence and evolution of cooperative relationships \citep{epstein1996growing,macy2002factors,tesfatsion2006ace,ilachinski2004artificial}. Fundamentally, multi-agent systems are distributed rather than centralized: no single authoritative entity governs the interactions among agents. Instead, coordination often emerges through decentralized communication and interaction, leading to self-organization and self-evolution \citep{bonabeau1999swarm,stone2000survey,shoham2009mas}. This decentralized structure resonates with the civil law principle of private autonomy.


At the same time, emerging legal issues such as data transactions, often characterized by regulatory uncertainty or even ``illegal emergence’’ \citep{ling2024dynamics}, lack clear normative guidance, rendering the strategic behavior of actors particularly crucial in shaping market outcomes. Although distributed, multi-agent systems are not fragmented. Their design and study seek to organize multiple agents in an integrated manner by specifying behavioral rules, interaction protocols, and coordination mechanisms, enabling collective problem-solving beyond the capacity of any individual agent or linear aggregation \citep{wooldridge2009imas,shoham2009mas}. Moreover, the growing literature on normative multiagent systems explicitly studies how norms (rules, sanctions, compliance, and institutional constraints) can be represented, communicated, and enforced in MAS \citep{boella2006normative}, which is especially congenial to legal-institutional analysis. This structure bears a striking resemblance to the market mechanism emphasized in microeconomics and the agent-based computational economics tradition \citep{tesfatsion2006ace}. Accordingly, multi-agent systems provide innovative perspectives, models, and methodologies for advancing research in law and economics.

Just as mainstream quantitative research methods in the social sciences must primarily rely on quasi-natural experiments aimed at causal inference---supplemented, at most, by rigorously ethically reviewed randomized controlled trials---rather than the large-scale laboratory experiments typical of the natural sciences, empirical examination of how legal systems shape data trading markets cannot be directly tested in the real world. On the one hand, modifications to legal rules or their application must adhere to strict constitutional and judicial procedures; on the other, legal norms and judicial authority carry profound societal implications and cannot be altered lightly. As a result, exploring complex social science questions---such as how legal institutions influence the structure and efficiency of data markets---proves difficult within conventional experimental paradigms. Conducting controlled, replicable experiments in real-world legal contexts is often impractical, unethical, or prohibitively costly. Consequently, mainstream legal research has tended to rely on qualitative reasoning and speculative analysis, complemented by quantitative studies grounded in econometric or behavioral approaches.

This paper seeks to ``return'' to the paradigm of experimental science by employing agent-based artificial intelligence models and computer-based simulations to construct a virtual experimental environment. Within this computational framework, \textit{ceteris paribus} conditions can be strictly maintained, allowing for the parallel testing of different legal systems and liability allocation schemes to observe their long-term impacts on data markets and social welfare. Through this approach, the paper aims to provide both conceptual and computational foundations for the design of legal regulatory frameworks and the prediction of evolutionary trends in data transactions.

\section{Background}
China’s current push to make data circulate is inseparable from the recent boom in foundation-model AI. As large models moved from research into products and industrial policy, the binding constraint became not only compute, but access to usable, high-quality, legally safe datasets at scale. This linkage is now explicit in official policy signaling. China’s 2025 government work report prepared for the National People’s Congress highlighted support for large-scale AI models while also calling to improve the basic data system and facilitate cross-border data flows, effectively pairing AI ambition with data-governance capacity-building.\footnote{Reuters, ``China says it will increase support for AI, science and tech innovation,'' 2025, \href{https://www.reuters.com/technology/china-says-it-will-increase-support-ai-science-tech-innovation-2025-03-05/}{\texttt{https://www.reuters.com/technology/china\-says\-it\-will-increase\-support\-ai\-science\-tech\-innovation\-2025\-03\-05/}}.}

Regulatory design has also begun to treat generative AI as a data-governance problem. The Cyberspace Administration of China and six other agencies issued the Interim Measures for the Management of Generative AI Services in July 2023, which frame the policy objective as promoting development while safeguarding public interests, and place lifecycle safety responsibilities on providers—centrally including obligations around data security and personal information protection in the provision of public-facing generative AI services.\footnote{Cyberspace Administration of China and Other Agencies, ``Interim Measures for the Administration of Generative Artificial Intelligence Services,'' issued July 2023, effective August 15, 2023, \url{https://www.pkulaw.com/en_law/6dc227b9153496c2bdfb.html}.} Read together, the AI surge and the emerging AI-specific compliance perimeter make the state’s enthusiasm for ``data flow'' legible: data are simultaneously the key input into model capability and a politically and legally sensitive object whose mishandling can trigger public backlash, administrative accountability, and enforcement risk.

Against this background, China has pursued an unusually explicit market-building agenda for data since 2021. The architecture has three visible layers. First, a dense compliance perimeter was consolidated around personal information and data security, with the Personal Information Protection Law (PIPL) supplying a high-penalty ceiling for serious violations.\footnote{Personal Information Protection Law of the People's Republic of China, 2021, \url{https://www.pkulaw.com/en_law/d653ed619d0961c0bdfb.html}.} Second, the central government elevated ``data as a factor of production'' into a national economic strategy and called for a basic institutional system that would support circulation, trading, and distribution of data-derived value.\footnote{The State Council of the People's Republic of China, ``China unveils measures to build basic systems for data,'' Dec 19, 2022, \url{https://english.www.gov.cn/policies/latestreleases/202212/19/content_WS63a17f7dc6d0a757729e49bd.html}; CPC Central Commitee and State Council of the People's Republic of China, ``Opinions on Establishing a Data Base System to Maximize a Better Role of Data Elements,'' Dec 2, 2022, \url{https://www.pkulaw.com/en_law/1b51343d19be0d11bdfb.html}.} Third, China expanded state capacity for coordination by inaugurating the National Data Administration in October 2023\footnote{The State Council of the People's Republic of China, ``China inaugurates national data administration,'' Oct 25, 2023, \url{https://english.www.gov.cn/news/202310/25/content_WS6538ae0dc6d0868f4e8e0a30.html}.} and subsequently issuing cross-ministerial action plans to operationalize ``data elements'' across sectors.\footnote{National Data Administration and other ministries, ``Notice by the National Data Bureau and Other Ministries and Commissions of Issuing the Three-Year Action Plan (2024-2026) for ``Data Elements X'','' No. 11 [2023] of the National Data Bureau, 2023, \url{https://www.pkulaw.com/en_law/8ffb95c3f760188ebdfb.html}.}

Within this policy frame, local governments and state-affiliated actors rapidly built transaction infrastructure. By late 2023, official reporting indicated that up to 48 data exchanges had been established nationwide. These exchanges are commonly tasked with market-design functions that resemble ``plumbing'', standardizing trading rules, vetting counterparties, and building security and governance mechanisms that make transactions legible to regulators and auditable after the fact.\footnote{The State Council of the People's Republic of China, ``Data trading expected to play bigger role,'' Nov 27, 2023, \url{https://english.www.gov.cn/news/202311/27/content_WS6563e9f9c6d0868f4e8e1a95.html}.} The Shanghai Data Exchange (SDE) has been one of the most prominent examples, and recent public communications emphasize efforts to extend exchange functions toward data-asset trading arrangements rather than mere cataloguing.\footnote{Shanghai Municipal People's Government, ``SDE officially launches data asset trading market,'' July 1, 2024, \url{https://english.shanghai.gov.cn/en-Latest-WhatsNew/20240701/0e47838b58ec4e109854df56c0e8b1bf.html}.}

However, the same landscape also exhibits a persistent tension that motivates this article. Exchange-building is not the same as deal completion, especially for institutional, high-stakes datasets such as clinical and financial data. Much of what is prominently listed and traded via exchanges is better described as data products, data services, or packaged access rights rather than raw, high-risk institutional datasets that would materially shift an AI developer’s training frontier. Official and semi-official descriptions of exchanges emphasize standardization, controllability, traceability, and safe trading environments.\footnote{Cooley LLP, ``PRC's New Efforts to Facilitate Data Trading: Shanghai Data Exchange Kicks Off Trading,'' Jan 12, 2022, \href{https://cdp.cooley.com/prcs-new-efforts-facilitate-data-trading-shanghai-data-exchange-kicks-off-trading/}{\texttt{https://cdp.cooley.com/prcs\-new\-efforts\-facilitate\-data\-trading\-shanghai\-data\-exchange\-kicks\-off\-trading/}}; Shanghai Municipal People's Government, ``SDE officially launches data asset trading market,'' July 1, 2024, \url{https://english.shanghai.gov.cn/en-Latest-WhatsNew/20240701/0e47838b58ec4e109854df56c0e8b1bf.html}.} However, our fieldwork in the exchange ecosystem (introduced in the next part of this chapter) aligns with a widely observed practical pattern: the market segment that attracts the strongest policy attention is precisely the segment where actors are most reluctant to transact.

A central reason is that the binding constraints in high-stakes data transactions are not merely informational or contractual, but are institutional and accountability-related. Even when parties can draft contracts, high-stakes sellers often face expected losses dominated by regulatory and reputational downside risk and by internal administrative accountability. PIPL’s sanction ladder makes the tail risk salient: in serious cases, penalties can reach up to RMB 50 million (\$7 million) or 5\% of annual turnover, accompanied by potential suspension of services and personal liability for responsible individuals.\footnote{Personal Information Protection Law of the People's Republic of China, 2021, \url{https://www.pkulaw.com/en_law/d653ed619d0961c0bdfb.html}.} In such an environment, a seller’s decision calculus is shaped not only by expected revenues but also by whether governance procedures, approvals, and downstream control are sufficiently credible to justify a signature and an audit trail. This helps explain why exchange platforms can be simultaneously visible as policy symbols and limited as clearing venues for the most valuable institutional datasets. Exchanges can subsidize standardization and provide compliance-facing services, but they do not, by themselves, resolve liability placement or create enforcement predictability.

This article is informed by multi-sited fieldwork conducted between 2022 and 2025, designed to identify the micro-level mechanisms that determine whether high-stakes data transactions in China actually clear. Public-facing exchange listings and aggregate indicators of platform activity reveal little about where deals fail or succeed, because the binding constraints often sit inside organizations and negotiation processes: who is willing and authorized to sign, how responsibility is allocated, what approvals are required, and which controls are regarded as credible. To study these margins, we draw on three complementary sources: participant observation inside an exchange ecosystem, transaction-level exposure through advising two AI start-ups, and semi-structured interviews across market participants \citep{EmersonFretzShaw2011Fieldnotes,KvaleBrinkmann2009InterViews, Collier2011ProcessTracing}. Throughout, we adopt a strict confidentiality posture. We do not disclose names or organizations and paraphrase potentially identifying details, because many respondents remain employed in the industry and spoke candidly on the understanding that their disclosures would not be traceable back to them.

From September to December 2023, one of the authors worked full-time as a researcher in the research institute affiliated with the Shanghai Data Exchange. The daily work included drafting rule proposals, participating in internal meetings, and interviewing market participants to diagnose why transactions were not closing. During this period, the author also completed a white paper based on internal discussions and interview-derived diagnostics. Materials generated in this window include contemporaneous meeting notes, iterative white paper drafts, and summaries of interviews with exchange-adjacent staff and counterparties.

A core stylized fact from this window is that completed transactions in the high-stakes segment were scarce. The most frequent content of discussions was not how to clear a steady pipeline of institutional data trades, but why buyers and sellers were reluctant to transact at all and why negotiations often stalled after substantial time was spent. This provides an empirical anchor for the paper’s central premise: exchanges can be prominent as policy objects while remaining limited as clearing venues for the most valuable and most legally sensitive datasets.

The author also served as legal advisor to two AI start-ups, which provided transaction-level exposure to how data are acquired when high-stakes exchange transactions do not clear. The first start-up operates in AI medical imaging and sought access to raw medical images with high metadata completeness. In this setting, the dominant bottleneck was trust building with hospital department heads who effectively controlled access, rather than the mechanical drafting of contracts. The contractual form most commonly used or proposed was framed as a research collaboration agreement, a structure intended to make the arrangement legible as research cooperation rather than a straightforward sale. This contractual choice is itself a micro-mechanism: it reflects how parties adapt transaction form to perceived compliance and accountability exposure. The second start-up operates in AI image generation and negotiated for platform-facing data access, typically via API access or bulk datasets. A salient friction in these negotiations was the platform’s concern that offered prices did not compensate prior costs (including acquisition, organization, and governance costs) associated with building and maintaining the underlying content repository. This window is used in the paper as complementary evidence that willingness to supply data depends on outside options and perceived opportunity costs, not merely on transfer costs.

Between 2022 and 2025, the author conducted over 30 semi-structured interviews with participants in China’s data economy, approximately half by phone and half in person. Interviews typically lasted one to three hours. Recruitment combined targeted outreach with network-based access through college-mate professional connections, which proved important for reaching practitioners who control operational knowledge of data acquisition and compliance. Interviews were recorded and transcribed, followed by analytic memos that distilled recurring mechanisms and cross-role contrasts. The interview sample spans: (i) data buyers (including large technology firms and small AI start-ups), (ii) intermediaries and compliance consultancies, (iii) exchange staff, and (iv) personnel in bank data departments and other institutional data holders. These interviews are used to triangulate claims from the other two windows and to discipline the mapping from qualitative mechanisms to model primitives.

The next chapter formalizes a market environment in which buyers and sellers face heterogeneous valuations, costs, and constraints. The fieldwork motivates treating these elements as first-order rather than residual. In particular:

\begin{itemize}[leftmargin=2em]
    \item High-stakes transactions are shaped by downside exposure and accountability, motivating primitives that distinguish intrinsic transaction risk from the enforcement environment that determines expected loss.
    \item Seller tier matters not only as a proxy for data capability, but also for governance overhead and sign-off burdens, motivating a seller-tier term that shifts reservation prices even holding volume and risk fixed.
    \item Buyer heterogeneity is visible in both capability (ability to extract value from data) and constraints (ability to sustain long negotiation cycles and cumulative spending), motivating buyer-tier and budget constraints in the model.
    \item Geographic separation raises the real cost of persuasion and coordination in practice (travel, repeated meetings, relationship building), motivating a distance-related term that depresses expected utility.
    \item Transaction form selection (for example, research-collaboration framing) is used to manage perceived compliance and accountability exposure, motivating modeling choices that do not presume frictionless spot-market exchange.
\end{itemize}

These three windows therefore supply the empirical basis for the primitives introduced in the next chapter and help ensure that the simulation tests institutions against mechanisms that practitioners actually confront.

\section{The model primitives}
This chapter operationalizes the preceding fieldwork as model primitives. Rather than treating the agent payoffs as purely theoretical, we translate the micro-level mechanisms observed in the Chinese high-stakes data environment, how parties assess value under limited verifiability, where negotiations stall, and which constraints dominate clearance, into a parsimonious set of state variables and parameters that can be varied in counterfactual institutional simulations. The chapter proceeds in two subsections: we first specify the buyer-side utility and willingness-to-pay, and then the seller-side utility and willingness-to-accept.

\subsection{Buyers}
We propose a buyer's WTP incorporates multiple factors, ensuring each element is grounded in economic theory and empirical findings. Following the literature on random utility maximization (RUM) framework \citep{McFadden1974logit,BLP1995demand,Train1998demand,McFadden2000mnl}, we set a mixed (random-coefficients) logit model with quasi-linear utility assumption
\begin{equation}\label{eq:buyer_utility}
\begin{split}
    U^{\text{Buyer}}_{ijt}= & f(x_{it})\cdot \beta x_{j}+\tau s_{j}+\gamma z_{i}+\phi(s_{j}\times z_{i})\\
     & -\alpha_{i}p_{ijt}-\kappa\ln(1+d_{ij})+\varepsilon_{ijt}
\end{split}
\end{equation}
 on buyer $i$'s utility $U^{Buyer}_{ijt}$ on dataset $x_j$ provided by seller $j$ at time $t$, with random coefficient $\alpha_{i}$ captures how the utility of dataset $x_j$ for buyer $i$ depends on the price $p_{ijt}$. $s_j$, $x_j$, $z_i$, and $d_{ij}$ represent observable features.

\subsubsection{Data quantity and diminishing marginal value}
A central driver of buyer willingness-to-pay is the quantity of usable data, but with diminishing marginal value as the buyer’s internal stock grows. This assumption is consistent with classic diminishing-returns logic and with empirical ``scaling law'' evidence in machine learning: performance improves predictably with more training data, yet the incremental gain from additional data shrinks as scale increases \citep{Hestness2017ScalingPredictable,Kaplan2020ScalingLaws,Hoffmann2022scal}. It is also strongly aligned with our fieldwork in data procurement for AI. Early-stage startups are typically data-scarce and operate in the steep region of the learning curve, where each additional tranche of images or studies materially improves development and validation. By contrast, mature enterprises with large in-house repositories face a flatter region of the curve and therefore exhibit a lower marginal willingness-to-pay for ``more of the same.''

In our application, $x_{j,t}$ denotes the number of data offered by seller $j$ at time $t$, and $x_i$ denotes the buyer’s in-house stock of usable data. We model the quantity benefit as $f(x_i)\cdot x_{j,t}$, where $f(x_i)$ is decreasing in $x_i$ (e.g., $f(x_i)=e^{-\rho x_i}$). Economically, $f(x_i)$ captures the buyer’s diminishing marginal value of additional volume once internal data holdings reach scale. Substantively, it maps onto two fieldwork-regularities. First, buyers commonly adopt a pilot-then-scale strategy. They start with a smaller dataset to test usability, internal integration, and model lift, and only then expand purchase volume. A concave $f(x_i)$ rationalizes this staged procurement as an optimal response to diminishing marginal returns under uncertainty. Second, in medical AI the ``binding constraint'' is often not raw volume, but coverage of rare and long-tail cases and the associated annotation or curation burden. Reviews of medical AI note that data scarcity persists precisely because rare conditions, costly labeling, and restricted population coverage limit the availability of clinically valuable samples \citep{Groger2025DataScarcity}. Moreover, large radiology datasets exhibit extremely long-tailed disease distributions, with some diagnostic categories having very few cases \citep{Zheng2024LongTailedRadiology}. Consistent with our interviews, this implies that adding another 10{,}000 routine cases from a low-tier source may deliver little incremental value if it does not expand long-tail coverage. By contrast, additional volume from top-tier institutions that concentrate diverse and rare cases can have much higher marginal value. Our specification keeps this idea within the quantity term in a reviewer-friendly way: $x_{j,t}$ measures the scale of usable data, while $f(x_i)$ ensures that the same quantity is valued more by data-scarce buyers and less by data-saturated ones.

\subsubsection{Seller's institutional strength}
A second core determinant of buyer willingness-to-pay is the seller’s institutional tier $s_j$, which we interpret as a market signal of expected data quality under severe pre-purchase valuation frictions. Data is an experience good. Its usefulness for model development depends on downstream cleaning, metadata completeness, protocol consistency, and task-specific fit. All attributes that determine the usefulness of a dataset, protocol consistency, metadata completeness, provenance, and long-tail coverage, are costly to verify \textit{ex ante}. This creates the classic ``inspection'' or ``information'' paradox: the buyer cannot fully evaluate the value of information without access, yet once the information is disclosed the seller’s bargaining position weakens \citep{Arrow1962Invention}. Importantly, in data transactions the resulting difficulty is often not a textbook lemons market in which sellers privately know quality while buyers do not. Instead, both sides face substantial valuation uncertainty. Sellers frequently do not know the dataset’s exact use value for a given buyer, and buyers cannot reliably infer that use value from surface attributes alone. In this environment, trade frictions arise because exchange requires contracting over information whose value is endogenously revealed only through use and integration \citep{JanssenRoy2023InfoUncertainty}, generating systematic price premia for trusted sources \citep{Cabral2012ReputationInternet}. 

Our fieldwork shows that buyers respond to this two-sided uncertainty by relying on institution-level signals rather than attempting to ``price the dataset'' directly. In practice, buyers largely \emph{price the dataset by pricing the seller}: observable seller tier (e.g., hospital rank) shifts beliefs about expected quality and reduces perceived variance about hidden dimensions such as protocol standardization, provenance integrity, and long-tail coverage. Consistent with signaling and reputation theory, such observable seller attributes can command systematic premia precisely when intrinsic quality is hard to verify \citep{Cabral2012ReputationInternet,Tadelis2016ReputationFeedback}.

This heuristic is not mere brand psychology. It is tethered to concrete quality channels that repeatedly surfaced in our interviews. First, hospital rank is strongly associated (in practitioners’ beliefs and procurement practice) with standardized imaging protocols and better equipment, which directly affects signal-to-noise properties and cross-site comparability. Second, higher-tier institutions are more likely to provide richer metadata and more reliable provenance (fewer missing fields, clearer cohort definitions, traceable extraction logic), which matters because data-quality dimensions such as completeness, consistency, and accuracy systematically influence machine-learning performance \citep{Mohammed2025DataQuality}. Third, and most importantly for medical AI, buyers treat tier as a proxy for coverage of rare and long-tail cases. In practice, only top-tier hospitals consistently concentrate the diversity of complex cases that buyers cannot easily assemble through many low-tier purchases. Consistent with this logic, large radiology datasets exhibit highly long-tailed disease distributions, making ``coverage'' rather than raw volume the binding constraint for many clinically relevant tasks \citep{Zheng2024LongTailedRadiology}.

These mechanisms also explain a recurring deal-making pattern observed in the AI medical startup setting. Buyers often attempt to secure one high-tier data source first, investing substantial time to persuade a department head, because a successful $S_5$ partnership is believed to deliver quality, diversity, and sufficient scale in one shot, thereby saving the time and coordination costs of stitching together many lower-tier sources. When persuasion fails, buyers frequently restart the search process, not because lower-tier volume is unavailable, but because it is perceived as an inferior substitute for top-tier coverage. In operational terms, buyers partially mitigate \textit{ex ante} uncertainty by requesting sample extracts before contracting. Nevertheless, our fieldwork suggests that sample-based validation rarely eliminates the need to rely on seller tier as the dominant quality signal, since samples do not reveal long-tail coverage or the full metadata reliability at scale. Accordingly, we include $s_j$ as a positive shifter of buyer utility. In the model, higher $s_j$ increases expected value holding quantity constant, capturing the empirically salient fact that buyers systematically pay more for data originating from higher-tier hospitals because tier aggregates otherwise hard-to-price attributes: protocol standardization, metadata reliability, and long-tail case coverage.

\subsubsection{Buyer capability tier and budget constraint}
We model each buyer’s tier $z_i\in\{1,\dots,5\}$ as a proxy for the buyer’s data-to-product capability. The ability to translate a dataset into deployable model performance and downstream revenue. Concretely, $z_i$ bundles (i) model-training and evaluation know-how, (ii) compute and engineering capacity, and (iii) complementary organizational assets needed to integrate data into production systems. This choice follows a standard economic insight: the private value of an input often depends on the buyer’s complementary assets and organizational complements, not merely on the input’s intrinsic characteristics. In Teece’s classic framework, innovators capture value only when they control or access complementary assets that enable commercialization \citep{Teece1986PFI}. Likewise, firm-level evidence on information technology emphasizes that productivity gains arise from complementarity between technical inputs and organizational change and skills, rather than from ``IT alone'' \citep{BrynjolfssonHitt2000Beyond,BresnahanBrynjolfssonHitt2002QJE}. Translating this logic to data markets, a higher-$z_i$ buyer can extract systematically higher returns from the same dataset because it can (a) train and iterate faster, (b) operationalize the model in real workflows, and (c) appropriate value through productization and distribution.

Importantly, our fieldwork suggests that this capability effect is not monotone in firm age or ``maturity''. Some smaller AI startups exhibit high project ambition and thus high marginal value of data even when they remain financially constrained. We therefore keep $z_i$ conceptually distinct from the price coefficient $\alpha_i$ and from the budget constraint below. A buyer can be capability-strong (high $z_i$) yet budget-limited, or capability-weak (low $z_i$) yet well-funded.

Separately, for the purposes of ABM experimentation, we impose an exogenous procurement budget $B_i$ that caps the buyer’s total spending across transactions over the simulation horizon. This is a modeling device to generate realistic market selection under scarcity and to allow counterfactual institutional comparisons: when a buyer’s remaining budget is depleted, it exits (or sharply reduces) demand regardless of its latent valuation. Budgeted acquisition is also a natural assumption in the broader data or machine learning literature, where ``data acquisition for model improvement'' is explicitly studied under a fixed budget constraint \citep{LiYuKoudas2021PVLDB}. To parameterize $B_i$ in tiers, we use order-of-magnitude bands (e.g., increasing by roughly a factor of ten across $z_1$ to $z_5$), not as a precise empirical claim about any particular firm, but as a tractable approximation consistent with the highly right-skewed (heavy-tailed) distribution of firm size observed in the economy \citep{Axtell2001Zipf,KondoLewisStella2018FirmSize}. This tiered-budget structure is especially useful for the ABM’s comparative statics: it lets us test whether particular institutional designs disproportionately benefit (or exclude) budget-constrained but high-need buyers, versus well-funded buyers with stronger outside options.

\subsubsection{Buyer-seller interaction}
In addition to the main effects of seller tier $s_j$ and buyer tier $z_i$, we include an interaction term $s_j\times z_i$ with coefficient $\phi$. The goal is not to impose a particular matching pattern \textit{ex ante}, but to add a minimal reduced-form flexibility that relaxes separability in valuations. The marginal value of contracting with a higher-tier seller may depend on the buyer’s capability tier, and the marginal value of a higher-tier buyer may depend on the seller’s tier.

Theoretically, this term corresponds to a complementarity (or mismatch) channel. When match surplus exhibits increasing differences (supermodularity) in partner types, standard matching theory predicts systematic sorting (e.g., positive assortative matching) in equilibrium \citep{Becker1973Marriage, ShimerSmith2000AssortativeSearch}. In our demand specification we do not assume such sorting. Instead, $s_j\times z_i$ simply allows the data market to display (or not display) the kind of type-dependent gains from trade that supermodularity would generate. Put differently, $s_j$ is interpreted as an average quality or reliability signal (holding buyer capability fixed), and $z_i$ as an average data-to-product capability or budget proxy (holding seller type fixed). The interaction captures whether ``quality'' and ``capability'' are complements in producing buyer-side surplus \citep{MilgromShannon1994MCS,Topkis1998Supermodularity}. Indeed, there is evidence of positive assortative matching by capability. Studies of exporter–importer matches finds that because of complementarity, only high-capability exporters can match with high-capability importers \citep{Benguria2021export, SugitaTeshimaSeira2023export}.

Crucially, we leave $\phi$ a priori ambiguous. While a positive $\phi$ would indicate that higher-tier buyers obtain disproportionate incremental value from higher-tier sellers (because they can exploit richer data through stronger compute/engineering and can absorb higher procurement/compliance burdens), a negative $\phi$ is also plausible. Top-tier sellers may come with heavier integration, governance, and contracting overhead, making high-$s_j$ datasets less attractive to low-capability buyers even after controlling for the main effects. And $\phi$ may simply be near zero, implying approximate separability. This interpretation also avoids double counting: $s_j$ captures the baseline premium associated with seller tier as a quality signal, $z_i$ captures baseline differences in buyer capability and scale, and $s_j\times z_i$ captures only the incremental ``fit'' component—how much extra value (or extra friction) arises when a particular buyer type meets a particular seller type.

\subsubsection{Geographical distance}
Even in ``digital'' data markets, distance remains a meaningful friction because high-stakes data procurement is rarely a click-to-buy transaction. It often requires repeated face-to-face persuasion, relationship building, and intensive coordination with the seller’s operational staff. A large literature in economic geography emphasizes that proximity facilitates rich communication, trust formation, and the transfer of tacit knowledge that is difficult to codify, making face-to-face contact an efficient coordination technology rather than a redundant relic of pre-internet commerce \citep{StorperVenables2004Buzz,GasparGlaeser1998ITCities,Boschma2005ProximityInnovation}. Empirical evidence also shows that geographic proximity fosters trust and information sharing, offering competitive advantages in business dealings \citep{PetersenRajan2002geoDistance,SorensonStuart2001VCGeography, AudretschFeldman1996GeographyInnovation}. By contrast, a large distance can introduce communication delays and unfamiliarity, effectively reducing a buyer’s expected utility from the deal \citep{Porter1998Clusters}.Consistent with this view, work exploiting quasi-exogenous reductions in travel costs finds that lowering travel frictions causally increases collaboration, underscoring that physical mobility remains complementary to remote communication \citep{CataliniFonsRosenGaule2020TravelCosts}.

Our fieldwork aligns closely with these mechanisms. In hospital-AI procurement, early-stage transactions are particularly distance-sensitive. Buyers commonly need on-site meetings with department heads and multiple rounds of follow-up to explain the intended cooperation, negotiate the initial deal structure, and coordinate extraction or formatting with the seller’s team. While buyers may request sample extracts remotely, the first deal still tends to hinge on relational work and iterative operational coordination, both of which become more costly when partners are in different cities and require flights and travel budgets. Distance therefore acts as a reduced-form proxy for (i) relationship-formation costs that fall with proximity and repeated interaction \citep{Uzzi1996Embeddedness}, and (ii) coordination costs in initiating and executing the first transaction. This is also consistent with the ``temporary geographical proximity'' argument. When collaboration requires trust and high-bandwidth communication, parties often substitute travel for permanent co-location, but travel itself is a real cost and constraint \citep{RalletTorre2009TGP}.

Accordingly, we include $-\kappa\ln(1+d_{ij})$ in buyer utility, where $d_{ij}$ is the straight-line geographic distance between buyer $i$ and seller $j$. The log form captures diminishing marginal distance effects: the first increments of distance (e.g., within-city to cross-city) can sharply raise the need for travel and slow coordination, while additional distance adds less incremental friction at already-long ranges. This choice is also in the spirit of gravity-style formulations where trade frictions rise with distance in a roughly log-linear way \citep{AndersonVanWincoop2003Gravity}.

\subsubsection{Price sensitivity heterogeneity}
Rather than assume all buyers respond identically to price, we allow the price coefficient $\alpha_i$ to vary across buyers. This is standard in random-coefficients (mixed) logit models and is often essential for realism. Different buyers face different opportunity costs of funds, different outside options, and different internal procurement frictions, so the same posted price can induce very different behavioral responses \citep{BLP1995demand,McFadden2000mnl}. Formally, price enters utility linearly as $-\alpha_i p_{ijt}$, so $\alpha_i$ is the marginal disutility of price for buyer $i$. Under a quasi-linear interpretation with a numeraire, it can be read as a constant multiple of buyer $i$'s marginal utility of income. A key implication is that marginal willingness to pay (MWTP) for any attribute is pinned down by the ratio of coefficients: for a characteristic with utility coefficient $\beta_k$, 
\begin{equation}
    \text{MWTP}_{ik}=\frac{\beta_{ik}}{\alpha_i},
\end{equation}
so $\alpha_i$ serves as the conversion rate between ``utils'' and money in welfare-relevant units \citep{Train2002}.

Importantly, we do not interpret $\alpha_i$ as a feasibility constraint (``can/cannot afford'') in the sense of an explicit budget set. In a standard random coefficient logit demand, the choice set is not restricted by budget. Instead, any affordability-driven non-participation is absorbed as high effective price sensitivity. Accordingly, heterogeneity in $\alpha_i$ should be understood as a reduced-form statistical summary of multiple underlying sources of price responsiveness---income and budget tightness, internal approval and contracting costs, search and switching costs, and unobserved preference or information differences---without committing to any single structural channel \citep{AllenbyRossi1998,DuvvuriAnsariGupta2007}. This separation is deliberate in our design: we model buyer tier $z_i$ and the ABM budget constraint as distinct objects, while $\alpha_i$ flexibly captures remaining unobserved heterogeneity in price sensitivity that improves fit and yields more credible WTP dispersion.

\subsection{Sellers}
On the supply side, we also specify a utility (profit) function for the seller and derive a corresponding willingness-to-accept (WTA)---essentially the minimum price the seller is willing to accept for the data. While the buyer’s utility above is the more complex part of the model, it is important to note how sellers’ considerations enter, especially via costs and risks. In our model, a seller $j$'s net utility from selling data (inside a logit framework) is modeled as
\begin{equation}\label{eq:seller}
    U_{ijt}^{\text{Seller}} = \alpha_j p_{it} - (c_0 + c_1 s_j+ c_2 x_{j} + \beta_R R_j + \beta_E E_{jt}) +\varepsilon_{ijt},
\end{equation}
where $p_{it}$ is the price paid by buyer $i$, and the terms in parentheses represent the seller’s costs or disutility from the transaction. Here $\alpha_j$ s a seller-specific random coefficient on price. Analogous to buyers’ random coefficient $\alpha_i$, it captures how strongly the seller values additional revenue. We generally assume sellers are fairly insensitive to a single transaction’s price beyond its profit, so $\alpha_j$ may not vary as much as buyers’ do. The cost terms are as follows.

\subsubsection{Baseline and volume-dependent cost}
Selling data is not costless for the seller \citep{Radauer2023TSDataSharing}. We therefore include a fixed cost term $c_0$ to capture the up-front organizational overhead required to make any transfer feasible, independent of the deal’s size. In practice, a major component of this overhead is governance and contracting work: negotiating and reviewing a data use/sharing agreement, aligning internal stakeholders, and producing the documentation needed to justify access and downstream constraints. Health data-sharing experience in particular emphasizes that drawing up and approving DUAs/DSAs consumes substantial time and administrative resources, and that efforts to standardize agreements are motivated precisely by the goal of reducing these recurring fixed burdens \citep{Allen2014BeaconDSA}.\footnote{See also, Center for Open Data Enterprise (CODE), in partnership with the U.S. Department of Health and Human Services Office of the Chief Technology Officer, ``Sharing and Utilizing Health Data for AI Applications: Roundtable Report,'' Roundtable held Apr 16, 2019; report published by CODE, \url{https://www.hhs.gov/sites/default/files/sharing-and-utilizing-health-data-for-ai-applications.pdf}; U.S. Department of Health \& Human Services, ``HHS Policy for the Common Data Use Agreement (DUA) Structure and Repository,'' Feb 13, 2025, \href{https://www.hhs.gov/web/governance/digital-strategy/it-policy-archive/hhs-policy-common-data-use-agreement-structure-repository.html}{\texttt{https://www.hhs.gov/web/governance/digital\-strategy/it\-policy\-archive/hhs\-policy\-common\-data\-use\-agreement\-structure\-repository.html}}.} More broadly, responsible data management and sharing is widely described as imposing an additional administrative workload (e.g., planning, documentation, legal/privacy review), which is borne as a fixed participation cost before any ``transaction'' can occur \citep{Pellen2025DataManagementSharing}.

We model $c_2 x_j$ as a variable cost that scales with the volume of data supplied. Larger transfers require more extraction, cleaning, formatting, and quality control, and—when personal or sensitive information is involved—more extensive application of de-identification workflows and related checks. Practical guidance on de-identification underscores that achieving an acceptable level of de-identification typically involves nontrivial operational steps rather than a costless switch, implying that the marginal effort rises with dataset scale and complexity.\footnote{U.S. Department of Health \& Human Services, ``Methods for De-identification of PHI,'' Feb 3, 2025, \url{https://www.hhs.gov/hipaa/for-professionals/special-topics/de-identification/index.html}.} In short, as $x_j$ grows, sellers face higher direct handling costs and a higher shadow cost of managing a larger release under governance and documentation constraints, implying a higher reservation price for larger transactions.

\subsubsection{Seller's own institutional tier}
In our model, the seller’s institutional tier $s_j\in\{1,\dots,5\}$ proxies for hospital rank and the associated bundle of institution-intrinsic factors that raise the shadow cost of supply even when volume $x_j$, deal-specific risk $R_j$, and ambient enforcement $E_{jt}$ are held constant. This mapping is consistent with academic descriptions of China’s tiered hospital system, where higher-tier (tertiary) hospitals are characterized by more comprehensive medical, teaching, and scientific research capabilities \citep{Shi2021TierHospitals}, and with peer-reviewed work explicitly noting that Grade-A tertiary hospitals sit at the top of the grading system and are evaluated along dimensions such as specialty departments, staff/facilities, patient flow, bed accessibility, and scientific research achievements \citep{Wang2025TCMInternetHospitals}. In addition, major hospital ranking systems in China place weight on scientific research competitiveness, reinforcing that “top tier” is a bundle of capability and institutional expectations rather than size alone \citep{Li2022HospitalRankingReview}.  

Interpreted this way, $c_1 s_j$ captures three mechanisms that emerged repeatedly in our fieldwork. First, outside options and foregone rents. Higher-tier hospitals have stronger outside options because they can internalize larger benefits from exclusive in-house use (e.g., clinical operations, internal analytics/AI, and research pipelines) and often prefer to wait until rules become clearer, so releasing raw data destroys more private rents and requires higher compensation. Second, reputational and accountability exposure. Top-tier hospitals are more visible and face greater professional and political scrutiny. In our interviews, they were distinctly more sensitive to being framed as “selling patient data,” so they demanded a premium for incremental exposure even holding $R_j$ and $E_{jt}$ fixed. Third, governance overhead and organizational friction. Higher-tier sellers typically combine greater compliance capability with more veto points and more formal review/sign-off, which inflates both fixed and effective variable costs beyond what $x_j$ alone captures (and often pushes deals toward pilot-first structuring). Under this interpretation, $c_1 s_j$ isolates institution-level opportunity cost, approval frictions, and governance intensity, while $R_j$ and $E_{jt}$ continue to represent transaction- and regime-level hazards, avoiding double counting. The specification therefore yields a clear empirical implication: conditional on $x_j$, $R_j$, and $E_{jt}$, higher-tier sellers should exhibit systematically higher reservation prices.

\subsubsection{Transaction risk}
We introduce $R_j$ to capture the seller’s perceived intrinsic hazard of supplying its data, the probability-weighted downside that ``something bad could happen'' and the seller will be expected to bear responsibility. This notion of risk is broader than technical leakage alone. In our fieldwork, sellers, especially hospitals, consistently framed risk in accountability terms: if a dataset is later leaked, re-identified, or misused, the organization anticipates not only legal exposure but also reputational and political fallout. We therefore operationalize $R_j$ as a reduced-form index of downside exposure that is largely determined by the data category and its misuse potential, holding the ambient enforcement environment fixed.

To keep the model tractable, we discretize $R_j$ into three levels (Low/Medium/High). A low-risk transaction may involve data that is relatively non-sensitive and easy to bound contractually, so that plausible misuse yields limited harm. A high-risk transaction, by contrast, involves data types where re-identification or harmful downstream use is plausible and the fallout is difficult to contain. For example, raw medical data (including images and longitudinal records) where auxiliary information can enable linkage and re-identification, and where an incident can trigger public controversy and third-party claims. Importantly, in our fieldwork sellers did not treat technical controls (e.g., clean rooms, on-prem arrangements, output takedowns) as fully credible in eliminating this hazard; rather, they viewed risk as essentially persistent because control is difficult once data is transferred and because re-identification can occur even after ``anonymization'' in practice \citep{Ohm2010anon,NarayananShmatikov2008Deanon}.

Higher $R_j$ increases the seller’s disutility from transacting, so we expect $\beta_1>0$: conditional on price, a seller requires greater compensation to accept a transaction with higher intrinsic hazard. This is a standard economics intuition: when an action exposes an actor to probabilistic but potentially severe losses, the actor demands a private risk premium. The fieldwork implication is that sellers rationally overweight downside relative to incremental deal revenue because breach-and-misuse events can generate large reputational and financial harms. Consistent with this, empirical studies show that publicly announced security breaches can generate negative market reactions for affected firms \citep{Campbell2003BreachCost}, and breach events can also trigger costly litigation dynamics \citep{RomanoskyHoffmanAcquisti2014BreachLitigation}. These facts make the seller-side participation constraint highly sensitive to risk: as $R_j$ rises, sellers demand a higher WTA or decline to transact, even in settings where buyers’ willingness-to-pay may be substantial \citep{Simpson2021lack, Li2020lack}. This is analogous to how a supplier of a hazardous product would charge more to offset liability risk. Although difficult to measure directly, we incorporate discrete risk levels to acknowledge this factor in data markets \citep{Meier2024lack}. It reflects findings that lack of trust and fear of negative outcomes can stymie data sharing, because sellers are more hesitant and demand higher WTA when the perceived risk is high \citep{Gefen2019lack, Skatova2023ValuationPersonalData, Wang2021lack}.

\subsubsection{Regulatory enforcement}
In tandem with intrinsic transaction hazard $R_j$, we include $E_{jt}$ to represent the intensity of regulatory enforcement faced by seller $j$ (i.e., the surrounding legal regime). In our fieldwork, sellers did not describe enforcement as a transparent, rule-bound mapping from conduct to sanction. Rather, it was experienced as unclear but scary: sellers worried that if ``anything bad happens,'' they would be asked to bear responsibility, and that scrutiny could be triggered by salient events (especially a leak incident or a viral social-media post). This is precisely why $E_{jt}$ matters independently of $R_j$: it captures how strongly the institutional environment converts latent hazards into expected private costs through audits, investigations, remediation orders, and sanctions.

We discretize $E_{jt}$ into three levels that correspond to qualitatively different sanction menus available to authorities. Under weak enforcement ($E_1$), regulators primarily rely on corrective and reputational tools, warnings, orders to rectify, and public exposure, with limited or no monetary penalties. This is consistent with the baseline penalty structure in Article 66 of China’s Personal Information Protection Law (PIPL), which authorizes orders to correct, warnings, confiscation of illegal gains, and orders to suspend/terminate app services, with monetary penalties escalating if rectification is refused.\footnote{See \textit{Personal Information Protection Law of the People’s Republic of China} art.~66 (adopted Aug.~20, 2021, effective Nov.~1, 2021) (P.R.C.) (establishing a graduated sanction scheme: competent authorities may order correction, issue warnings, and confiscate unlawful gains; order apps unlawfully processing personal information to suspend or terminate service; impose a fine up to RMB 1 million (\$143,000) for refusal to correct; and, for ``serious circumstances,'' impose fines up to RMB 50 million or 5\% of prior-year turnover, with possible suspension for rectification or permit/license revocation, plus individual fines up to RMB 1 million and potential disqualification from serving as directors, supervisors, senior managers, or the person in charge of personal information protection for a period). An English translation is available at \url{https://en.spp.gov.cn/2021-12/29/c_948419.htm} (last visited Dec.~31, 2025).} Under moderate enforcement ($E_2$), sellers anticipate material administrative consequences such as fines/confiscation and operational disruptions (e.g., app store removal or suspension of functions/registrations), consistent with the PIPL’s escalating sanctions and commonly observed compliance interventions. Under strong enforcement ($E_3$), sellers anticipate (i) the credible possibility of top-of-range corporate fines and business disruption (suspension for rectification or revocation of permits/licenses) and (ii) personal exposure for executives or responsible signatories. This expectation is grounded in the ``serious circumstances'' clause of PIPL Article 66, which permits fines up to RMB 50 million or 5\% of prior-year turnover and also authorizes individual penalties (including fines and role bans). Recent high-profile enforcement further makes the upper tail salient: China’s Didi decision culminated in a RMB 8.026 billion (\$1.15 billion) penalty and individual fines for top executives.\footnote{Reuters, ``China fines Didi \$1.2 bln for violating data security laws,'' July 21, 2022, \href{https://www.reuters.com/technology/china-fines-didi-global-12-bln-violating-data-security-laws-2022-07-21/}{\texttt{https://www.reuters.com/technology/china\-fines\-didi\-global\-12\-bln\-violating\-data\-security\-laws\-2022-07-21/}}.}

In addition, enforcement intensity increasingly operates through routinized compliance mechanisms rather than only \textit{ex post} punishment. The CAC’s Measures on Personal Information Protection Compliance Audits (effective May 1, 2025) institutionalize both regular audits and for-cause audits, strengthening the perceived probability of detection and the expected cost of non-compliance in higher-enforcement regimes.\footnote{KPMG China, ``Personal Information Protection Compliance Audit: Management and Response,'' 2025, \href{https://assets.kpmg.com/content/dam/kpmg/cn/pdf/en/2025/03/personal-information-protection-compliance-audit-management-and-response.pdf}{\texttt{https://assets.kpmg.com/content/dam/kpmg/cn/pdf/en\-/2025\-/03\-/personal\-information\-protection\-compliance\-audit\-management\-and\-response.pdf}}; Mayer Brown, ``China finalises the Measures for Personal Information Protection Compliance Audits,'' 2025, \href{https://www.mayerbrown.com/en/insights/publications/2025/04/china-finalises-the-measures-for-personal-information-protection-compliance-audits}{\texttt{https://www.mayerbrown.com/en/insights/publications\-/2025\-/04\-/china\-finalises\-the\-measures\-for\-personal\-information\-protection\-compliance\-audits}}.} In law and economics terms, higher $E_{jt}$ raises the seller’s expected cost of supplying data by increasing the probability and/or severity of adverse legal outcomes conditional on controversy or an incident. Accordingly, we expect $\beta_2>0$: stronger enforcement reduces seller utility and raises WTA. Conversely, when enforcement is weak, sellers anticipate lower expected sanctions and can rationally accept lower reservation prices, holding $R_j$ fixed.

\subsubsection{Risk-enforcement interaction}
To capture how regulatory regimes transform latent hazards into expected private costs, we augment the seller’s reservation price with a risk-enforcement interaction term $R_jE_{jt}$. The intuition is standard in law and economics: expected sanctions depend on both the underlying harmfulness of conduct (here, the intrinsic hazard of the dataset) and the enforcement technology that makes that harm privately consequential, probability of detection/audit and the severity of available sanctions \citep{Becker1968Crime,PolinskyShavell2000Enforcement}. In data transactions, many harms (re-identification, leakage, unlawful reuse) are probabilistic and often materialize only through downstream events; absent enforcement, these hazards can remain partly externalized. Stronger enforcement increases the expected private cost of the same hazard by (i) raising the likelihood that incidents trigger scrutiny (e.g., audits or investigations following a leak or viral exposure) and (ii) increasing the sanction menu and its severity, including operational restrictions and individual accountability. Conversely, under weak enforcement, even intrinsically risky transfers may not translate into comparable expected private cost. The interaction therefore captures a key empirical regularity suggested by our fieldwork: what deters supply is not “risk” in isolation, but the expectation that if something goes wrong the seller will be held responsible in a high-stakes enforcement environment. Finally, this specification also helps avoid double counting: $R_j$ captures the transaction’s inherent hazard, $E_{jt}$ captures the surrounding enforcement intensity, and $R_jE_{jt}$ captures how the \emph{same} hazard is priced differently across enforcement regimes.

\section{Calibration}
\subsection{LLM-based discrete choice experiment}
\subsubsection{Data constraints}
Calibrating an agent-based model of data transactions requires preference primitives that are empirically defensible and behaviorally interpretable. In our setting, these include WTP on buyer and WTA on seller side, along with how they shift with data scale $x$, seller capability $s_j$, and institutional conditions $(R,E)$. In mature product markets, such parameters can be disciplined by transaction microdata, field experiments, or regulatory disclosures. By contrast, data brokerage is structurally opaque. Transactions are bilateral and private, pricing is often bundled or NDA-constrained, and there is no public panel of prices/quantities or counterpart attributes at scale. As a result, there is no direct market evidence against which to anchor the key elasticities and interaction terms that drive welfare and policy counterfactuals in an ABM of the data economy.

Substitutes perform poorly. Hand-tuned parameters or ``calibration by convenience'' invite researcher degrees of freedom. Conclusions risk reflecting priors rather than behavior, weakening external validity and legal-policy relevance. Brute-force parameter sweeps are not a remedy. High-dimensional ABMs make exhaustive search computationally prohibitive, and a large share of the parameter space is economically nonsensical, which is far from plausible institutional or technological regimes. 

Due to data constraints, we need to find a new method that is transparent (auditable prompts and seeds), reproducible (rerunnable instruments), and structurally informative (attributes mapped to interpretable primitives), without presupposing unavailable market data. This motivates our choice to generate data for demand estimation through a discrete choice experiment (DCE) embedded in a RUM model \citep{Louviere1983Design,Revelt1998choice}. In a general DCE setting, each respondent completes $T$ choice tasks. In task $t$, respondent $i$ faces a choice set $C_{it}$ containing $J$ product profiles plus an outside option. Product profiles are defined by a vector of attributes $x_{jt}$ and a price $p_{jt}$, with attribute levels varied according to an experimental design. Respondents are instructed to choose the option they would select under the described conditions, with the opt-out capturing non-participation and helping anchor substitution to alternatives outside the designed set.

Under the RUM assumption, respondent $i$’s utility from alternative $j$ in task $t$ is $U_{ijt}=V(x_{jt},p_{jt};\theta_i)+\varepsilon_{ijt}$, and the observed choice $y_{it}$ is the alternative maximizing $U_{ijt}$ over $j\in C_{it}$. We stack the data in panel format at the respondent–task–alternative level, recording the attributes shown and an indicator for the chosen option. The experimental variation in $(x_{jt},p_{jt})$ provides identification of preference parameters, and the repeated-choice structure allows us to estimate heterogeneity using a mixed (random-coefficients) logit specification.

However, assembling large human samples of relevant actors (e.g., hundreds of AI firms as buyers and hospitals/banks as sellers) for controlled experimentation is practically and ethically infeasible. This is because even expert elicitations seldom yield the volume or structure needed to identify interaction effects such as $R \times E$. Therefore, we precede simulation with a DCE conducted on a LLM treated as a proxy subject. The DCE yields internally consistent choice data that discipline signs, elasticities, and interaction terms before those parameters are embedded in agents and propagated through the ABM.

\subsubsection{LLM model choice}
A rapidly accumulating literature now shows that state-of-the-art LLMs are competent ``silicon participant'' in exactly the kinds of scenario and discrete-choice tasks we use. In psychology and management, a large replication study re-ran 156 published vignette experiments on three frontier LLMs and recovered 73–81\% of main effects and 46–63\% of interaction effects (with the well-noted caveat of effect-size inflation), demonstrating that randomized factorial manipulations are reliably reflected in LLM responses \citep{Cui2025naturereplication}. In political science, Argyle et al. establish ``algorithmic fidelity,'' showing that LLMs can be conditioned to emulate specific human subpopulations—enabling the heterogeneity analyses our persona-based WTA/WTP design requires \citep{Argyle2023politicalllm}. In economics, Horton formalizes homo silicus and demonstrates that LLMs, given endowments and constraints, reproduce classic experimental regularities and can stand in as simulated subjects for ex-ante design and calibration \citep{horton2023llmecon}. In computational social science, \citet{ziems2024can} conclude that LLMs can augment experimental pipelines under clear prompting and evaluation protocols. Beyond replication, HCI and simulation work demonstrates that LLM-driven agents generate believable individual decisions and emergent social behavior in multi-agent environments, which aligns with our pipeline of calibrate via DCE, then simulate via ABM under legal-institutional treatments \citep{park2023llm,Gao2023llmreview}.

We field the DCE on DeepSeek because recent peer-reviewed evaluations document frontier-level reasoning and competitive performance with proprietary models in tasks that are structurally analogous to social-science vignette and choice experiments. First, the DeepSeek-R1 program introduced a reinforcement-learning framework that materially improves multi-step reasoning—with evidence published in Nature—thereby supporting reliability on multi-attribute trade-offs central to discrete-choice designs \citep{Guo2025nature}. Second, independent studies like decision-support evaluations show DeepSeek-V3 and R1 performing on par with, and in some settings better than, GPT-4o and Gemini on decision tasks. These prove that DeepSeek is an applied proxy for the type of structured judgment we elicit in WTA and WTP experiments \citep{Sandmann2025natureclinical,zhang2025euro}.

We selected DeepSeek as the primary instrument for the DCE due to a combination of its architectural design, demonstrated performance, and methodological transparency. The model's core strength lies in its ``reasoning-first'' orientation, which is a direct result of a reinforcement learning pipeline explicitly engineered to enhance multi-step logical inference for tasks in mathematics, coding, and structured problem-solving. This training encourages the model to generate explicit ``chain-of-thought'' processes, effectively ``thinking aloud'' as it works through a problem, which provides a transparent window into the decision-making calculus for our multi-attribute choice tasks. This inherent reasoning capability translates to highly competitive performance, with independent benchmarks showing DeepSeek performing on par with, and in some cases exceeding, leading proprietary models on complex decision-making tasks. Its proficiency in applied domains, such as clinical decision support where it must navigate multi-stage reasoning under constraints, serves as an informative proxy for its ability to handle the structured economic trade-offs in our experimental setting. Finally, from a legal-methodological standpoint, the availability of detailed technical reports and model checkpoints enhances the replicability of our study, allowing for independent verification and stress. 

The LLM-enabled DCE offers a suite of methodological advantages that directly address the challenges of calibrating agent-based models in data-scarce environments, aligning with the evidentiary standards of law and economics. Foremost among these is its purpose-built replicability and auditability. The entire experimental protocol, including prompts, randomization seeds, and model parameters, is fixed \textit{ex ante} and logged. This yields a complete calibration audit trail, allowing any researcher to re-field the survey and reproduce the estimated preference primitives. Such procedural transparency is rarely attainable with ad-hoc parameter tuning or opaque expert elicitation.

Furthermore, the experimental design provides strong identification and structural interpretability. Because attributes such as price, data scale $x$, and the institutional environment $(L,E)$ are experimentally manipulated, the recovered parameters are directly interpretable in welfare-economic terms. The design isolates the specific effects of price sensitivity, capability gradients, scale effects (including diminishing-returns curvature), and legal-enforcement interactions—precisely the primitives the ABM requires to generate policy-relevant counterfactuals. This approach is also highly efficient and scalable. LLM respondents enable the rapid, large-$N$ collection of large, internally consistent datasets at a low cost, avoiding the recruitment and compliance frictions of large human panels. This allows for the estimation of full parameter distributions for heterogeneous agent types, a critical feature for realistic simulation.

From a diagnostic perspective, the methodology offers transparency for methodological scrutiny. The use of structured outputs and the elicitation of concise, free-text rationales for each choice permit a granular, post-hoc audit of the model's decision logic. This ``reasoning trace'' strengthens the credibility of the calibration in legal and policy settings that demand explainability. The estimated primitives are also directly portable to the simulation, mapping one-to-one into agent decision rules and endowing the ABM with empirically-grounded, heterogeneous preferences. In an opaque market where observational microdata are unavailable, this approach provides a proportionate and documentable alternative to speculative parameterization. The net effect is a calibration method that replaces subjective priors with experimentally disciplined, reproducible primitives, thereby improving the credibility of downstream legal-economic counterfactuals.

\subsection{Experimental design}
This section details the stated-preference DCE used to generate empirically disciplined primitives for calibrating the ABM of hospital--AI-firm data transactions. The DCE treats a LLM as a proxy respondent under tightly controlled conditions. By systematically varying legally and economically salient attributes of hypothetical transactions, the design identifies marginal effects that parameterize (i) seller-side compliance and organizational costs, (ii) buyer-side acquisition utilities and frictions, and (iii) the sensitivity of both sides to institutional environments.

\subsubsection{Two-sided elicitation}
We implement two parallel DCE instruments. The buyer module elicits WTP: in each of 24 rounds, the LLM chooses among two purchase options A/B and an outside option as an AI firm procurement lead:
\begin{equation}
\text{option}=
    \begin{cases}
        \text{A: bundle A}\\
        \text{B: bundle B}\\
        \text{C: no purchase}
    \end{cases}.
\end{equation}
Each purchase option is described by a bundle of attributes: data scale $x$ (in $10^4$ records), seller capability tier $s_j$, price $P$, and geographic distance $d$, which proxies cross-jurisdictional coordination and compliance frictions. To align the instrument with the ABM, distance is encoded as a smooth friction index $\ln(1+d)$ where $d$ is mapped to kilometers by distance tier.

The seller module elicits willingness-to-accept (WTA): in each of 24 rounds, the LLM chooses between options
\begin{equation}
\text{option}=
    \begin{cases}
        \text{A: sell}\\
        \text{B: not sell}
    \end{cases}
\end{equation}
given a single offer as a hospital decision-maker. The offer varies by data scale $x$, offered payment $p$, and two institutional attributes capturing the regulatory environment: risk level $L$ (\textit{ex ante} ambiguity and potential externalities) and enforcement intensity $E$ (\textit{ex post} supervisory intensity and expected penalties). The instrument covers the full $3\times 3$ interaction space of $(L,E)$ and includes additional ``stretch'' tasks with extreme $(x,p)$ combinations to improve identification of slopes and corner behavior under selected institutional cells.

\subsubsection{Persona conditioning and baseline heterogeneity}
Before any choice tasks, the LLM receives a concise baseline persona by the system prompt. In the buyer module, personas specify the firm's capability tier $z_i$, baseline in-house data stock $x_i$, and price attitude anchor (strong/weak/low price sensitivity). The system prompt assigns the role (AI-firm procurement lead), states the objective (purchasing hospital data for medical foundation-model training), specifies the number of rounds (24), and enforces a strict JSON-only response policy:

\begin{prompt*}{Buyer module}{}
You are now acting as the head of data procurement at an AI technology company. Your goal is to purchase medical data from hospitals for training a medical foundation model. I will present 24 rounds of choice tasks. In each round, there are two data offers (A/B), each with an offered data scale of $\{x_j\}$, and an outside option C = no purchase. Your task is: in each round, choose exactly one option (A/B/C) that best matches your preferences, and provide a brief reason of no more than 20 words. Output strictly in JSON format. Do not provide any additional explanation.

\ 

Your company’s baseline information is as follows: (1) existing usable data stock $x_i=\{x_i\}$ (in $10^4$ records); (2) company capability tier $z_i=\{z_i\}$; (3) price-sensitivity anchor: \{price\_attitude\}. Please answer the following questions according to your true preferences.
\end{prompt*}

In the seller module, personas specify the hospital capability tier $s_j$ and risk preference (risk-averse/neutral/seeking). The LLM is role-conditioned as a hospital decision-maker and is required to output JSON-only responses. The system prompt specifies the scenario (a hospital with accumulated patient data facing a purchase request by a medical AI company), the number of rounds (24), and the decision rule (binary choice: A = sell vs.\ B = do not sell):

\begin{prompt*}{Seller module}{}
    You are now acting as the head of a hospital. The hospital has accumulated patient data over many years, and a medical AI company is seeking to purchase the data you hold. I will present 24 rounds of choice tasks. In each round, you will be given: (i) the selling risk level (low/medium/high), (ii) the enforcement intensity (weak/moderate/strong), and (iii) the buyer’s offered price $p$ and the data scale $x_j$ you offer. Your task is: in each round, choose one of the two options, A = sell or B = do not sell, and provide a brief reason of no more than 20 words. Output strictly in JSON format. Do not provide any additional explanation.

    \ 

    Your hospital’s baseline information is as follows: (1) Hospital capability tier $s_j = \{S_j\}$. (2) Risk-preference anchor: \{risk\_attitude\}. Please answer the following questions according to your true preferences.
\end{prompt*}

Both modules also include a short price-attitude anchor to stabilize preference orientation across rounds without revealing hypotheses. These baseline variables map directly into ABM state variables, enabling type-specific response functions (e.g., curvature in buyer value conditional on $x_i$ and capability-dependent seller reservation behavior conditional on $s_j$). The description of prompting variables are shown in Table~\ref{tab:dce_levels}. 

\subsubsection{Randomization and reproducibility}
In the experiment, each round is presented as a vignette that lists attribute values for the relevant options and requires a strict JSON output with fields \texttt{round}, \texttt{choice}, and a brief free-text \texttt{reason} constrained to $\leq 20$ words. This ``reasoning trace'' provides a lightweight, auditable diagnostic record (e.g., whether risk/enforcement, distance, or price is cited) while limiting narrative drift.

To ensure consistent role conditioning and machine-readable outputs, we use fixed prompt templates. In the buyer (WTP) module, each round is instantiated with a round instruction that lists Option A, Option B, and the outside option C (no purchase), and requires a single JSON object with fields \texttt{round}, \texttt{choice}, and \texttt{reason}. Finally, before the choice tasks, a baseline persona introduction provides firm-specific state variables (baseline data stock $x_i$ and capability tier $z_i$) and a price-sensitivity anchor, so that across-round choices reflect stable, type-specific preferences. In the seller (WTA) module, each round is instantiated with an instruction that presents the institutional environment (risk level and enforcement intensity) and the economic terms (offered price $p$ and data scale $x$), and requires a single JSON object with fields \texttt{round}, \texttt{choice}, and \texttt{reason}. Before the tasks, a baseline persona introduction fixes the hospital capability tier $S_j$ and provides an attitude anchor (risk preference) to stabilize cross-round decision patterns.

Within each module, the 24 tasks are administered in randomized order using a fixed random seed. We also implement paraphrase randomization and attribute-order randomization to reduce sensitivity to framing and template artifacts. All prompts, randomization seeds, model identifiers, and response schemas are version-controlled, and all outputs are stored in a structured machine-readable format alongside the encoded covariates. This creates a fully auditable calibration trail that can be re-run on the same model release to reproduce the estimated marginal effects used to parameterize the ABM.

\begin{table*}[ht]
\centering
\small
\caption{DCE attributes, levels, and operational descriptions}
\label{tab:dce_levels}
\resizebox{1.0\textwidth}{!}{
\begin{tabularx}{\linewidth}{llllX}
\toprule
Module & Attribute & Symbol & Level & Level description / operationalization \\
\midrule

\multirow{5}{*}{\makecell{Buyer\\(WTP)}} &
\multirow{5}{*}{Seller capability} &
\multirow{5}{*}{$s_j$} &
$S_1$ & \textbf{Primary-care integration hospital:} a frontline provider for first-contact care, chronic-disease management, public health, and rehabilitation/continuity of care. \\
& & & $S_2$ & \textbf{County anchor hospital:} the ``first terminal'' for county-wide comprehensive services and tiered referral; largely achieves the goal of within-county hospitalization. \\
& & & $S_3$ & \textbf{Regional backbone hospital:} a prefecture-level referral endpoint covering common medium/large surgeries and comprehensive acute care; exerts strong capability spillovers to downstream hospitals. \\
& & & $S_4$ & \textbf{Provincial flagship hospital:} a provincial hub for critical-care treatment and specialty integration; provincial-level quality-control responsibilities. \\
& & & $S_5$ & \textbf{National hub hospital:} a national center for complex/rare critical cases and original clinical innovation; cross-regional referral coordination and public-health command functions. \\
\addlinespace
\multirow{4}{*}{\makecell{Buyer\\(WTP)}} &
\multirow{4}{*}{Distance tier} &
\multirow{4}{*}{$d$} &
$D_0$ & \textbf{Same city:} Mapped to $d=5$ km; friction covariate $\ln(1+d)$. \\
& & & $D_1$ & \textbf{Same province:} Mapped to $d=50$ km; friction covariate $\ln(1+d)$. \\
& & & $D_2$ & \textbf{Cross-province:} Mapped to $d=300$ km; friction covariate $\ln(1+d)$. \\
& & & $D_3$ & \textbf{Cross-region:} Mapped to $d=1000$ km; friction covariate $\ln(1+d)$. \\
\addlinespace
\multirow{5}{*}{\makecell{Buyer\\(WTP)}} &
\multirow{5}{*}{Buyer capability} &
\multirow{5}{*}{$z_i$} &
$Z_1$ & \textbf{Technology follower / systems integrator:} primarily provides AI-related system integration, consulting, and secondary development; lacks original core technology and platform capability. \\
& & & $Z_2$ & \textbf{Application innovator / vertically focused:} builds on open-source or upstream foundation models; focuses on a specific vertical scenario, toolchain, or industry service to innovate and productize rapidly. \\
& & & $Z_3$ & \textbf{Technical innovator in scaling-up:} relatively strong technical capability with breakthroughs in certain verticals or algorithmic directions, but not yet industry-dominant; currently expanding. \\
& & & $Z_4$ & \textbf{Regional/industry leader:} strong technological and market dominance within a particular industry or provincial region; operates a stable technical--commercial closed loop. \\
& & & $Z_5$ & \textbf{National strategic leader:} plays a national-level leading role in core technologies, compute infrastructure, and ecosystem standards; shapes industry structure and policy direction. \\
\midrule
\multirow{3}{*}{\makecell{Seller\\(WTA)}} &
\multirow{3}{*}{Risk level} &
\multirow{3}{*}{$R$} &
$R_1$ & \textbf{Low:} localized and controllable social disturbance/externality. \\
& & & $R_2$ & \textbf{Medium:} visible unfairness and order disruption, which can be corrected through regulation. \\
& & & $R_3$ & \textbf{High:} systemic spillovers at the level of public safety. \\
\addlinespace
\multirow{3}{*}{\makecell{Seller\\(WTA)}} &
\multirow{3}{*}{Enforcement intensity} &
\multirow{3}{*}{$E$} &
$E_1$ & \textbf{Weak:} mainly guidance and corrective measures---regulatory interviews, orders to rectify, time-limited remediation, and public notice/exposure; little or no fines. \\
& & & $E_2$ & \textbf{Moderate:} fines/forfeiture of illegal gains, app-store delisting, suspension of related functions/registrations, and compliance re-audits. \\
& & & $E_3$ & \textbf{Strong:} maximum statutory fines, suspension of operations or license revocation, and personal liability for senior managers. \\
\bottomrule
\end{tabularx}}
\end{table*}

\subsection{Posterior estimation}
We use Markov chain Monte Carlo (MCMC) to draw from the joint posterior of the structural parameters and individual-level random coefficients, because the mixed logit likelihood $\mathcal{L}_{i}(\theta,\mu_{\alpha_i},\sigma_{\alpha_i})$ and $\mathcal{L}_{j}(\theta,\mu_{\alpha_j},\sigma_{\alpha_j})$ involves high-dimensional integrals over random coefficients that are analytically intractable \citep{HuberTrain2001Similarity,RossiAllenbyMcCulloch2005BayesianMarketing,train2009DCE,RegierRyanPham2009BayesianClassicalMXL}. The estimated coefficients on buyer's WTP and seller's WTA are reported in Table~\ref{tab:calibration_buyer} and Table~\ref{tab:calibration_seller}. On the buyer side, the 95\% HDI for the coefficients $\rho$, $\beta$, $\tau$, and $\kappa$ do not include zero or narrowly touching zero, suggesting positive directions of these coefficients, while the directions of $\gamma$ and $\phi$ are uncertain. On the seller side, the posterior distributions of $c_1$ and $\delta$ include zero within their 95\% HDI, indicating uncertainty about the direction of effects. Therefore, the buyer's utility function effectively collapses to the simpler form
\begin{equation}\label{eq:buyer_simple}
    U^{\text{Buyer}}_{ijt}=e^{-\rho x_{it}}\beta x_j +\tau s_j-\alpha_i p_{ijt} -\kappa\ln(1+d_{ij}),
\end{equation}
with estimated coefficients $\hat{\rho}$, $\hat{\beta}$, $\hat{\tau}$, $\hat{\mu}_{\alpha}$, $\hat{\sigma}_{\alpha}$, and $\hat{\kappa}$, while the seller's utility function is practically indistinguishable from the model
\begin{equation}\label{eq:seller_simple}
    U^{\text{Seller}}_{ijt}=\alpha_{j}p_{ijt}-(c_0+c_2 x_j +\beta_{R}R_{j}+\beta_{E}E_{jt}),
\end{equation}
with estimated coefficients $\hat{c}_0$, $\hat{c}_2$, $\hat{\beta}_R$, and $\hat{\beta}_E$.
\begin{table}[ht]
    \centering
    \caption{Calibration on buyer's WTP}
    \resizebox{1.0\linewidth}{!}{
    \begin{tabular}{lccccccc}
    \toprule
    Coefficient & Mean & 95\% HDI & $\text{MCSE}_\text{Mean}$ & $\text{MCSE}_\text{SD}$ & $\text{ESS}_\text{Bulk}$ & $\text{ESS}_\text{Tail}$ & $\hat{R}$\\
    \midrule
    $\rho$ & \makecell{0.0087\\(0.0068)} & [0,0.0207] & 0.0001 & 0.0001 & 10743 & 6010 & 1.0005\\
    $\beta$ & \makecell{0.8093\\(0.0329)} & [0.7463,0.8703] & 0.0003 & 0.0003 & 15028 & 9440 & 1\\
    $\tau$ & \makecell{0.454\\(0.0378)} & [0.3807,0.5234] & 0.0004 & 0.0003 & 8482 & 9443 & 1.0007\\
    $\gamma$ & \makecell{0.0698\\(0.0755)} & [-0.0736,0.2095] & 0.0008 & 0.0006 & 8081 & 9516 & 1.0008\\
    $\phi$ & \makecell{-0.0282\\(0.0183)} & [-0.0628,0.0052] & 0.0002 & 0.0001 & 9051 & 9868 & 1.0006\\
    $\mu_{\alpha}$ & \makecell{0.6461\\(0.0615)} & [0.5297,0.7595] & 0.0005 & 0.0005 & 15547 & 9762 & 1\\
    $\sigma_{\alpha}$ & \makecell{0.0204\\(0.0153)} & [0,0.0485] & 0.0002 & 0.0001 & 7691 & 6167 & 0.9999\\
    $\kappa$ & \makecell{1.2212\\(0.0252)} & [1.1751,1.2702] & 0.0002 & 0.0002 & 18875 & 10182 & 1.0002\\
    \bottomrule
    \end{tabular}}
    \begin{tablenotes}
        The table reports posterior estimates obtained via Markov Chain Monte Carlo sampling. The discrete choice model was estimated using 4 parallel chains, with 3,000 tuning iterations followed by 3,000 posterior draws per chain (for a total of 12,000 retained samples). Standard deviations are in parentheses.
    \end{tablenotes}
    \label{tab:calibration_buyer}
\end{table}

\begin{table}[ht]
    \centering
    \caption{Calibration on seller's WTA}
    \resizebox{1.0\linewidth}{!}{
    \begin{tabular}{lccccccc}
    \toprule
    Coefficient & Mean & 95\% HDI & $\text{MCSE}_\text{Mean}$ & $\text{MCSE}_\text{SD}$ & $\text{ESS}_\text{Bulk}$ & $\text{ESS}_\text{Tail}$ & $\hat{R}$\\
    \midrule
    $c_0$ & \makecell{4.883\\(0.327)} & [4.255,5.536] & 0.004 & 0.002 & 8580 & 8656 & 1\\
    $c_1$ & \makecell{0.007\\(0.02)} & [-0.033,0.045] & 0 & 0 & 23219 & 10280 & 1\\
    $c_2$ & \makecell{0.32\\(0.033)} & [0.253,0.385] & 0 & 0 & 23219 & 10280 & 1\\
    $\beta_{R}$ & \makecell{0.708\\(0.105)} & [0.498,0.911] & 0.001 & 0.001 & 9038 & 8950 & 1\\
    $\beta_{E}$ & \makecell{2.976\\(0.17)} & [2.637,3.303] & 0.002 & 0.001 & 8851 & 9212 & 1\\
    $\delta$ & \makecell{-0.161\\(0.083)} & [-0.325,0.003] & 0.001 & 0.001 & 8732 & 8750 & 1\\
    $\mu_{\alpha}$ & \makecell{0.727\\(0.015)} & [0.697,0.754] & 0 & 0 & 21112 & 10019 & 1\\
    $\sigma_{\alpha}$ & \makecell{0.432\\(0.012)} & [0.41,0.456] & 0 & 0 & 6118 & 8825 & 1\\
    \bottomrule
    \end{tabular}}
    \begin{tablenotes}
        The table reports posterior estimates obtained via Markov Chain Monte Carlo sampling. The discrete choice model was estimated using 4 parallel chains, with 3,000 tuning iterations followed by 3,000 posterior draws per chain (for a total of 12,000 retained samples). Standard deviations are in parentheses.
    \end{tablenotes}
    \label{tab:calibration_seller}
\end{table}

\section{The model}
\subsection{The geographical sandbox}
\begin{figure}
    \centering
    \includegraphics[width=1.0\linewidth]{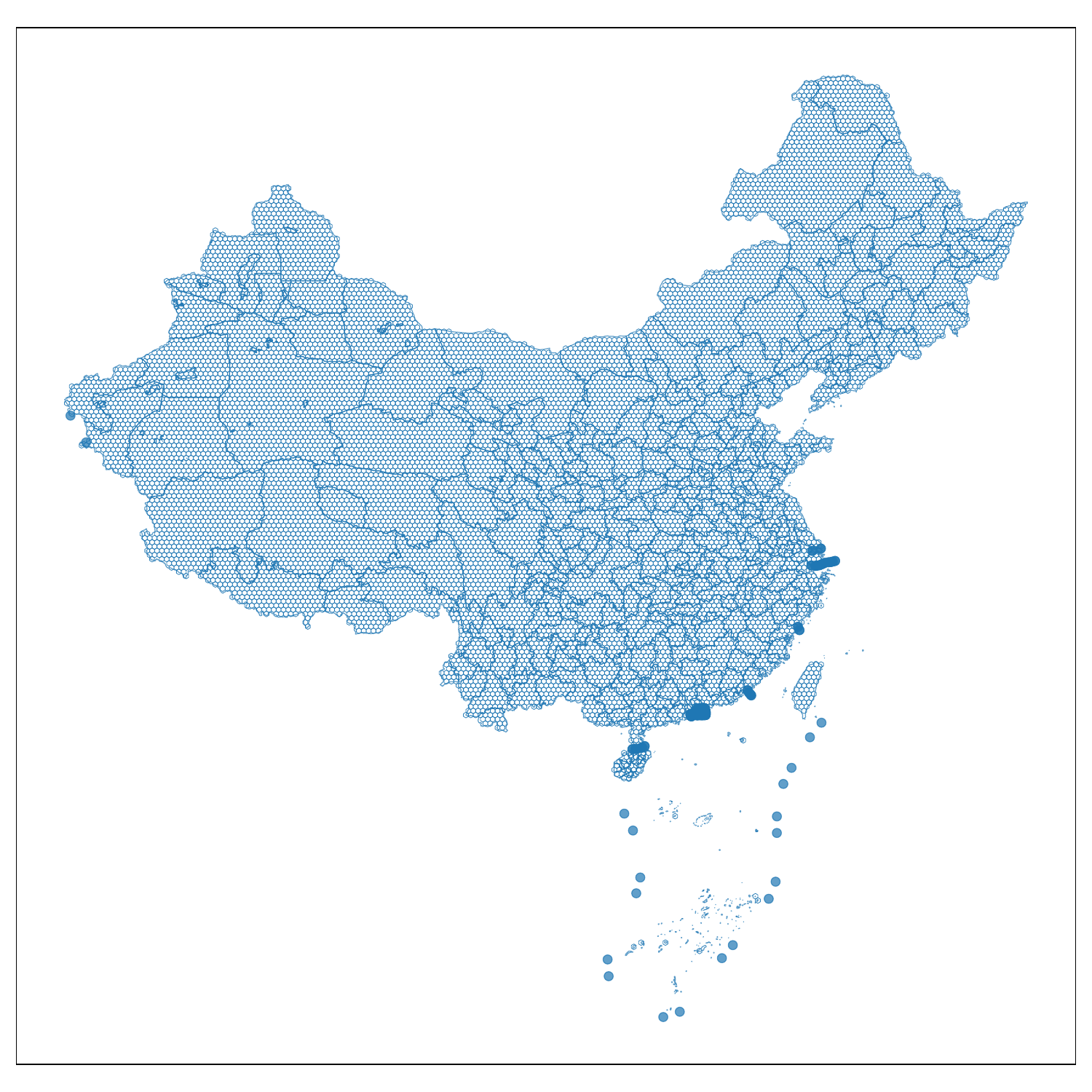}
    \caption{Hexagonal grid with 20 km radius}
    \begin{figurenotes}
        The dark blue lines indicate the boundaries of municipal administrative units.
    \end{figurenotes}
    \label{fig:hex_grid}
\end{figure}

Our first step is to discretize China’s economic space---excluding Hong Kong, Macao, and Taiwan due to the distinctiveness of their legal jurisdictions---into a nationwide lattice of 14,526 hexagonal cells with a 20-kilometer radius (center-to-vertex), each intended to represent a potential industrial cluster (Fig.~\ref{fig:hex_grid}). We generate the grid in an equal-area projection to ensure that cell areas are comparable across latitudes, then clip it to the national land boundary and drop cells with negligible land share (e.g., open water). Each cell $i$ is indexed by its centroid $c_{i}$ and treated as a stable unit over time, so the empirical panel becomes $\{\text{cell}\times\text{time}\}$. This choice deliberately avoids the complexity and changes in administrative boundaries and reduces the extent to which results hinge on idiosyncratic local jurisdictional shapes, which is a common concern in spatial analysis.

Compared with triangle or square tessellations, hexagons provide the most balanced and isotropic discretization of continuous space: every cell has up to six first-order neighbors at the same edge-to-edge distance, which minimizes directional artifacts when modeling spillovers, diffusion, and commuting. Formally, we define a contiguous adjacency graph on the grid, where $j\in\mathcal{N}(i)$ if cells $i$ and $j$ share a border, and use this to construct spatial adjacency and centroid-to-centroid distance $d_{ij}=||c_{i}-c_{j}||$. This makes it straightforward to operationalize ``local'' versus ``regional'' interactions (e.g., within one cell vs. within one ring of neighbors) while keeping the meaning of ``neighbor'' constant nationwide. 

In economic geography, hex grids are also consistent with the intuition from central place theory that, under uniform population density and transport costs, market areas tend to approximate hexagonal service regions \citep{Meeteren2018cpt}. The 20-kilometer radius is chosen to approximate a typical sub-hour commuting shed, aligning the cell with a plausible local labor-market and product-market catchment \citep{marchetti1994travel}. Accordingly, each hexagonal cell can be interpreted as a spatial unit within which industries predominantly operate under intra-city (or intra-urban-area) dynamics, while interactions across cells capture inter-city spillovers---such as worker mobility, supplier–customer linkages, knowledge diffusion, and policy externalities---implemented transparently through the neighbor structure and spatial weights.

\subsection{Spatial distribution of agents}
To ensure that our agent-based model reflects the real-world industrial landscape and preserves the practical relevance of its geographic sandbox, we allocate agents across the hexagonal grid using the observed spatial distribution of economic activity. Concretely, all real-world observations on (i) artificial intelligence enterprises and (ii) top-tier hospitals are geocoded and then assigned to the unique hexagonal cell that contains their coordinates (i.e., a point-in-polygon mapping), yielding cell-level proxies for the density of buyers and sellers.

On the buyer side, we compiled administrative-region data on the number of AI enterprises registered in each county/district $r$ in mainland China. Because enterprise counts are observed at the county/district level rather than as point locations, we project them onto the hex grid using an equal-within-region allocation rule. Let $B_{r}$ denote the number of AI enterprises in region $r$ and let $\mathcal{J}(r)$ be the set of hex cells whose centroids fall within $r$, with cardinality $N_{r}=|\mathcal{J}(r)|$. We assign each cell $i\in\mathcal{J}(r)$ an expected buyer intensity
\begin{equation}
    b_{i}=\frac{B_{r}}{N_{r}},
\end{equation}
which preserves the regional total $\sum_{i\in\mathcal{J}(r)}b_{i}=B_{r}$ while avoiding strong parametric assumptions about within-county concentration.

On the seller side, we identify top-tier hospitals using their points-of -interest (POI) coordinates and compute a direct cell-level count. Let $S_{i}$ be the number of top-tier hospital POIs falling inside cell $i$; we use $S_{i}$ as a proxy for the local availability of high-end medical resources, and thus proxy for the spatial distribution of seller intensity $s_i$.

To translate these continuous and quasi-continuous cell attributes into a parsimonious discrete agent landscape, we apply the \citet{Jenks1963breaks} natural breaks algorithm separately to the buyer intensity $b_{i}$ and the seller intensity $s_{i}$, partitioning each into six tiers (Tier 0 through Tier 5). Jenks classification chooses breakpoints that minimize within-tier dispersion (equivalently, minimize within-class sum of squared deviations) while maximizing between-tier separation, producing categories that align with the empirical clustering of the underlying spatial distributions rather than imposing equally spaced or quantile-based bins.

\begin{figure*}
    \centering
    \includegraphics[width=1.0\linewidth]{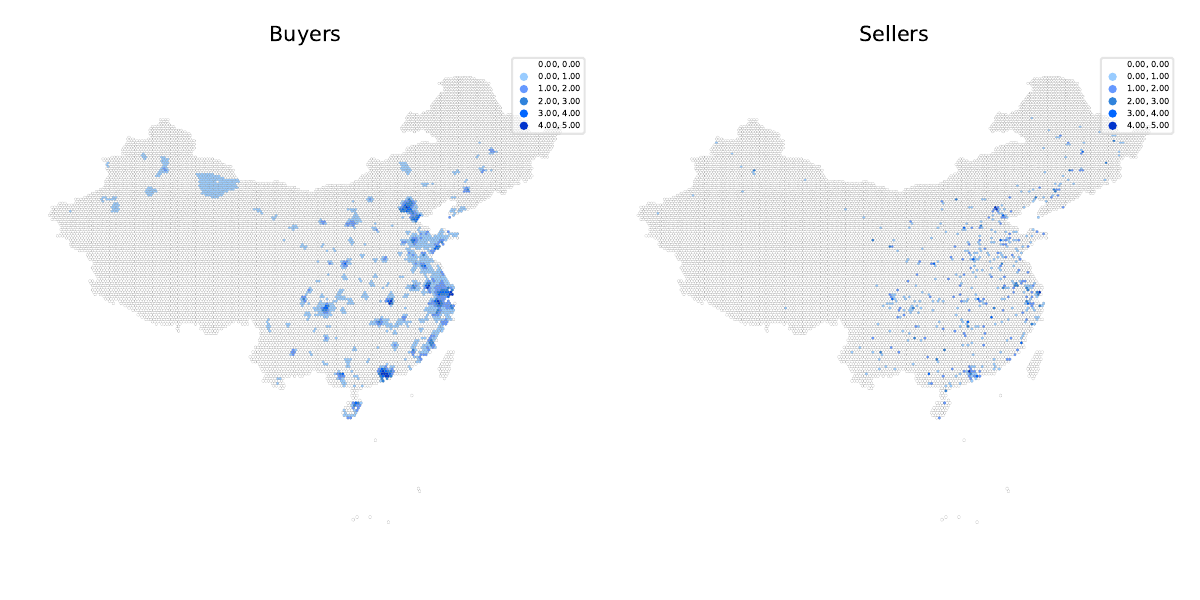}
    \caption{Distribution of agents}
    \label{fig:agents}
    \begin{figurenotes}
        The graphs illustrate the spatial distribution of buyer and seller agents, with colors indicating the number of agents within each cell, respectively.
    \end{figurenotes}
\end{figure*}

Without loss of generality, we then map these tiers into a simplified discrete agent configuration. Tier 0 contains no agents. Tier $t\in\{1,2,3,4,5\}$ contains exactly $t$ agents---one at each capacity ``level'' from Level 1 up to Level $t$. Equivalently, a Tier-$t$ cell contains agents $\{\text{Level 1},\cdots,\text{Level}\ t\}$, so Tier 5 contains one agent from every level (Level 1 through Level 5). This monotone tier-to-level mapping preserves the empirical ordering of cell intensities while keeping the ABM tractable: higher-tier cells represent systematically richer local demand (buyers) or supply (sellers) environments, and heterogeneity across tiers enters the model through the presence of higher-level agents. The resulting spatial distributions of buyer and seller agents are shown in Fig.~\ref{fig:agents}.

\subsection{Agents' action strategies}
\subsubsection{Buyer}
From Eq.(\ref{eq:buyer_simple}), the buyer's willingness to pay on dataset $x_j$ at time $t$ is given as
\begin{equation}
    \text{WTP}_{ijt} = \frac{f(x_{it})\cdot\beta x_{j} + \tau s_j - \kappa \ln(1 + d_{ij})}{\alpha_i}.
\end{equation}
In each period $t$, buyers iterate through all the unconnected sellers and compute a willingnness to pay ($\text{WTP}_{ijt}$) for each seller $j$. Each buyer then selects the seller $j$ with the highest $\text{WTP}_{ijt}$ among all potential sellers and attempts to establish a connection. Once connectionns are formed---according to the action strategy outlinend in Algorithm~\ref{alg:seller}---the paired agents negotiate a transaction price $p_{ijt}$. A deal is executed if the negotiated price does not exceed the buyer's budget constraint. The detailed action strategy of buyers is presented in Algorithm~\ref{alg:buyers}.
\begin{algorithm}
    \caption{Buyer's action strategy}\label{alg:buyers}
    \begin{algorithmic}
        \STATE STEP\_BUYER(\textbf{b}, \textbf{s})
        \STATE \hspace{0.5cm} $\text{WTP}_{ijt}\leftarrow U^{\text{Buyer}}(b_i,\ s_j)$ \textbf{for} $s_j\in\{\textbf{connected}=0\}$
        \STATE \hspace{0.5cm} \textbf{select} $s_j$ \textbf{if} $\text{WTP}_{ijt}=\max_{j\in\textbf{Seller}}(\text{WTP})$
        \STATE
        \STATE \hspace{0.5cm} \textbf{connected} $\leftarrow$ \hyperref[alg:seller]{\text{STEP\_SELLER(\textbf{b}, \textbf{s})}}
        \STATE
        \STATE \hspace{0.5cm} \textbf{if connected:}
        \STATE \hspace{1.0cm}\textbf{if} $\text{WTP}_{ijt}\leq\text{WTA}_{jt}$\textbf{:}
        \STATE \hspace{1.5cm} \textbf{return}
        \STATE \hspace{1.0cm} \textbf{else:}
        \STATE \hspace{1.5cm} $p_{ijt}\leftarrow p\in [\text{WTA}_{jt},\text{WTP}_{ijt}]$
        \STATE \hspace{2.0cm} \textbf{if} $p_{ijt}>m_{it}$\textbf{:}
        \STATE \hspace{2.5cm} \textbf{return}
        \STATE \hspace{2.0cm} \textbf{else:}
        \STATE \hspace{2.5cm} $x_{it}=x_{i,t-1}+x_{j}$
        \STATE \hspace{2.5cm} $m_{it}=m_{i,t-1}-p_{ijt}$
        \STATE \hspace{2.5cm} \textbf{return}
        \STATE \hspace{0.5cm} \textbf{else:}
        \STATE \hspace{1.0cm} \textbf{return}
    \end{algorithmic}
    \begin{tablenotes}
        $\textbf{b}$ and $\textbf{s}$ denote individual buyer and seller agents, respectively. $\textbf{Seller}$ denotes the set of sellers.  $\textbf{connected}$ is a Boolean indicator representing the connection status between a buyer and a seller. $m_{it}$ denotes the remaining budget of buyer $i$ at time $t$.
    \end{tablenotes}
\end{algorithm}

\subsubsection{Seller}
From Eq.(\ref{eq:seller_simple}), the seller’s minimum acceptable price (willingness to accept, WTA) can be derived from the condition $U_{jt}^{\text{Seller}} = 0$ (indifference to trading or not). Solving
\begin{equation}
    \alpha_j p_{ijt} = c_0 + c_2 x_{j} + \beta_R R_j + \beta_E E_{jt}
\end{equation} gives
\begin{equation}
\text{WTA}_{jt} = \frac{c_0 + c_2 x_{j} + \beta_R R_j + \beta_E E_{jt}}{\alpha_j}.
\end{equation}

This formula shows that a seller will set a higher floor price for transactions that are costly, risky, or legally perilous. It resonates with the intuition that data suppliers need to be compensated for both direct costs and indirect expected costs (like risk of sanctions). Notably, many of the seller-side coefficients mirror those in the buyer utility. (e.g., $x_{j}$ appears for both sides, reflecting that volume affects both benefit to buyer and cost to seller). However, the enforcement ($E$) and risk ($R$) variables are uniquely pertinent to the legal-economic environment of the seller, underlining our paper’s focus on law and economics interplay in data markets.

Through interactions with buyers and exogenous market environment, each seller's action strategy involves sensing, decision-making, and updating, as outlined in Algorithm~\ref{alg:seller}. In each period $t$, sellers update their WTA by recalculating it under the renewed enforcement intensity $E_{jt}$ and comparing the result with the previous transaction price $p_{j,t-1}$; the higher value is adopted as the new $\text{WTA}_{jt}$. Each seller then selects, from the set of received buyer offers, the buyer with the highest WTP and communicates this choice back to the buyers.
\begin{algorithm}
    \caption{Seller's action strategy}\label{alg:seller}
    \begin{algorithmic}
        \STATE STEP\_SELLER(\textbf{b}, \textbf{s})
        \STATE \hspace{0.5cm} $\text{WTA}_{jt}\leftarrow U^{\text{Seller}}(b_i,\ s_j)$ \textbf{with} $E_{jt}$
        \STATE \hspace{0.5cm} $\text{WTA}_{jt}\leftarrow\max\{\text{WTA}_{jt},\ p_{j,t-1}\}$
        \STATE \hspace{0.5cm} \textbf{select} $b_i$ \textbf{if} $\text{WTP}_{ijt}=\max_{b_i\in\textbf{Received}}(\text{WTP})$
        \STATE
        \STATE \hspace{0.5cm} \textbf{return} ($b_i$, \textbf{connected})
    \end{algorithmic}
    \begin{tablenotes}
        $\textbf{b}$ and $\textbf{s}$ denote individual buyer and seller agents, respectively. $E_jt$ denotes the enforcement intensity on seller $j$ at time $t$. $\textbf{Received}$ denote the set of potential buyers. $\textbf{connected}$ is a Boolean indicator representing the connection status between a buyer and a seller.
    \end{tablenotes}
\end{algorithm}

\subsection{External environment}
\subsubsection{Volume-indexed assignment of transaction risk.}
In the model, the risk class attached to a data sale is not a fixed attribute of the seller or the contract. Instead, it is an environmental label that reflects how regulators, counterparties, and technical auditors would probabilistically perceive the hazard of a contemplated transfer given the scale of data involved. Formally, we treat the seller’s risk level $R \in \{1, 2, 3\}$ as a discrete, ordered outcome that is randomly drawn conditional on the seller’s data volume $x_j$. The economics intuition is straightforward: larger datasets expand the attack surface through greater likelihood and scope of leakage, and raise the third-party exposure from re-identification. Hence, holding all else equal, larger $x_j$ should be associated with a higher probability of being classified as medium or high risk. We implement this mapping with an ordered logit. Let
\begin{equation}
z_R(x_j) = \ln\left(1 + \frac{x_j}{\text{scale}}\right), \quad \text{scale} > 0
\end{equation}
a monotone, concave transform that tempers extreme volumes and yields stable probabilities across heterogeneous sellers. The ordered logit then assigns probabilities to $R = 1, 2, 3$ using a slope (intensity) parameter $\gamma = R\_logit\_gamma > 0$ and two cut points $c_1 < c_2$ (R\_logit\_cuts):
\begin{equation}
    \begin{split}
    & \Pr(R=3 \mid x_j) = \sigma(\gamma[z_R(x_j) - c_2]), \\
    & \Pr(R=2 \mid x_j) = \sigma(\gamma[z_R(x_j) - c_1]) - \Pr(R=3 \mid x_j), \\
    & \Pr(R=1 \mid x_j) = 1 - \Pr(R=2 \mid x_j) - \Pr(R=3 \mid x_j),
\end{split}
\end{equation}
where $\sigma(u) = \frac{1}{1 + e^{-u}}$. 

Two design choices make this assignment data-adaptive rather than \textit{ad hoc}. First, the scaling constant $scale = R\_logit\_scale$ is calibrated from the cross-section of sellers, using the median of positive $x_j$, anchoring $z_R$ near zero for a typical seller and preventing extreme volumes from mechanically saturating the top risk class. Second, the cut points $(c_1, c_2)$ are set endogenously to the market’s volume distribution. We take the 33rd and 67th percentiles of $z_R$ across all sellers. This yields an baseline: in an otherwise uninformative environment, roughly one third of sellers would fall into each risk tier; as volumes shift (e.g., because entry brings many small sellers, or consolidation yields a few very large ones), the induced risk mix adjusts mechanically rather than by manual re-tuning.

Economically, the specification implies testable predictions. Holding seller tier $s_j$ and enforcement $E_jt$ fixed, thicker right tails in the market’s $x_j$ distribution push mass toward higher $R$, raising average WTA and reducing marginal trade at the top end. Tightening $\gamma$ (a steeper slope) amplifies this effect, yielding more convex risk premia in volume. Conversely, if market development lowers typical package sizes (a left shift in $x_j$), the induced downgrading of $R$ relaxes sellers’ reservation prices and expands feasible trades—precisely the margin along which data minimization, sampling, or purpose-limited access policies are expected to improve welfare.

\subsubsection{Activity-indexed enforcement intensity.}
Whereas the intrinsic transaction risk $R_j$ is an exogenous label tied to a seller’s own data volume and assigned once at entry, enforcement intensity $E_{jt} \in \{1, 2, 3\}$ is modeled as a dynamic, locally endogenous state that evolves with market activity in the seller’s surroundings. The economic intuition is that regulators, auditors, platforms, and public scrutiny allocate attention where the probability and salience of violations are highest. In data markets, that attention is not uniformly distributed: spatial and temporal ``hot spots'' emerge as volumes concentrate, raising detection probabilities, tightening audits, and increasing the expected private cost of non-compliance. We capture this by letting $E_jt$ be re-classified each period via an ordered logit that depends on recent neighborhood trade volume.

At initialization $(t = 0)$, there is no history, so all sellers start at the baseline $E_{j0} =1$. At the beginning of each subsequent period, the model computes, for every seller $j$, a windowed neighborhood total of traded data: the sum of all package sizes $x$ transacted in $j$'s own cell plus its geometrically adjacent cells over the last $E_{\text{window}}$ periods. If this total is zero, enforcement remains at the baseline $E_{jt} = 1$. Otherwise, the total is converted into a concave, monotone score
\begin{equation}
z_E = \ln\left(1 + \frac{\text{total}}{\text{scale}}\right),
\end{equation}
which tempers extreme spikes while preserving rank. This score is mapped to probabilities for $E \in \{1, 2, 3\}$ through an ordered-logit link with slope $\gamma_E > 0$ and cut points $c_1^{(E)} < c_2^{(E)}$:
\begin{equation}
\begin{split}
& \Pr(E=3 \mid z_E) = \sigma(\gamma_E[z_E - c_2^{(E)}]), \\
& \Pr(E=2 \mid z_E) = \sigma(\gamma_E[z_E - c_1^{(E)}]) - \Pr(E=3 \mid z_E), \\
& \Pr(E=1 \mid z_E) = 1 - \Pr(E=2 \mid z_E) - \Pr(E=3 \mid z_E),
\end{split}
\end{equation}
with $\sigma(u) = (1 + e^{-u})^{-1}$. A seller's $E_jt$ for the current step is a single draw from this categorical distribution.This adaptive calibration lets the enforcement landscape track endogenous changes in congestion. As activity clusters or disperses, the implied thresholds shift without manual tuning. Economically, this specification formalizes a local feedback mechanism between market conduct and legal exposure. Concentrated trading raises $z_E$, which increases the probability of being classified at $E = 2\ \text{or}\ E = 3$, thereby internalizing a greater portion of social monitoring and sanction risk into sellers’ reservation prices in the next round. The model thus embeds the comparative statics that policy makers care about: increases in neighborhood volume raise expected enforcement, and the risk–enforcement complementarity $(\beta_3 > 0)$ ensures that this bite is strongest precisely where intrinsic hazards are high. Unlike a fixed-penalty assumption, the activity-indexed $E_jt$ captures how regulatory capacity, platform audits, and public attention endogenously follow the market, yielding a lawful channel through which local congestion begets stricter oversight and, in turn, reshapes equilibrium prices, participation, and spatial patterns of trade.

\subsection{Interaction cycles}
At time $t=0$, the spatial positions and attributes of all agents are initialized according to the predefined rules. The exogenous enforcement intensity is initially set at a low level ($E=1$). In each subsequent period $t$, the system evolves following the process outlined in Algorithm~\ref{alg:market_model}. Periodically, the enforcement intensity $E$ of each cell is probabilistically updated based on the transaction frequency within that cell and its six adjacent cells during the preceding period---the greater the local transaction volume, the higher the likelihood that the enforcement intensity will be elevated. Based on the updated enforcement intensity, each seller recalculates its transaction cost $c_t$ at time $t$ and compares it with the previous transaction price $p_{j,t-1}$, adopting the higher value as its new willingness to accept (WTA). Each buyer $i$ then initiates the offer process by selecting, from all unengaged sellers, the seller $j$ with the highest willingness to pay ($\text{WTP}_{jit}$) and issuing an offer. Each seller, in turn, chooses from the received offers the buyer with the highest $\text{WTP}_{ijt}$ to enter negotiation. The transaction price $p_{ijt}$ is determined as a convex combination
\begin{equation}
    p_{ijt}=\frac{s_{j}}{z_{i}+s_{j}}\times \text{WTP}_{ijt}+\frac{z_{i}}{z_{i}+s_{j}}\times \text{WTA}_{jt}
\end{equation}
of the buyer's and seller's valuations, reflecting their relative bargaining power
\begin{equation}
    \lambda=\frac{s_j}{z_i+s_j},\ 1-\lambda=\frac{z_i}{z_i+s_j}.
\end{equation}
If the negotiated price does not exceed the buyer's remaining budget constraint $m_{it}$, the transaction is executed.
\begin{algorithm}
    \caption{Process of the multi-agent system}\label{alg:market_model}
    \begin{algorithmic}
        \STATE
        \STATE {\textsc{STEP\_MODEL}}(\textbf{Buyer}, \textbf{Seller})
        \STATE \hspace{0.5cm} $E\leftarrow E\in\{1,2,3\}$ \textbf{if} $t=t_{E}\times k,\ k\in \textbf{Z}$
        \STATE \hspace{0.5cm} $\text{WTA}_{jt}\leftarrow\max\{c_{jt},\ p_{j,t-1}\}$ \textbf{for} $j\in\textbf{Seller}$
        \STATE
        \STATE \hspace{0.5cm} \hyperref[alg:buyers]{\textsc{STEP\_BUYER}(\textbf{b}, \textbf{s})};
        \STATE \hspace{0.5cm} \hyperref[alg:seller]{\textsc{STEP\_SELLER}(\textbf{b}, \textbf{s})} \textbf{for} $\textbf{b},\ \textbf{s}\in\textbf{Buyer},\ \textbf{Seller}$
        \STATE
        \STATE \hspace{0.5cm} \textsc{DATA\_COLLECT}($t$)
        \STATE \hspace{0.5cm} $t=t+1$
        \STATE \hspace{0.5cm} \textbf{return}
    \end{algorithmic}
    \begin{tablenotes}
        \textbf{Buyer} and \textbf{Seller} denote the sets of buyer and seller agents in the model, respectively. The parameter $t_{E}$ represents the frequency with which enforcement intensity is recalculated. \textsc{DATA\_COLLECT} is a function used to record measurement indicators throughout the simulation process.
    \end{tablenotes}
\end{algorithm}

\subsection{Key performance indicators}
We use trades, volume, buyer surplus, seller surplus, externality, and total welfare as the ABM’s key performance indicators (KPIs) because, taken together, they span the full efficiency–distribution–harm triad that law and economics evaluation typically requires. Trades capture the extensive margin---how often the market clears under a given legal regime---while volume captures the intensive margin---how much is transacted conditional on clearing, and thus whether a regime changes not only whether parties transact but how much they transact.

Because prices are endogenous in our model (emerging from matching and bargaining rather than being imposed), private benefits are most transparently summarized by buyer surplus (BS) and seller surplus (SS) in a single trade. For a completed transaction $k$ at price $p_k$, buyer willingness-to-pay $\text{WTP}_k$, and seller willingness-to-accept $\text{WTA}_k$, we define
\begin{equation}
    BS_{k}=\max\{\text{WTP}_k-p_k,\,0\}
\end{equation}
and
\begin{equation}
    SS_{k}=\max\{p_k-\text{WTA}_k,\,0\}.
\end{equation}
Aggregating across all trades in a period yields a consumer surplus $CS=\sum_k BS_k$ and a producer surplus $PS=\sum_k SS_k$. These two measures decompose who captures the gains from trade and therefore allow us to detect rule-induced transfers (e.g., regimes that increase seller bargaining power without changing allocative efficiency). They also distinguish ``more trade'' from ``better trade'': a rule can increase throughput while compressing one side’s surplus, or expand surplus mainly through price changes rather than improved matching.

Because many legal interventions are motivated by spillovers rather than bilateral efficiency, we explicitly track externality---the quantified harm imposed on third parties not internalized by the contracting pair. Let $Ext_k\ge 0$ denote the harm generated by transaction $k$ (potentially a function of quantity, risk type, information, or enforcement). We then compute total externality as $Ext=\sum_k Ext_k$. This component is crucial because legal rules can raise private surplus while worsening social outcomes (e.g., by expanding activity that produces uncompensated harm).

Finally, total welfare aggregates the model’s normative components into a Kaldor–Hicks benchmark
\begin{equation}
    W \;=\; CS + PS - Ext,
\end{equation}
with the understanding that it captures net efficiency rather than equity. Importantly, separating $CS$, $PS$, and $Ext$ prevents ``welfare improvements'' from being a black box: it allows us to identify whether a legal regime raises $W$ by (i) improving matching and enabling mutually beneficial trades, (ii) reallocating surplus between buyers and sellers, or (iii) reducing external harms via deterrence or compliance.

This KPI set also maps cleanly onto the ABM's core mechanisms---matching (affecting trades), bargaining/pricing (shaping the split between $CS$ and $PS$), budgets and constraints (limiting volume), risk and information (driving both surplus and harms), and enforcement (altering incentives and the externality term). As a result. it supports principled comparisons across regimes that may increase throughout without increasing welfare, reduce harm at the cost of volume, or shift surplus without meaningfully changing total output.

\section{Validation}
We validate our agent-based model (ABM) of the data-transaction market through external validation, assessing whether the simulated market reproduces (i) stylized facts emphasized in the economic geography and platform-economy literatures, (ii) spatial–temporal regularities that emerge in observed AI and data activity, and (iii) macro-level empirical benchmarks on real-world data trading. Rather than fitting the model to micro-transaction records (which are rarely observable for data trades), we require that the model—given empirically seeded geography and the institutional rules of each regime—generates aggregate patterns that are qualitatively and quantitatively consistent with externally documented evidence.

\begin{figure*}[ht]
  \centering
  \begin{subfigure}[t]{0.48\textwidth}
    \centering
    \includegraphics[width=\linewidth]{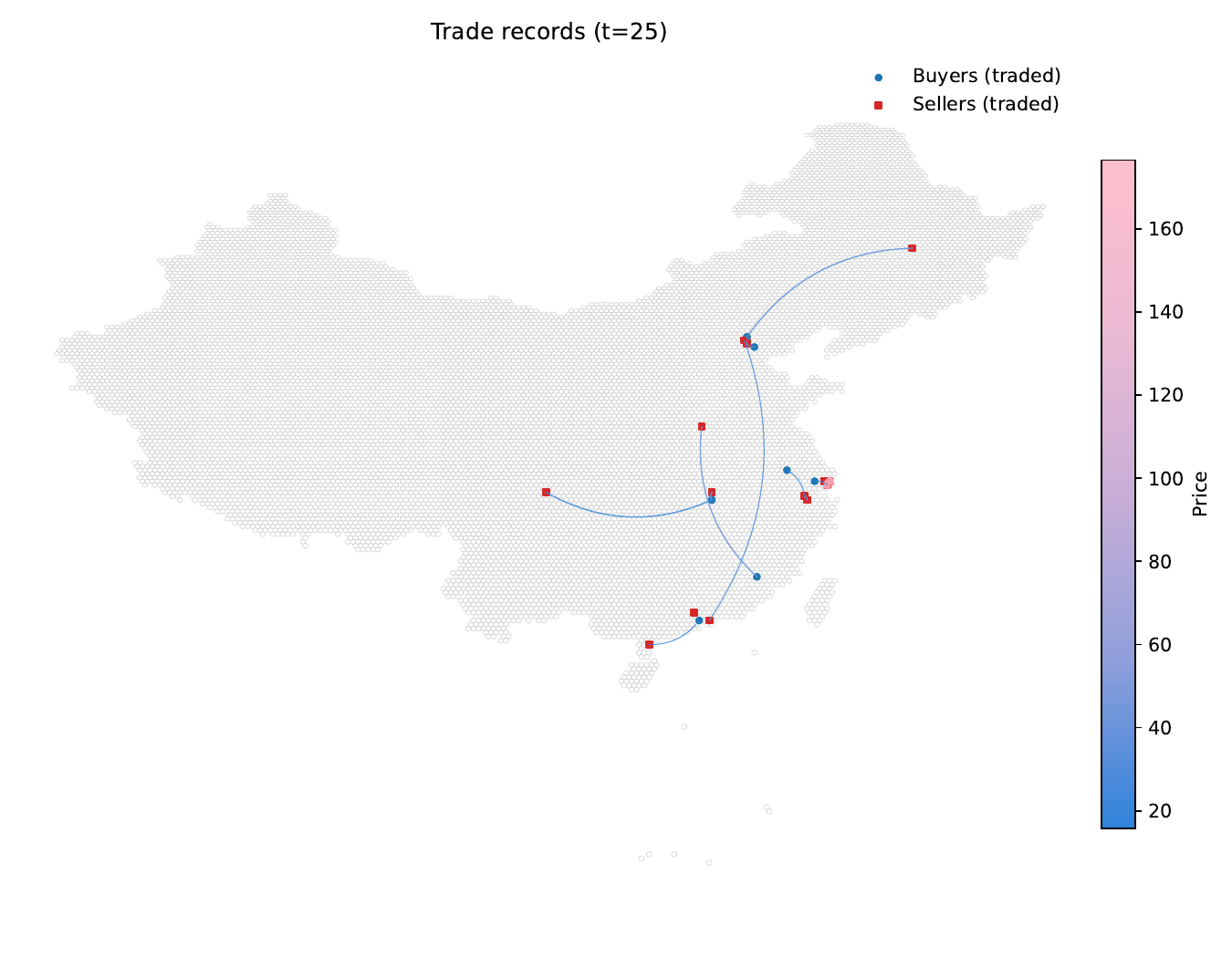} 
    \caption{}\label{fig:a}
  \end{subfigure}\hfill
  \begin{subfigure}[t]{0.48\textwidth}
    \centering
    \includegraphics[width=\linewidth]{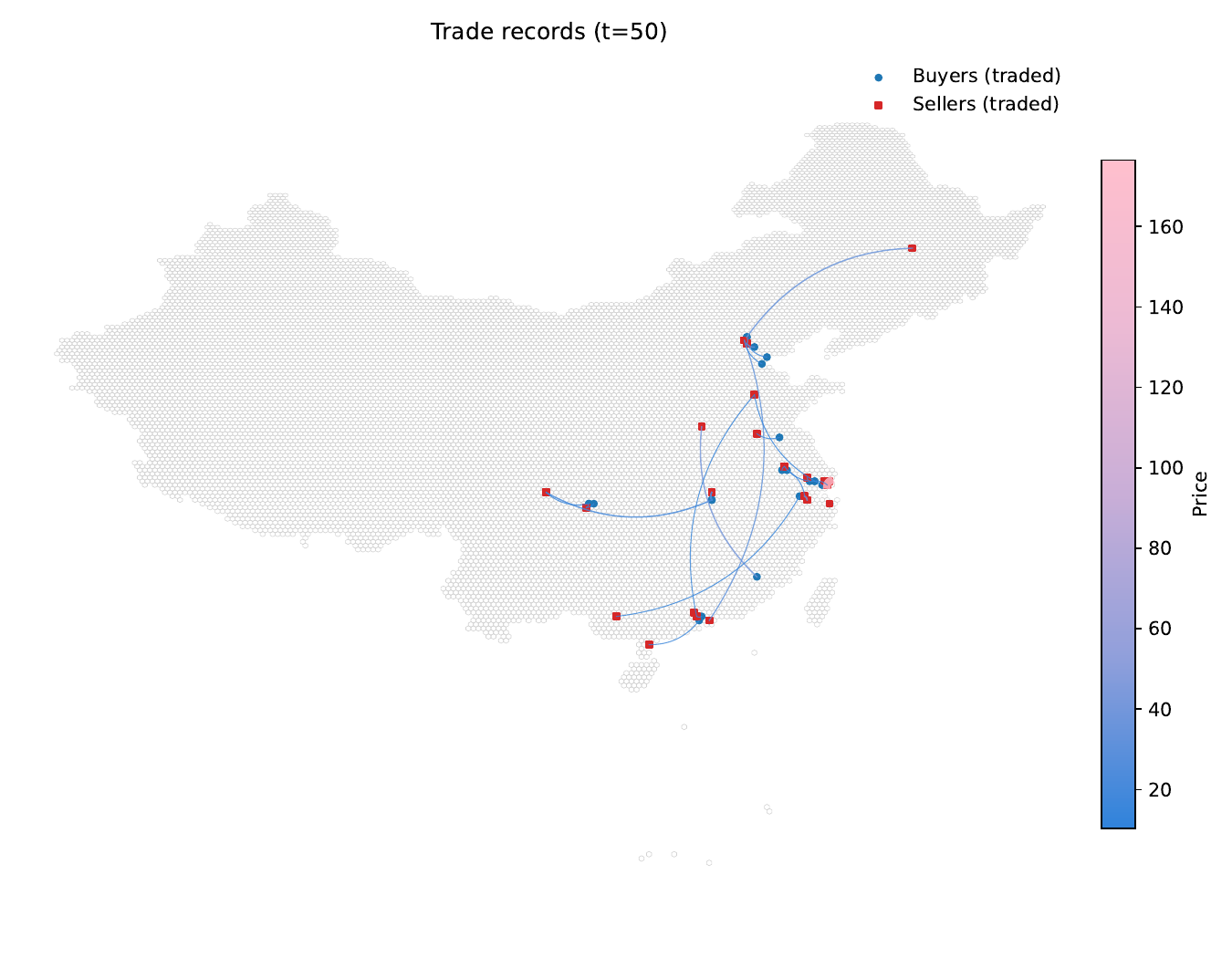}
    \caption{}\label{fig:b}
  \end{subfigure}
  \vspace{0.8em}
  \begin{subfigure}[t]{0.48\textwidth}
    \centering
    \includegraphics[width=\linewidth]{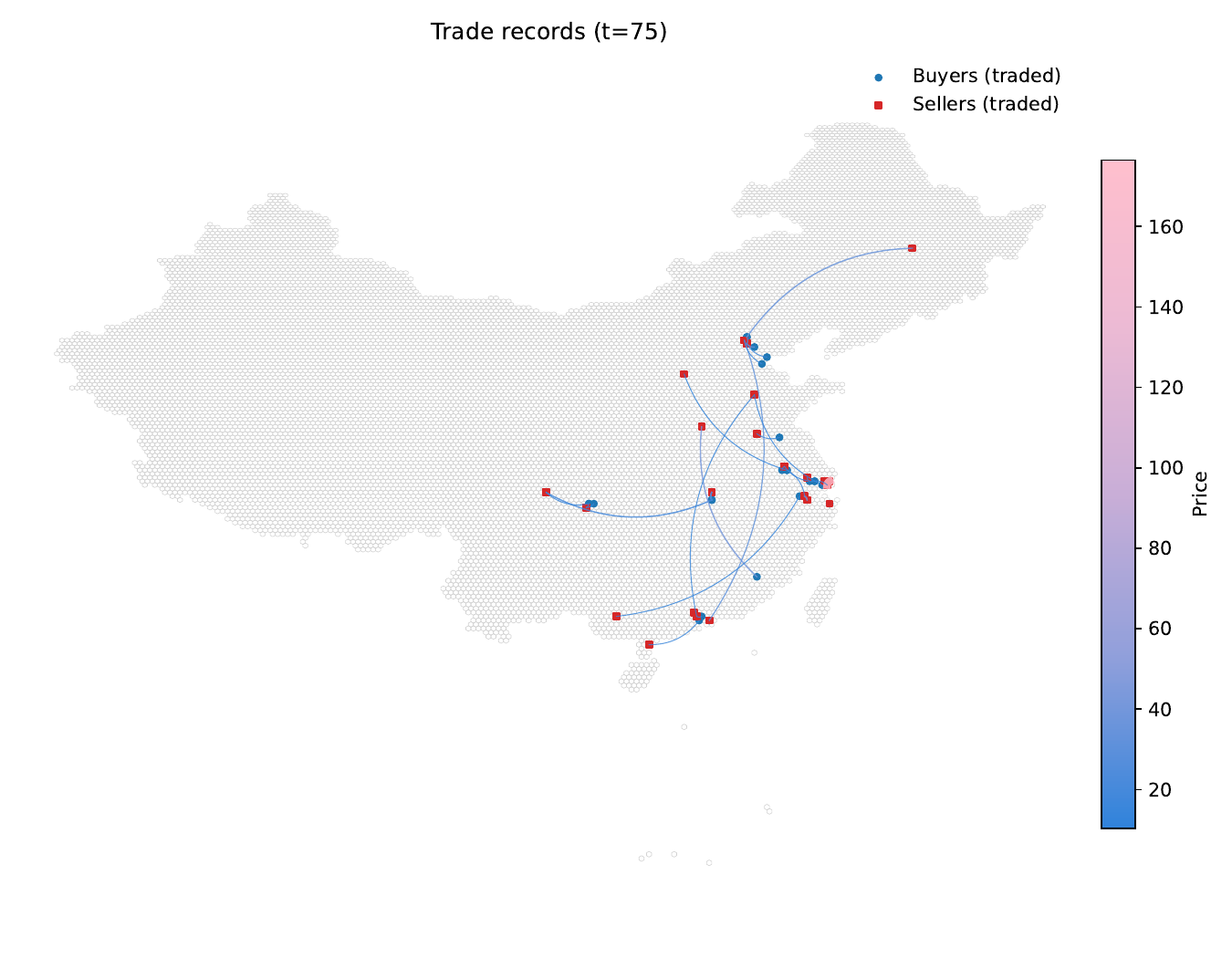}
    \caption{}\label{fig:c}
  \end{subfigure}\hfill
  \begin{subfigure}[t]{0.48\textwidth}
    \centering
    \includegraphics[width=\linewidth]{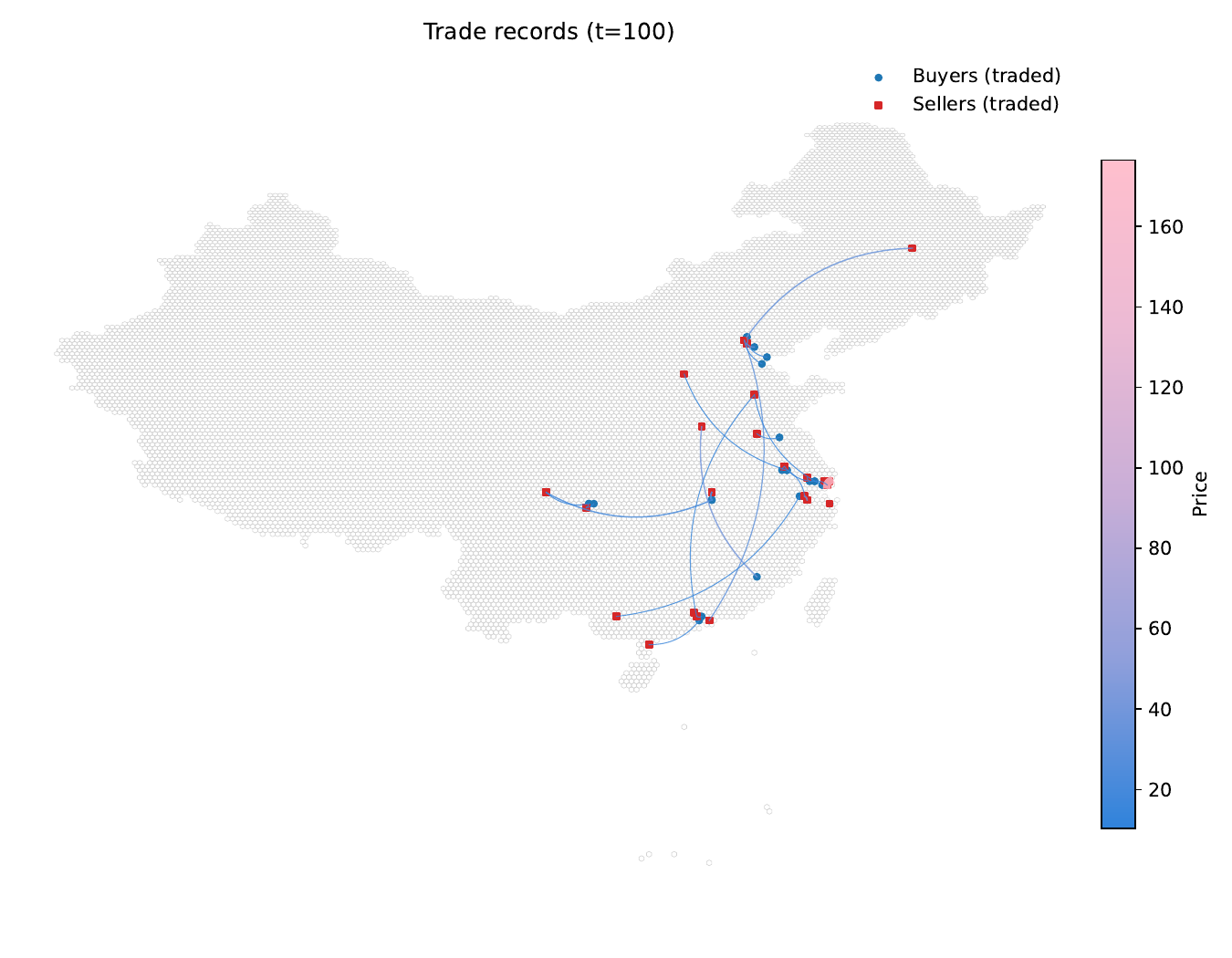}
    \caption{}\label{fig:d}
  \end{subfigure}
  \caption{Trade arcs between buyer and seller cells in the baseline market model}
  \begin{figurenotes}
      Arcs connect the centroids of the buyer and seller hexagons for each realized deal in the simulation. Line width scales with the traded data volume $x_j$; color denotes the price as configured. Only records with a ``deal'' status within the specified time filter are drawn.
  \end{figurenotes}
  \label{fig:baseline_map}
\end{figure*}

Fig.~\ref{fig:baseline_map} reproduces three salient spatial regularities that are widely documented for China’s AI and data economy. (1) Coastal concentration: transaction arcs concentrate in major coastal clusters, including the Yangtze River Delta (Shanghai–Hangzhou–Nanjing–Suzhou), the Pearl River Delta (Guangzhou–Shenzhen), and the Beijing–Tianjin corridor. (2) Distance decay: within-region trades systematically dominate across-region trades, consistent with the standard gravity intuition that frictions rise with distance. (3) Hub formation: cells seeded with higher initial levels accumulate disproportionately large flows, producing a hub-and-spoke structure visible as dense bundles of arcs converging on a small number of coastal nodes. In the simulation, a small set of coastal buyer cells acts as national demand aggregators, sourcing from a geographically dispersed set of sellers; visually, this appears as star-shaped “inbound” bundles. This pattern is consistent with a core–periphery demand structure in which capital- and capability-dense coastal markets intermediate demand, while peripheral buyers transact locally or remain inactive when fixed costs and enforcement frictions are binding.

The same figure also illustrates a coherent temporal regularity: the market exhibits early exploration followed by later consolidation. In early periods (e.g., $t=25$), the network is dominated by numerous short-range, intra-region exploratory links; by $t=50$ to $t=100$, a smaller set of longer-range inter-region arcs persists and thickens, reflecting the endogenous emergence of stable cross-regional relationships and routing through hubs. This transition---from many tentative local links to a more consolidated inter-regional network---is consistent with canonical mechanisms in geographical economics and networked trade, where repeated interactions, information accumulation, and reputation/relational contracting shift transactions toward more efficient but more spatially extensive linkages over time \citep{Rauch1999network}.

\begin{figure}
    \centering
    \includegraphics[width=1.0\linewidth]{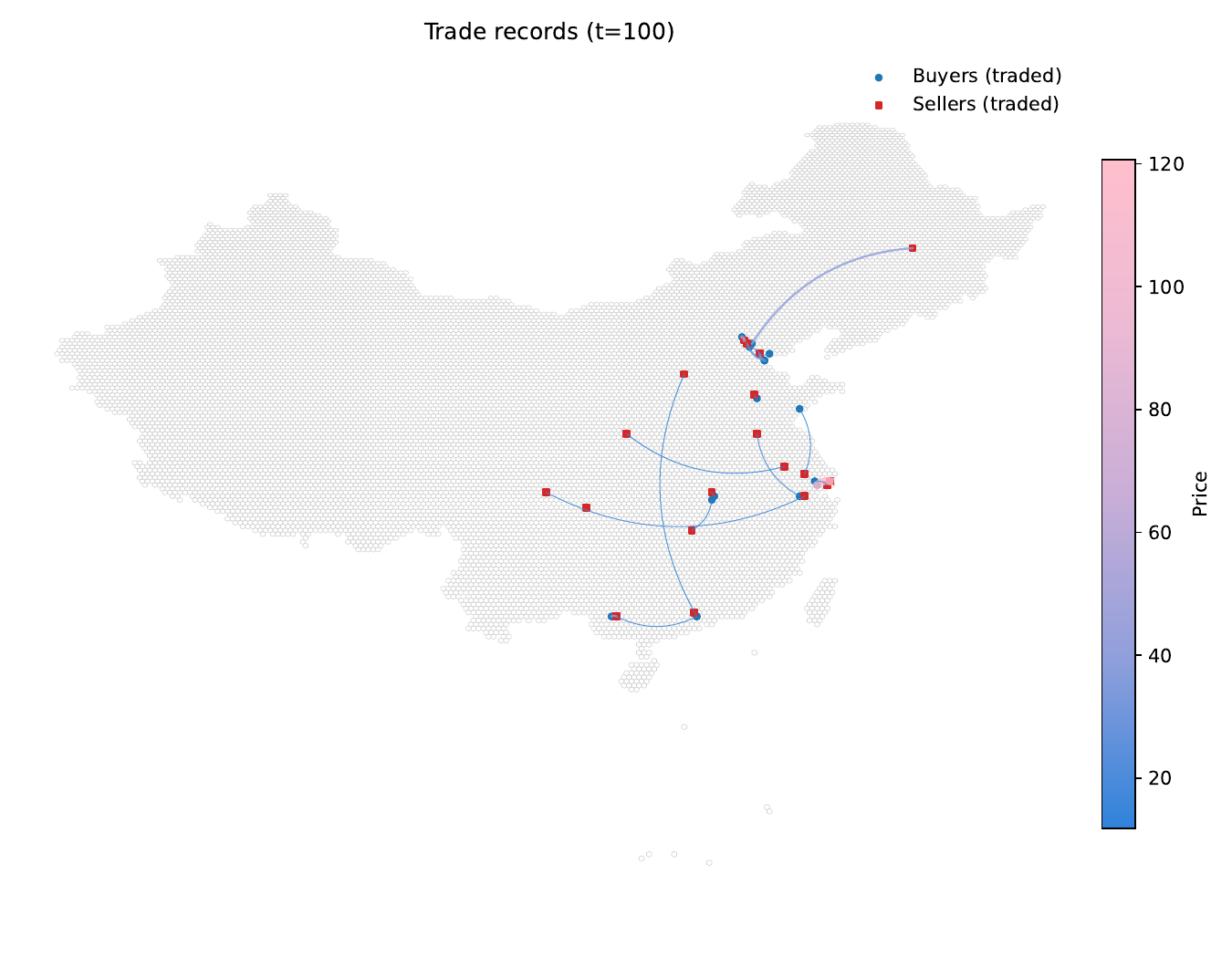}
    \caption{Trade arcs in platform-mediated exchange ($t=100$)}
    \begin{figurenotes}
        Arcs connect the centroids of the buyer and seller hexagons for each realized deal in the simulation. Line width scales with the traded data volume $x_j$; color denotes the price as configured. Only records with a ``deal'' status within the specified time filter are drawn.
    \end{figurenotes}
    \label{fig:exchange_map}
\end{figure}

Fig.~\ref{fig:exchange_map} reports results from another business model replicated by our simulations---the platform-mediated exchange model. The primary effects of platform mediation are to improve market efficiency and governance while maintaining the existing liability structure. Platforms enhance matching efficiency and contractability by providing standard forms, identity management, and continuous auditing. They also induce greater observability and traceability of conduct, which can tighten expected enforcement even without a formal change in liability rules. Crucially, however, the incidence of liability remains with the providers and subscribers, a principle confirmed by both marketplace terms and neutrality-based intermediary statutes. In welfare terms, the platform’s value is derived from shrinking coordination costs and information asymmetries, not from offloading legal risk. Consequently, while prices still embed the seller’s risk and enforcement premia, the standardization of search, screening, and assurance at scale can materially shift market participation, transaction volumes, and the spatial patterns of trade. This type of business model significantly reduces matching distance costs of buyers, but typically charges an about 3\% service fee on the sellers.\footnote{For example, Shanghai Data Exchange charges a transaction service fee of 2.5\% of the transaction amount. See \url{https://www.chinadep.com/bulletin/notices/CTC_20240301150654222852}.}

\begin{figure*}[ht]
    \centering
    \includegraphics[width=1.0\linewidth]{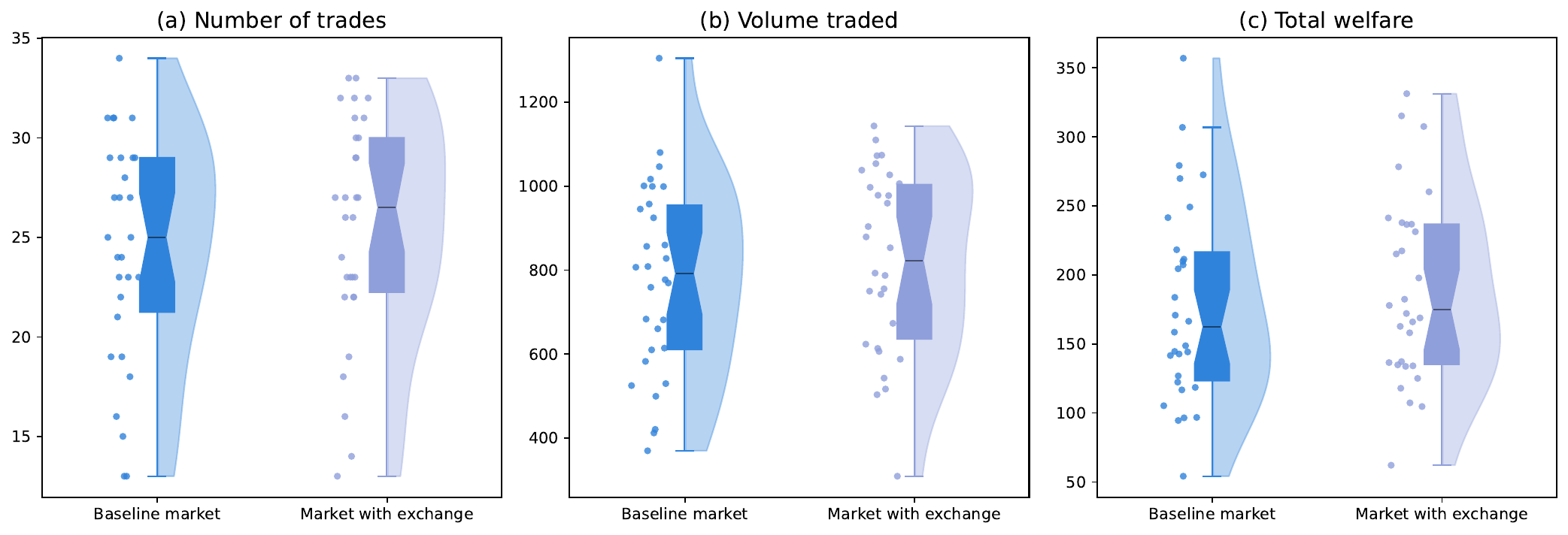}
    \caption{Models with liability on sellers ($t=100$)}
    \begin{figurenotes}
        Each panel shows a raincloud for the two groups on the indicated metric. For each group, the half-violin (right side) depicts the kernel density (probability distribution) of observations; width is proportional to estimated density. The boxplot (centered, notched) overlays the median (line), interquartile range (box), and whiskers extending to 1.5$\times$IQR; notches provide an approximate 95\% confidence interval for the median. Jittered points (left) display individual observations to show sample size and dispersion. Colors are consistent across panels to identify groups.
    \end{figurenotes}
    \label{fig:raincloud_seller_liability}
\end{figure*}

Although public micro-data on on-exchange transactions are limited, multiple external discussions place the share of formal on-exchange trading in the low single digits---often around 2–3\% in earlier market estimates,\footnote{\url{https://pdf.dfcfw.com/pdf/H3_AP202212011580714667_1.pdf?1669889783000.pdf}} and around 5\% in some policy-facing summaries,\footnote{\url{https://www.gzw.sh.gov.cn/shgzw_gqdj_szxc/20240307/befebe0df5294be69a2dd32d9233e770.html}} and a widely cited external estimate suggests that exchange-mediated transactions account for less than 4\% of all data trades \citep{daixin2023harbor}. This benchmark implies that the on-exchange channel is likely constrained in practice, which is consistent with the broad visual similarity between the aggregate arc patterns in Fig. \ref{fig:baseline_map} and Fig. \ref{fig:exchange_map}. Crucially, our simulations deliver an average on-exchange trading share of 3.58\%, closely matching the external estimate. Besides, we also validate our model by estimating an average treatment effect of the platform-mediated exchange business model by regressing each outcome on a PME dummy indicator with random seed and time fixed effects,
\begin{equation}
    Y_{st}=\alpha+\beta\cdot PME+\upsilon_{s}+\xi_{t}+\varepsilon_{st},
\end{equation}
interpreting $\beta$ as the within-seed, within-time change relative to the baseline rule. Table \ref{tab:PME} shows that PME's coefficients are small and statistically indistinguishable from zero for all five metrics---number of trades, volume traded, buyer surplus, seller surplus, and total welfare. The 95\% confidence intervals are tight around zero, implying that any systematic effect, if present, is economically modest. Fixed effects absorb seed heterogeneity and common time shocks, so the non-results are not an artifact of cross-sectional composition. The raincloud plots in Fig.~\ref{fig:raincloud_seller_liability} provide a distributional cross-check. Across panels (a)-(c), the half violins for the PME condition appear slightly right-shifted with fatter upper tails, but the boxplot notches---an approximate 95\% interval for the median---substantially overlap with the baseline. Jittered points confirm that extreme realizations occur under both rules, with PME producing a few very high-volume or high-welfare runs but no robust median gain. Visually, therefore, the distributional evidence coheres with the regression results: PME does not lift central tendency, even if it occasionally enables large matches.

\begin{table}[ht]
    \centering
    \caption{Effect of the platform-mediated exchange}
    \resizebox{1.0\linewidth}{!}{
    \begin{tabular}{lccccc}
    \toprule
     & \makecell{(1)\\Trades} & \makecell{(2)\\Volume traded} & \makecell{(3)\\ Buyer surplus} & \makecell{(4)\\Seller surplus} & \makecell{(5)\\Total welfare}\\
     \midrule
     \multicolumn{6}{l}{\textbf{Rule}}\\
     \ \ \textit{PME} & \makecell{0.01\\(0.014)} & \makecell{0.424\\(0.522)} & \makecell{0.052\\(0.076)} & \makecell{0.057\\(0.085)} & \makecell{0.109\\(0.161)}\\
     \\
     Constant & \makecell{0.245***\\(0.007)} & \makecell{7.776***\\(0.261)} & \makecell{0.858***\\(0.038)} & \makecell{0.929***\\(0.042)} & \makecell{1.787***\\(0.080)}\\
     \\
     $F-$value & 0.53 & 0.66 & 0.46 & 0.46 & 0.46\\
     $R^{2}$ & 0.213 & 0.181 & 0.135 & 0.136 & 0.136\\
     Observations & 6,000 & 6,000 & 6,000 & 6,000 & 6,000\\
     Time FE & $\checkmark$ & $\checkmark$ & $\checkmark$ & $\checkmark$ & $\checkmark$\\
     Seed FE & $\checkmark$ & $\checkmark$ & $\checkmark$ & $\checkmark$ & $\checkmark$\\
     \bottomrule
    \end{tabular}}
    \begin{tablenotes}
        \textit{PME} is a dummy variable for the platform-mediated exchange rule. The baseline rule is dropped as reference. Robust standard errors, clustered at seed level, are reported in parentheses. *, **, *** denote significance level 10\%, 5\%, and 1\%.
    \end{tablenotes}
    \label{tab:PME}
\end{table}

These findings suggest that, instituting a platform intermediary---holding risk and enforcement parameters constant---does not by itself expand the feasible contract set in a way that predictably increases trade counts or surplus. Put differently, information and matching services supplied by the platform do not relax the binding constraints that matter most for average outcomes. The platform seems to help a small set of pairs consummate large deals (upper-tail mass), but those episodic gains are offset elsewhere, leaving mean welfare unchanged. These results also strongly mirrors the inactivity observed in real-world data exchanges.

These macro-level replications---together with the spatial regularities reproduced under the baseline---reinforces the claim that the model’s geographic sandbox and institutional mechanisms generate empirically plausible market outcomes, even in settings where transaction-level ground truth is unavailable.

\section{Comparative institutional analysis}
Here we treat institutional design as a policy experiment in a simulated market. Agent-based computational economics provides a natural ``digital policy laboratory'' for this task: it models decentralized search, matching, bargaining, and compliance as they actually unfold among heterogeneous actors, and it lets us observe emergent market formation under alternative legal rules that would be analytically brittle in closed-form models. We follow the policy-analysis strand of the ABM literature in using the platform to run controlled counterfactuals, holding primitives fixed while switching the governing rule, so that differences in outcomes can be causally attributed to institutional design rather than to shocks or composition effects \citep{tesfatsion2002ABM,tesfatsion2006ABM,Arthur2021foundations}.

\subsection{Mapping legal frameworks}
Our comparison spans the baseline seller-centric liability regime and five legal rules that reassign entitlement and liability in different ways. By holding agent preferences fixed while systematically altering the governing rules, we can causally attribute emergent differences in market outcomes: total welfare, trade volume, and match frequency to institutional design. Crucially, our welfare metric is adapted for information goods by explicitly deducting the externalized harms of data misuse, providing a more accurate measure of social value than laissez-faire price signals alone. This approach provides a transparent bridge from the doctrinal choices available to policymakers, framed in the \citet{calabresi1972property} tradition (See Subection \ref{sec:frameworks}), to their measurable market consequences, revealing which rules genuinely improve efficiency and which merely reassign costs without creating value.


Methodologically, the chapter proceeds as a sequence of policy counterfactuals. For each framework above, we simulate market dynamics to steady-state (or a long finite horizon), then compute welfare and participation metrics, comparing them to the baseline. ABM is suited to this because it accommodates heterogeneous beliefs about value and risk, networked matching, endogenous enforcement intensity, and feedback between governance and participation, features that standard equilibrium-first approaches often must suppress. The goal is pragmatic, to reveal where institutional re-design improves efficiency and robustness in data trade, and where it merely reassigns incidence without efficiency gains, providing a transparent bridge between doctrinal choices and measurable market consequences. We operationalize the legal landscape by mapping six distinct institutional regimes into our agent-based model. Drawing on the framework established in the previous section, we classify these regimes not by their doctrinal labels, but by how they allocate the objective parameters of risk—specifically the expected cost of harm ($R$) and the administrative or criminal liability from regulatory enforcement ($E$). All the legal frameworks to be compared and their respective features are outlined in Table \ref{tab:all_models}.

\textbf{1. Controller-centric liability (baseline).} 
Our baseline models the default seller-centric regime. Here, the party determining the purpose of the seller fully internalizes the legal exposure. We implement this by embedding legal risk entirely into the supply side. The seller’s WTA includes an intrinsic hazard term capturing expected private costs ($\beta_R R$) and an enforcement intensity term capturing regulatory sanctions ($\beta_E E$). Buyers do not model liability directly, they face it only indirectly via the price. Consequently, under credible enforcement, all liability is internalized by the seller. Crucially, this means there is no liability externality in our welfare accounting, total welfare remains the sum of producer and consumer surplus.

\textbf{2. Buyer-shared risk.} 
Departing from the baseline, this regime allocates the objective harm risk ($R$) across both buyer and seller, while the seller continues to bear the sole enforcement-intensity channel ($E$). Economically, this shifts the pricing calculus. The seller’s WTA internalizes only its contracted share of $R$ (lowering the ask), while the buyer discounts its WTP by the remainder. This structure mirrors contractual indemnities supported by public law floors. By internalizing a portion of $R$ on the demand side, this regime creates a direct economic incentive for buyers to invest in downstream safeguards, dynamically reallocating market participation based on the negotiated split of risk.

\textbf{3. Two-sided liability split.} 
The most extensive deviation is the joint liability regime, where both substantive harm risk ($R$) and enforcement exposure ($E$) are allocated across the dyad. Economically, the seller’s WTA embeds only its contracted shares of $R$ and $E$, while the buyer discounts WTP by its complementary shares. This captures statutory architectures where downstream acquirers are directly reachable by regulators. This regime maximizes incentive alignment, as buyers bearing a portion of $E$ have a motivation to invest in compliance to mitigate fines rather than free-riding on seller precautions. Like the baseline, the full liability mass ($R+E$) remains internal to the bargain, so welfare is confined to surplus without externalities.

\textbf{4. Informed consent.} 
We implement informed consent as a strict property-rule gate. Reflecting its role in regimes like the GDPR, consent functions as a binary access constraint, only sellers who possess valid consent are tradable in the model. Analytically, this is a filter that imposes a non-trivial transaction cost on the seller. Once this cost is paid and the gate is passed, the legal nature of the risk changes: the seller’s objective risk loading for substantive harm ($R$) is reduced because the processing is authorized. However, because regulators actively police the validity of consent, the seller remains fully exposed to the enforcement intensity term ($E$). Thus, consent acts as a costly signal that lowers substantive liability risk but preserves procedural oversight.

\textbf{5. Low-risk carve-out.} 
The Low-risk carve-out (e.g., anonymization) operationalizes a discontinuous boundary in the legal landscape. Unlike liability rules that scale continuously with risk, this regime exempts datasets that meet an \textit{ex ante} safety threshold from the liability regime entirely. In the simulation, this is modeled as a status decision. If a dataset’s intrinsic risk profile falls below a specified threshold, both the substantive harm premium ($R$) and enforcement exposure ($E$) are removed from the seller’s pricing function. This functions as a ``regulatory exit,'' converting the transaction into a risk-free exchange from a legal perspective, provided the technical criteria are met.

\textbf{6. Provider immunization.} 
Finally, we model provider immunization as a property-rule entitlement shifted to the seller. In this regime, the provider is shielded from \textit{ex post} liability to encourage supply. We implement this by structurally removing the objective risk premium ($R$) from the seller’s WTA. The critical distinction between this regime and the others is the non-internalization of risk. Because the risk premium is stripped from the price to incentivize trade, the expected cost of harm is neither paid by the seller nor transferred to the buyer. Instead, it becomes a negative externality. Consequently, in our welfare accounting for this specific regime, we explicitly subtract the realized harm from the total surplus to capture the social cost of uncompensated privacy loss.

\newcolumntype{L}[1]{>{\raggedright\arraybackslash}p{#1}}     
\begin{table*}[ht]
        \centering
        \caption{Legal frameworks to be compared}
        \resizebox{1.0\linewidth}{!}{
        \begin{tabular}{>{\raggedright\arraybackslash}L{3.5cm}L{3.5cm}L{3cm}L{3cm}L{3cm}L{3cm}L{3cm}}
        \toprule
        & \multicolumn{1}{c}{Baseline} & \multicolumn{3}{c}{Public / third-party externalization} & \multicolumn{2}{c}{Buyer-shared liability}\\
        \cmidrule(lr){2-2}\cmidrule(lr){3-5}\cmidrule(lr){6-7}
        \textbf{Rule} & \hyperref[sec:baseline]{Seller-centric liability} & \hyperref[sec:lrco]{Low-risk carve-out} & \hyperref[sec:ic]{Informed consent} & \hyperref[sec:ri]{Risk immunity} & \hyperref[sec:divide]{Dividing risk liability} & \hyperref[sec:radical_divide]{Dividing all liability}\\
        \cmidrule(lr){1-7}
        \textbf{Feature} & Seller bears 100\% liability for Risk (R) \& Enforcement (E). & Liability depends on data risk (R). Low-risk (R=1) sellers are exempt from R \& E costs. & Sellers must get user consent for market access; consent exempts from R cost. & Sellers get blanket immunity from R liability; still bear E liability. & Liability for Risk (R) divided. Seller bears 100\% of E liability. & Liability for both R and E divided.\\
        \\
        \textbf{Risk cost ($R$)} & Internalized by Seller. & Externalized if $R=1$; Internalized if $R>1$. & Externalized if consent. & Fully externalized. & Divided between Buyer \& Seller. & Divided between Buyer \& Seller.\\
        \\
        \textbf{Enforcement cost ($E$)} & Internalized by Seller. & Externalized if $R=1$; Internalized if $R>1$. & Internalized by Seller. &Internalized by Seller. & Internalized by Seller. & Divided between Buyer \& Seller.\\
        \\
        \textbf{Seller WTA} & $\text{WTA}_{jt}$ & $R=1$: $(c_0+c_1 x_j)/\alpha_i$ & $\text{WTA}_{jt}-\beta_R R$ & Risk removed. & Risk divided. & $R$ \& $E$ divided.\\
        \\
        \textbf{Buyer WTP} & $\text{WTP}_{ijt}$ & $\text{WTP}_{ijt}$ & $\text{WTP}_{ijt}$ & $\text{WTP}_{ijt}$ & $\text{WTP}_{ijt}$ $-$ $share\%\times R$ & $\text{WTP}_{ijt}$ $-$ $share\%\times (R+E)$ \\
        \\
        \textbf{Market access} & All sellers. & All sellers. & Sellers with consent. & All sellers. & All sellers. & All sellers.\\
        \\
        \textbf{Externality} & No. & Yes, when $R=1$. & Yes. & Yes. & No. & No.\\
        \\
        \textbf{Total welfare} & CS + PS & CS + PS - Ext & CS + PS - Ext & CS + PS - Ext & CS + PS & CS + PS\\
        \bottomrule
        \end{tabular}}
        \label{tab:all_models}
\end{table*}

\subsection{Consent, carve-out, and immunity}
This family of regimes includes \textbf{informed consent}, \textbf{low-risk carve-out}, and \textbf{immunization}, which seeks to expand data transactions by removing or relaxing a seller's legal exposure when certain \textit{ex ante} conditions are met. These conditions, such as a dataset qualifying as anonymous or the presence of valid informed consent, function as legal safe harbors that allow transactions to proceed without the price embedding the full expected cost of potential harm. In practice, the law creates bright-line distinctions for this purpose. For example, GDPR's Recital 26 takes genuinely anonymized data outside the scope of data protection law, while U.S. health privacy law permits the use of de-identified data under HIPAA’s Safe Harbor provisions. Analytically, however, these designs shift a portion of the liability mass onto third parties or the public, as the contracting parties do not internalize the full harm term once these status or consent-based gates are satisfied.

\begin{figure}
    \centering
    \includegraphics[width=1.0\linewidth]{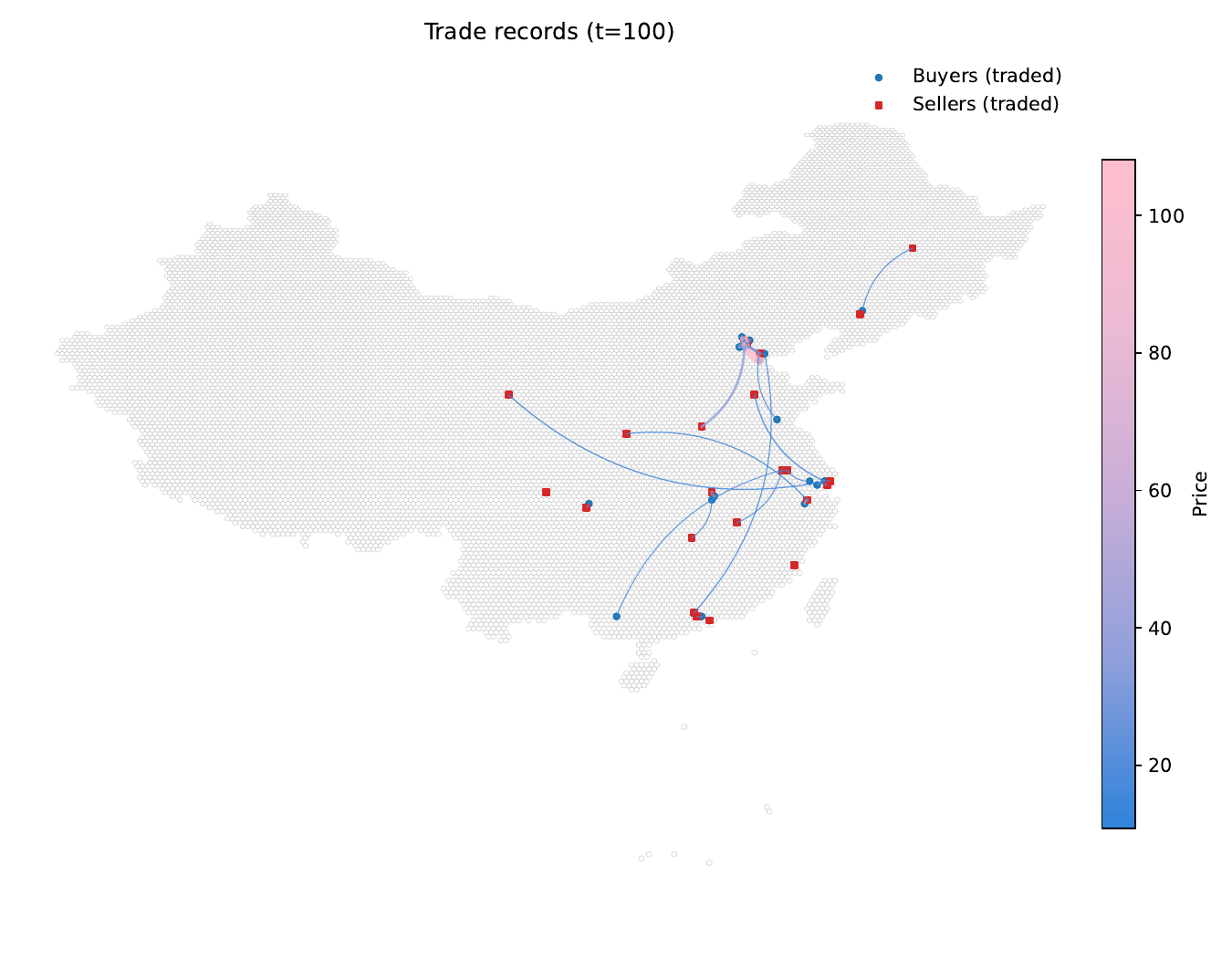}
    \caption{Trade arcs in low-risk carve-out ($t=100$)}
    \begin{figurenotes}
        Arcs connect the centroids of the buyer and seller hexagons for each realized deal in the simulation. Line width scales with the traded data volume $x_j$; color denotes the price as configured. Only records with a ``deal'' status within the specified time filter are drawn.
    \end{figurenotes}
    \label{fig:lrco_map}
\end{figure}

\begin{figure}
    \centering
    \includegraphics[width=1.0\linewidth]{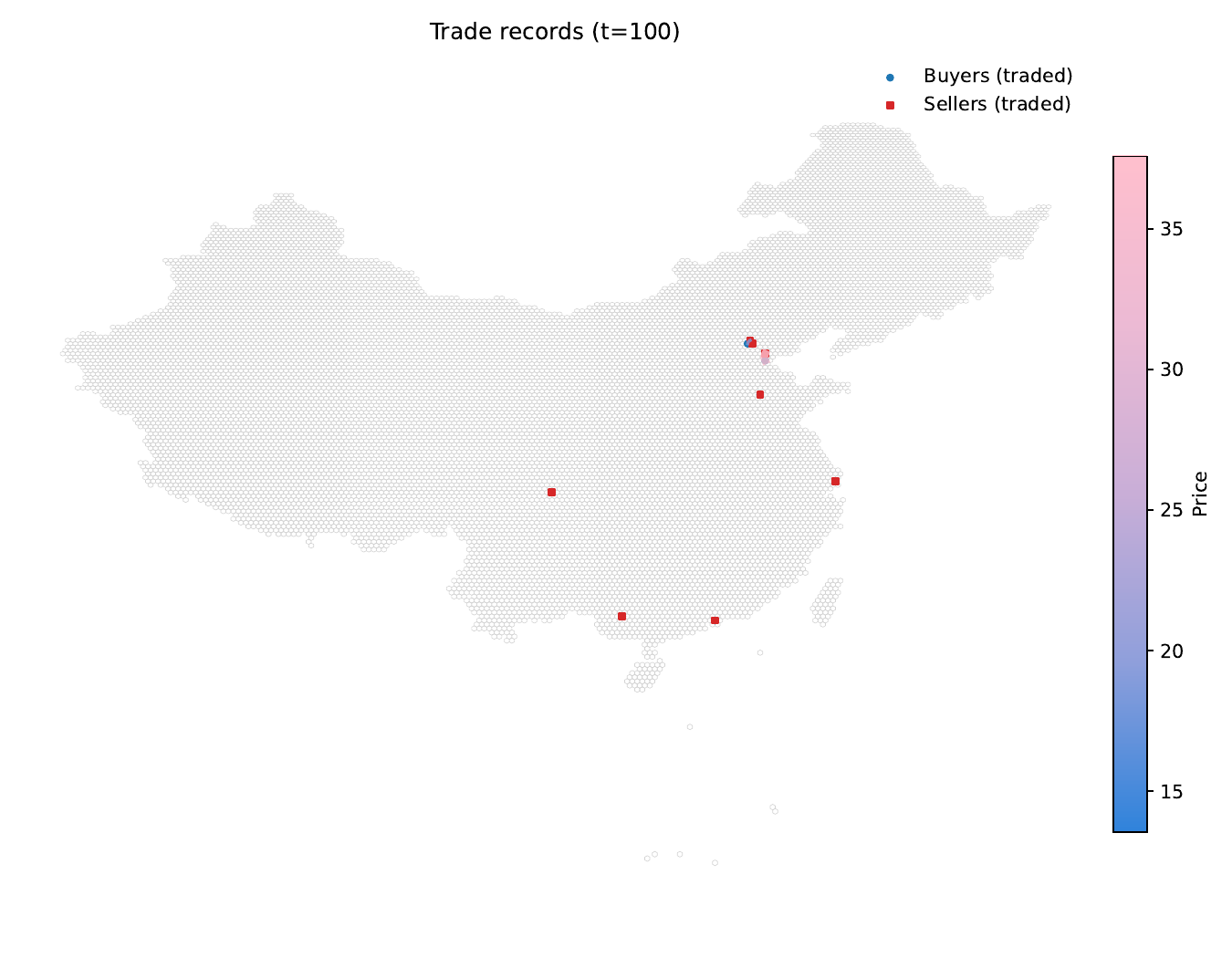}
    \caption{Trade arcs in informed consent ($t=100$)}
    \begin{figurenotes}
        Arcs connect the centroids of the buyer and seller hexagons for each realized deal in the simulation. Line width scales with the traded data volume $x_j$; color denotes the price as configured. Only records with a ``deal'' status within the specified time filter are drawn.
    \end{figurenotes}
    \label{fig:ic_map}
\end{figure}

\begin{figure}
    \centering
    \includegraphics[width=1.0\linewidth]{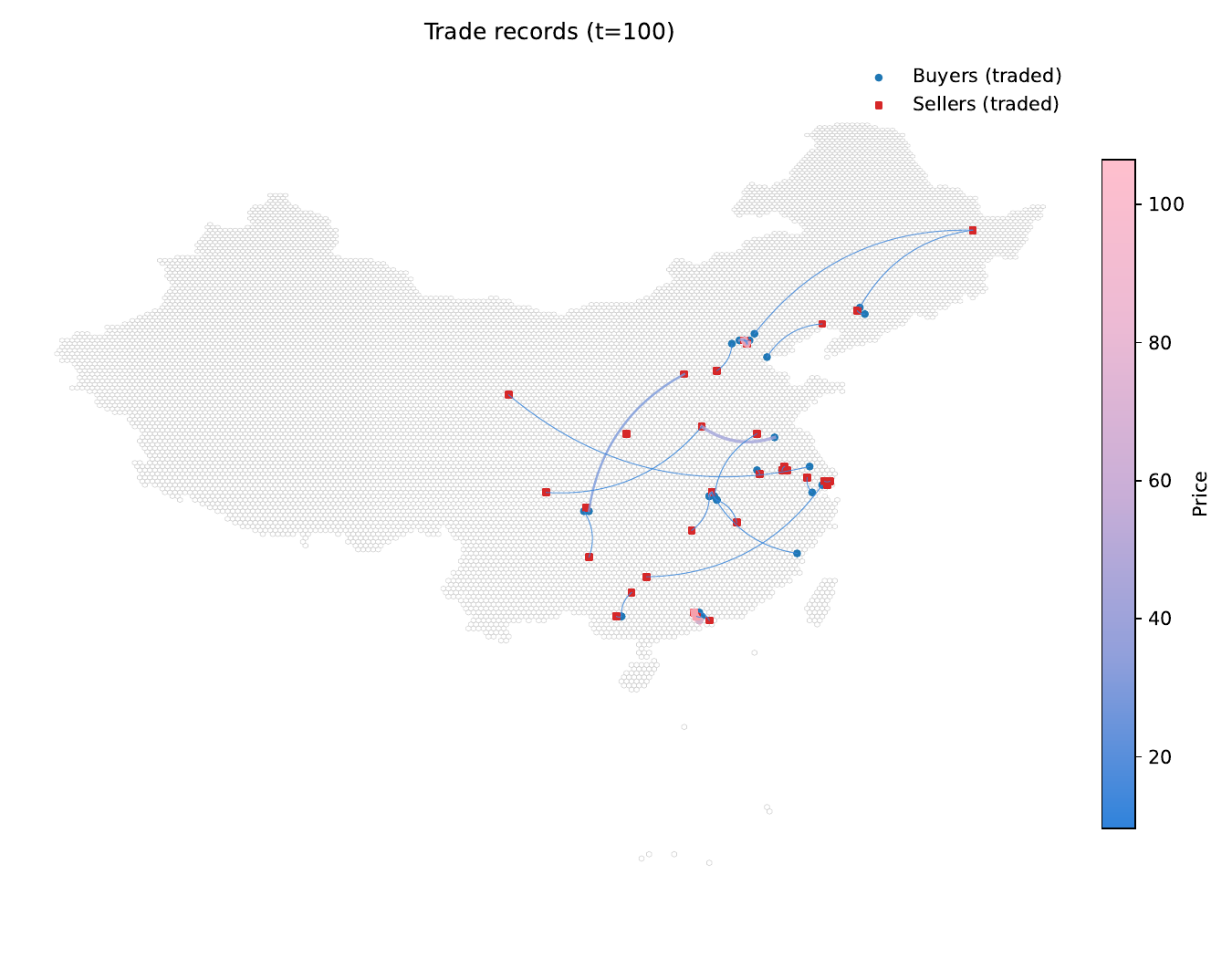}
    \caption{Trade arcs in risk immunity ($t=100$)}
    \begin{figurenotes}
        Arcs connect the centroids of the buyer and seller hexagons for each realized deal in the simulation. Line width scales with the traded data volume $x_j$; color denotes the price as configured. Only records with a ``deal'' status within the specified time filter are drawn.
    \end{figurenotes}
    \label{fig:ri_map}
\end{figure}

We estimate the within-seed, within-time average treatment effect of each rule---informed consent (IC), low-risk carve-out (LRCO), and risk immunity (RI)---relative to the baseline by
\begin{equation}
    Y_{st}=\alpha+\beta_{1}\cdot IC+\beta_{2}\cdot LRCO + \beta_{3}\cdot RI+\upsilon_s+\xi_t+\varepsilon_{st}.
\end{equation}
Coefficients $\beta$ are therefore causal contrasts within the simulated economies, purged of time shocks and seed heterogeneity. One representative simulation run for each model is visualized in Fig.~\ref{fig:lrco_map},~\ref{fig:ic_map}, and~\ref{fig:ri_map}. The statistical results are shown in Table \ref{tab:third_party}.

\begin{table}[ht]
    \centering
    \caption{Effects of third-party externalization rules}
    \resizebox{1.0\linewidth}{!}{
    \begin{tabular}{lcccccc}
    \toprule
     & \makecell{(1)\\Trades} & \makecell{(2)\\Volume traded} & \makecell{(3)\\ Buyer surplus} & \makecell{(4)\\Seller surplus} & \makecell{(5)\\Externality} & \makecell{(6)\\Total welfare}\\
     \midrule
     \multicolumn{7}{l}{\textbf{Rule}}\\
     \ \ \textit{IC} & \makecell{-0.145***\\(0.012)} & \makecell{-5.663***\\(0.432)} & \makecell{-0.606***\\(0.061)} & \makecell{-0.706***\\(0.073)} & \makecell{0.205***\\(0.013)} & \makecell{-1.517***\\(0.133)}\\
     \\
     \ \ \textit{LRCO} & \makecell{0.015\\(0.014)} & \makecell{0.528\\(0.601)} & \makecell{0.009\\(0.074)} & \makecell{0.018\\(0.097)} & \makecell{0.010***\\(0.002)} & \makecell{0.016\\(0.170)}\\
     \\
     \ \ \textit{RI} & \makecell{0.078***\\(0.016)} & \makecell{2.290***\\(0.613)} & \makecell{0.359***\\(0.104)} & \makecell{0.407***\\(0.128)} & \makecell{0.663***\\(0.027)} & \makecell{0.103\\(0.225)}\\
     \\
     Constant & \makecell{0.245***\\(0.009)} & \makecell{7.776***\\(0.353)} & \makecell{0.858***\\(0.052)} & \makecell{0.929***\\(0.064)} & \makecell{0.000\\--} & \makecell{1.787***\\(0.115)}\\
     \\
     $F-$value & 98.92 & 157.96 & 131.09 & 142.48 & 326.57 & 162.63\\
     $R^{2}$ & 0.194 & 0.158 & 0.107 & 0.103 & 0.177 & 0.101\\
     Observations & 12,000 & 12,000 & 12,000 & 12,000 & 12,000 & 12,000\\
     Time FE & $\checkmark$ & $\checkmark$ & $\checkmark$ & $\checkmark$ & $\checkmark$ & $\checkmark$\\
     Seed FE & $\checkmark$ & $\checkmark$ & $\checkmark$ & $\checkmark$ & $\checkmark$ & $\checkmark$\\
     \bottomrule
    \end{tabular}}
    \begin{tablenotes}
        \textit{IC}, \textit{LRCO}, and \textit{RI} are dummy variables for the informed consent, low-risk carve-out, and risk immunity rule, respectively. The baseline rule is dropped as reference. Robust standard errors, clustered at seed level, are reported in parentheses. *, **, *** denote significance level 10\%, 5\%, and 1\%.
    \end{tablenotes}
    \label{tab:third_party}
\end{table}

\begin{figure*}[ht]
    \centering
    \includegraphics[width=0.9\linewidth]{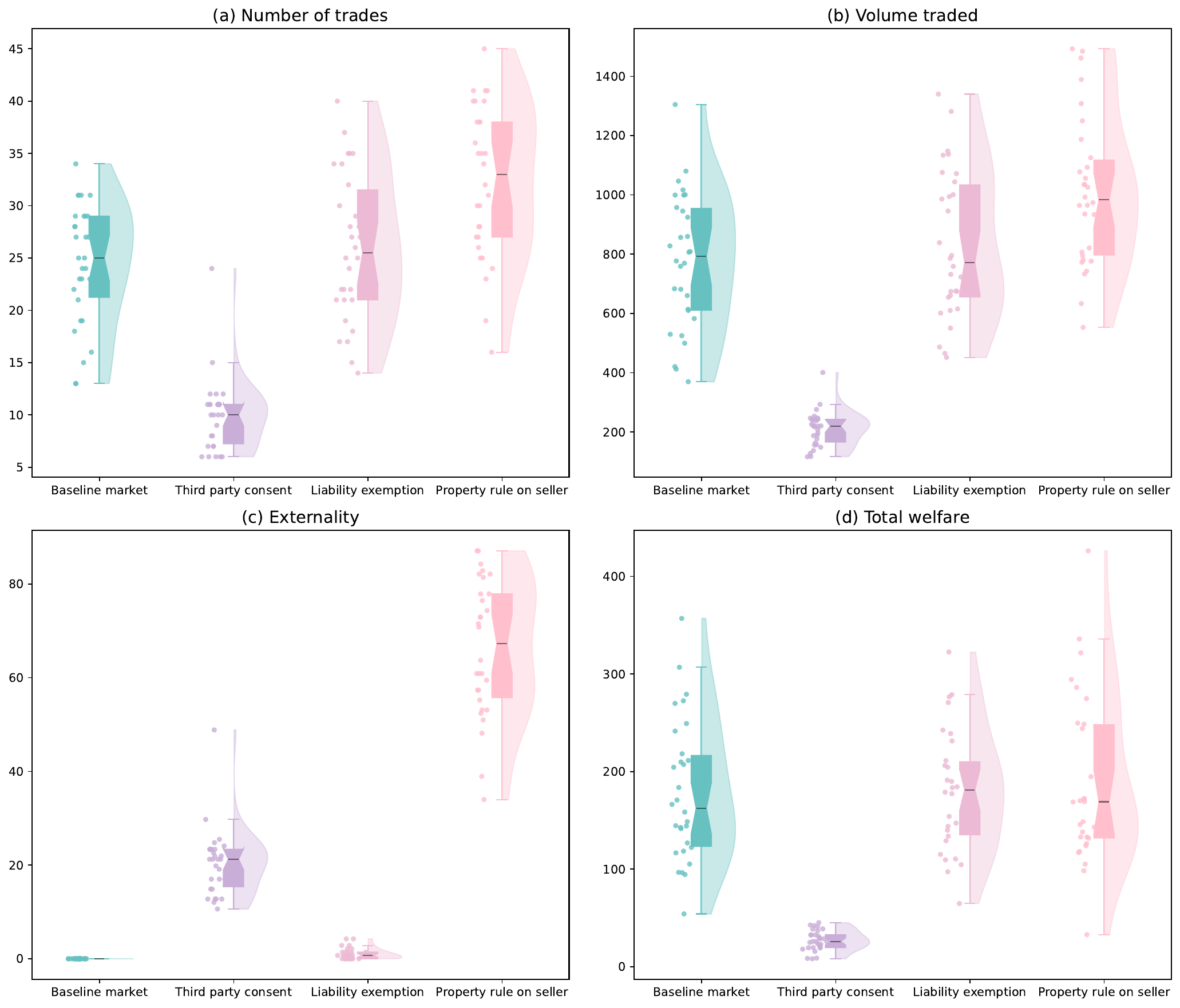}
    \caption{Models with externality on third party ($t=100$)}
    \begin{figurenotes}
        Each panel shows a raincloud for the four groups on the indicated metric. For each group, the half-violin (right side) depicts the kernel density (probability distribution) of observations; width is proportional to estimated density. The boxplot (centered, notched) overlays the median (line), interquartile range (box), and whiskers extending to 1.5$\times$IQR; notches provide an approximate 95\% confidence interval for the median. Jittered points (left) display individual observations to show sample size and dispersion. Colors are consistent across panels to identify groups.
    \end{figurenotes}
    \label{fig:third_party}
\end{figure*}

The informed consent rule reduces market activity and welfare across the board. The average treatment effects on all of the six metrics are significantly negative. Fig.~\ref{fig:third_party} echoes this: the IC rainclouds are sharply left-shifted with compressed IQRs, indicating both lower central tendency and reduced dispersion. Economically, this stringent \textit{ex ante} consent condition tightens the feasible-contract set, therefore lower both sides' surplus.

For the low-risk carve-out rule, the coefficients are small and statistically indistinguishable from zero on all outcomes. The rainclouds sit close to baseline, with overlapping notches and similar tails. Substantively, targeting low-risk transactions for lighter treatment does not move the aggregate needle: most gains available at low risk were already realizable under baseline matching and budgets; any incremental matches are offset by selection and price adjustments elsewhere. From a regulatory design perspective, LRCO appears least distortive---it avoids IC's output losses without generating much additional harm---but it also does not deliver systematic average gains.

The risk immunity rule produces broad-based increases on trades, volume, buyer surplus and seller surplus. Fig.~\ref{fig:third_party} shows clear right-shifts with fatter upper tails, especially for trades and volume. The same treatment also raises measured externalities, yielding statistically indistinguishable total welfare and indicating a classic scale-harm trade-off: removing liability frictions expands the contracting set and deepens matches, but at the cost of greater third-party exposure. In \citet{calabresi1972property} terms, RI mimics a strong property-rule protection for sellers: it enlarges the size of the pie and the frequency of trades, while shifting part of the expected loss to parties outside the contract.

\subsection{Buyer-shared liability}
In this family of regimes, liability is no longer concentrated solely on the seller. Instead, the buyer is treated as a legally reachable actor who shares in the potential costs of data-related harms. This model directly reflects contemporary legal doctrine, where ``Business Associate'' under HIPAA all bear direct liability. Economically, this approach keeps the full liability mass inside the transaction, reassigning the incidence of risk toward the party best placed to mitigate it. This creates powerful incentives for buyers to invest in post-acquisition governance, as they now internalize a meaningful share of both the substantive risk of harm and the exposure to enforcement.

\begin{figure}
    \centering
    \includegraphics[width=1.0\linewidth]{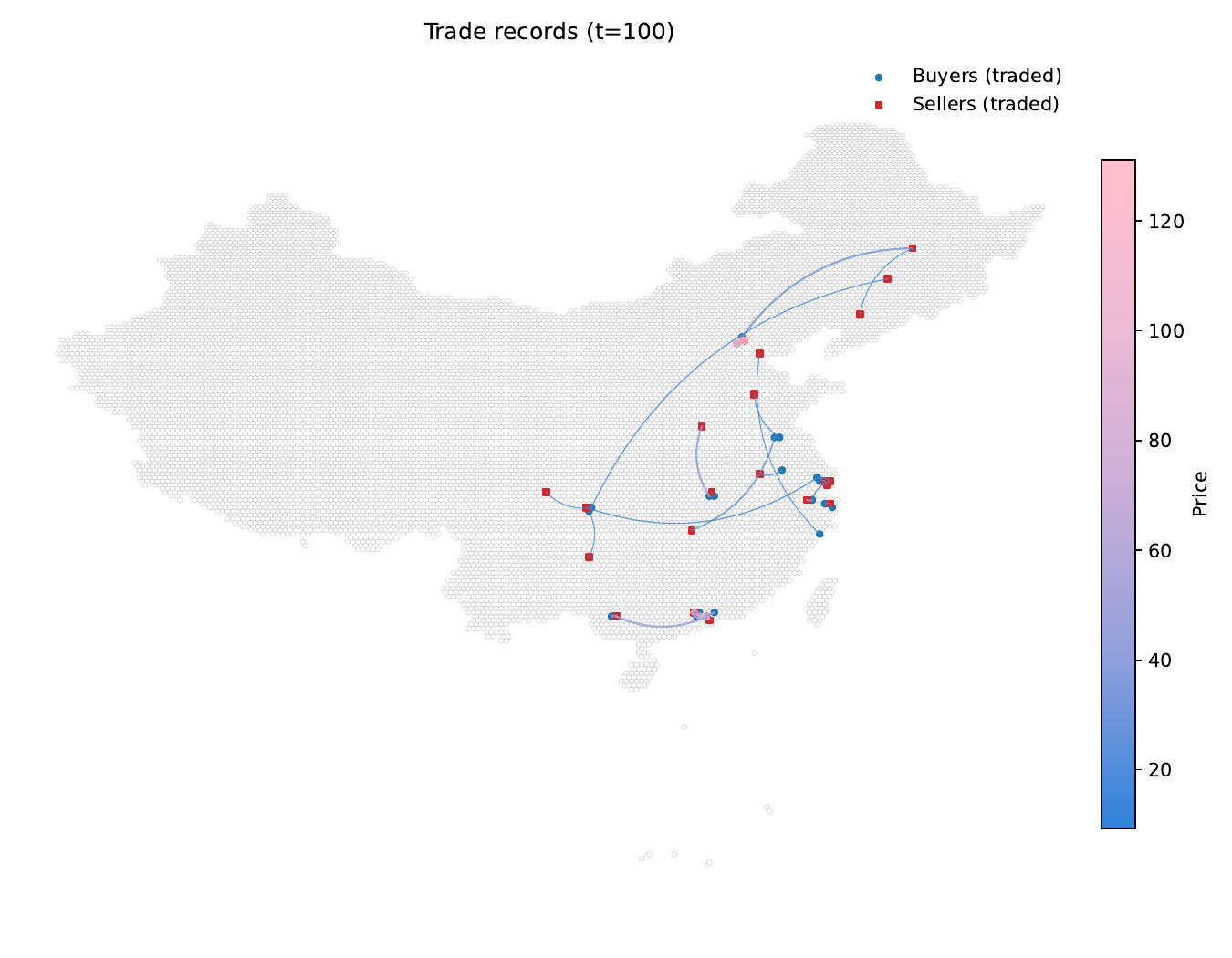}
    \caption{Trade arcs in buyer-shared risk ($t=100$)}
    \begin{figurenotes}
        Arcs connect the centroids of the buyer and seller hexagons for each realized deal in the simulation. Line width scales with the traded data volume $x_j$; color denotes the price as configured. Only records with a ``deal'' status within the specified time filter are drawn.
    \end{figurenotes}
    \label{fig:divide_map}
\end{figure}

\begin{figure}[ht]
    \centering
    \includegraphics[width=1.0\linewidth]{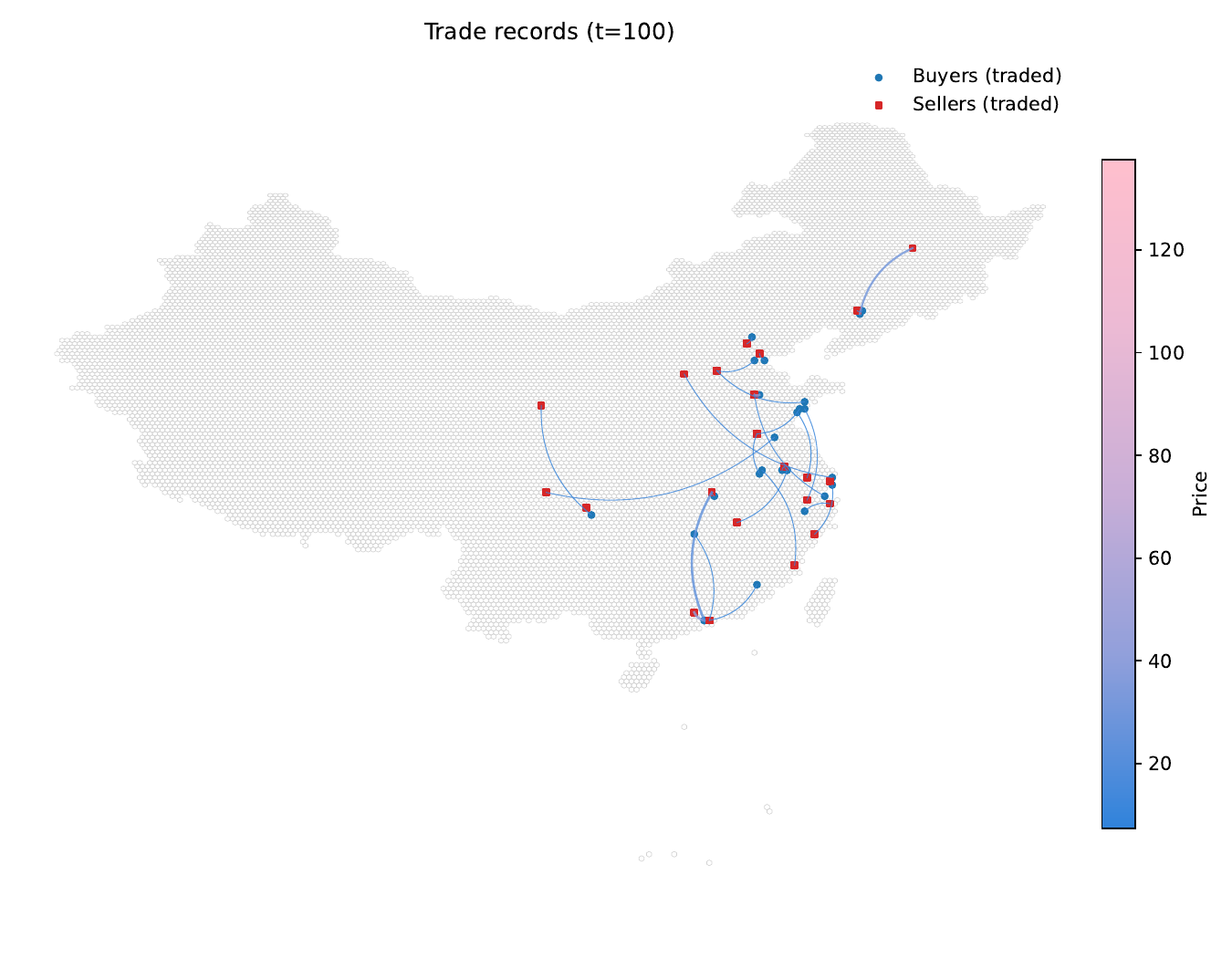}
    \caption{Trade arcs in two-sided liability split ($t=100$)}
    \begin{figurenotes}
        Arcs connect the centroids of the buyer and seller hexagons for each realized deal in the simulation. Line width scales with the traded data volume $x_j$; color denotes the price as configured. Only records with a ``deal'' status within the specified time filter are drawn.
    \end{figurenotes}
    \label{fig:radical_divide_map}
\end{figure}

We exploit within-seed and within-time variation in the buyer's liability share, $share\in[0,1]$ and estimate
\begin{equation}
    Y_{st}=\alpha+\beta\cdot share +\upsilon_{s}+\xi_{t}+\varepsilon_{st}.
\end{equation}
The two regimes discussed above are studied: (i) Buyer-shared risk ($R$ only) and (ii) Buyer-shared risk and enforcement ($R$ \& $E$). The coefficient $\beta$ identifies the average marginal effect of moving the buyer's share from 0 to 1 on the outcome $Y$, holding seed and time factors constant. One representative simulation run for each model is visualized in Fig.~\ref{fig:divide_map}, and~\ref{fig:radical_divide_map}.

\begin{figure*}
    \centering
    \includegraphics[width=1.0\linewidth]{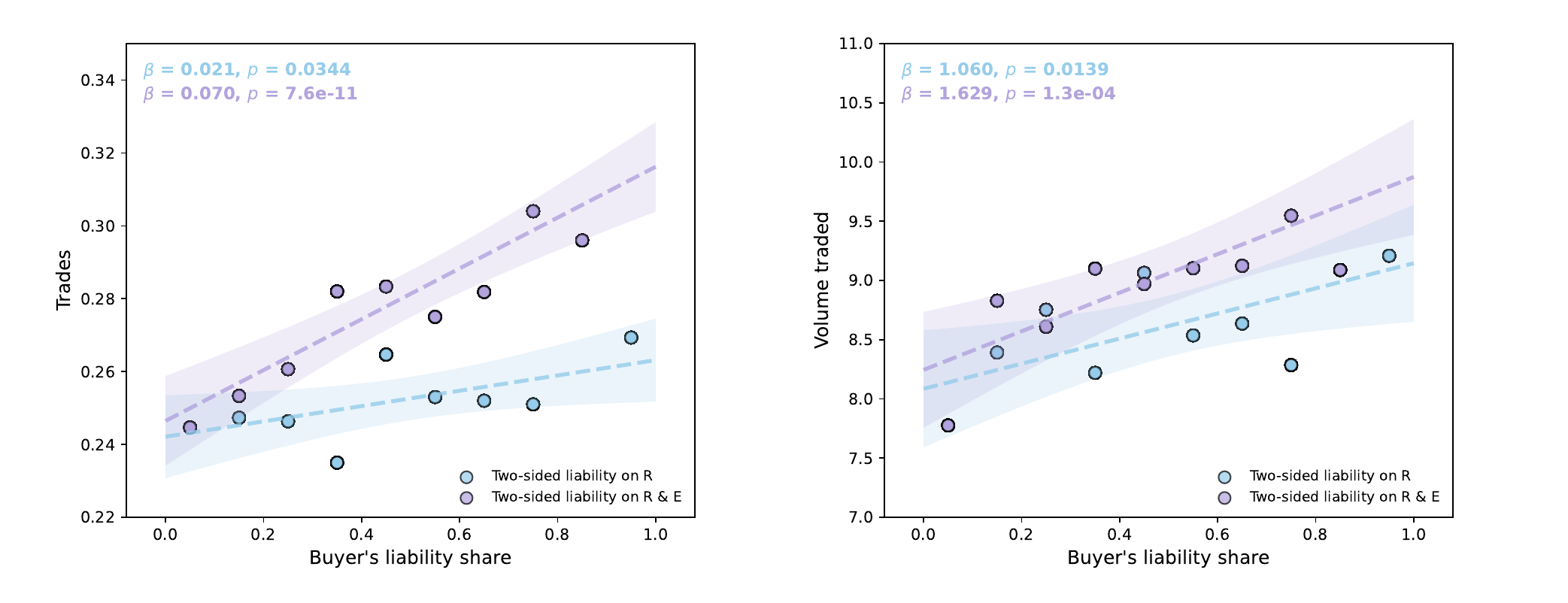}
    \caption{Effects of buyer-side liability splits on throughput indicators}
    \begin{figurenotes}
        Each panel plots binned scatter points of the indicated metric against the mid-bin center of buyer liability share. In both panels, the fitted lines and shaded 95\% OLS confidence intervals are estimated on the full sample using buyer's liability share as the regressor and the corresponding outcome. Colors denote policy regimes and are consistent across panels.
    \end{figurenotes}
    \label{fig:share_throughput}
\end{figure*}

According to Table \ref{tab:share}, both regimes show statistically positive throughput elasticities, but the effect is stronger when buyers also share enforcement. A full shift of share from $0\rightarrow 1$ under the risk-only regime raises trades by 0.021 ($p<0.05$) and volume traded by 1.060 ($p<0.01$). Under the $R$ \& $E$ rule, the corresponding effects are 0.070 ($p<0.01$) and 1.629 ($p<0.01$). The fitted lines in Fig.~\ref{fig:share_throughput} reproduce these slopes with tight 95\% bands through regressions without fixed effect and cluster-robust standard errors, visually confirming a monotone increase. Economically, asking buyers to internnalize a larger slice of downside (and, under $R$ \& $E$, enforcement frictions as well) screens in higher-type buyers and reassures sellers, reducing bargaining failure and expanding the feasible-contract set.

\begin{table*}[ht]
    \centering
    \caption{Effects of buyer-side liability splits}
    \resizebox{1.0\linewidth}{!}{
    \begin{tabular}{lcccccccccc}
     \toprule
     & \multicolumn{5}{c}{Buyer-shared risk} & \multicolumn{5}{c}{Buyer-shared risk and enforcement}\\
     \cmidrule(lr){2-6}\cmidrule(lr){7-11}
     & \makecell{(1)\\Trades} & \makecell{(2)\\Volume traded} & \makecell{(3)\\ Buyer surplus} & \makecell{(4)\\Seller surplus} & \makecell{(5)\\Total welfare} & \makecell{(6)\\Trades} & \makecell{(7)\\Volume traded} & \makecell{(8)\\ Buyer surplus} & \makecell{(9)\\Seller surplus} & \makecell{(10)\\Total welfare}\\
     \midrule
    \textit{Share} & \makecell{0.021**\\(0.009)} & \makecell{1.060**\\(0.440)} & \makecell{0.128*\\(0.069)} & \makecell{0.177**\\(0.082)} & \makecell{0.305*\\(0.150)} & \makecell{0.070***\\(0.011)} & \makecell{1.629***\\(0.387)} & \makecell{0.035\\(0.065)} & \makecell{0.078\\(0.080)} & \makecell{0.113\\(0.145)}\\
    \\
    Constant & \makecell{0.242***\\(0.004)} & \makecell{8.086***\\(0.220)} & \makecell{0.922***\\(0.035)} & \makecell{0.998***\\(0.041)} & \makecell{1.920***\\(0.075)} & \makecell{0.246***\\(0.006)} & \makecell{8.245***\\(0.194)} & \makecell{0.948***\\(0.033)} & \makecell{1.040***\\(0.040)} & \makecell{1.988***\\(0.072)}\\
    \\
    $F-$value & 5.60 & 5.80 & 3.46 & 4.70 & 4.13 & 38.49 & 17.69 & 0.29 & 0.95 & 0.61\\
    $R^{2}$ & 0.211 & 0.163 & 0.134 & 0.131 & 0.133 & 0.213 & 0.166 & 0.136 & 0.133 & 0.135\\
    Observations & 33,000 & 33,000 & 33,000 & 33,000 & 33,000 & 33,000 & 33,000 & 33,000 & 33,000 & 33,000\\
    Time FE & $\checkmark$ & $\checkmark$ & $\checkmark$ & $\checkmark$ & $\checkmark$ & $\checkmark$ & $\checkmark$ & $\checkmark$ & $\checkmark$ & $\checkmark$\\
    Seed FE & $\checkmark$ & $\checkmark$ & $\checkmark$ & $\checkmark$ & $\checkmark$ & $\checkmark$ & $\checkmark$ & $\checkmark$ & $\checkmark$ & $\checkmark$\\
    \bottomrule
    \end{tabular}}
    \begin{tablenotes}
        \textit{Share} is a continuous variable representing the buyer's share of liability. In the baseline model, \textit{share} is set to be 0. Robust standard errors, clustered at seed level, are reported in parentheses. *, **, *** denote significance level 10\%, 5\%, and 1\%.
    \end{tablenotes}
    \label{tab:share}
\end{table*}

For surplus distribution, the two rules exhibit heterogeneity on treatment effects. Under the risk-only rule, $share$ is associated with higher private surplus: a 0.128 ($p<0.10$) increase on buyer surplus and a 0.177 ($p<0.05$) increase on seller surplus. Once enforcement is also shifted to buyers (the $R$ \& $E$ rule), these surplus gains attenuate and lose significance. A natural interpretation is that while greater buyer skin-in-the-game induces more trades, cost pass-through and price adjustment transfer part of the gains, leaving neither side's per-trade surplus systematically higher. In other words, enforcement sharing acts like a tax on the matched pair that is offset by increased match frequency rather than larger rents.
\begin{figure}[ht]
    \centering
    \includegraphics[width=1.0\linewidth]{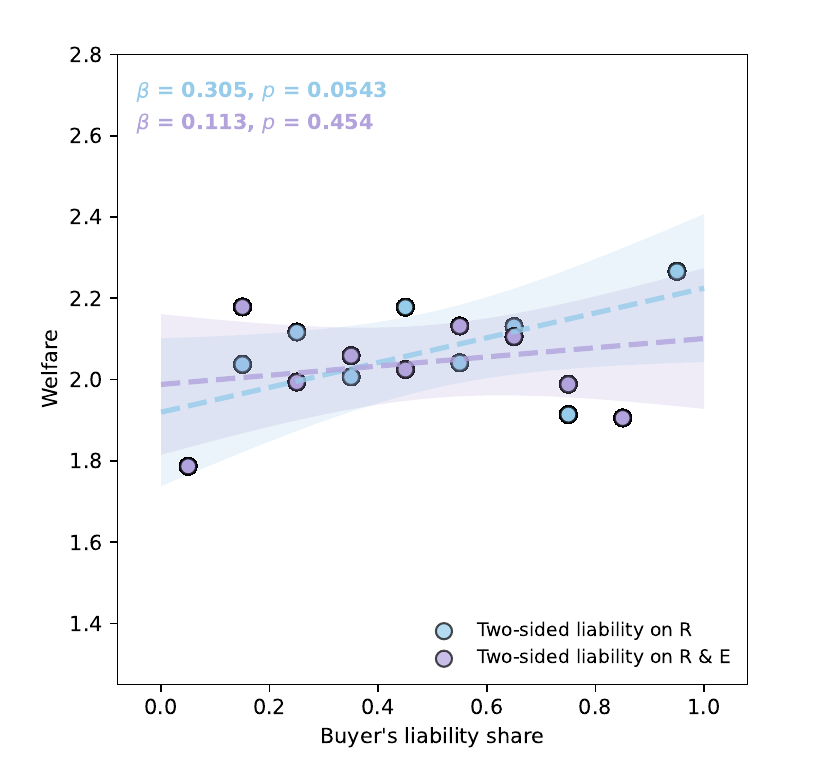}
    \caption{Effects of buyer-side liability splits on welfare}
    \begin{figurenotes}
        The figure plots binned scatter points of total welfare against the mid-bin center of buyer liability share. The fitted lines and shaded 95\% OLS confidence intervals are estimated on the full sample using buyer's liability share as the regressor and total welfare as outcome. Colors denote policy regimes and are consistent across panels.
    \end{figurenotes}
    \label{fig:share_welfare}
\end{figure}

The total welfare regression mirrors this tradeoff. The risk-only rule yields a positive welfare slope of 0.305 ($p\approx 0.05$), consistent with more matches and modest surplus increases on both sides. Under the $R$ \& $E$ rule, however, the welfare slope drops to be indistinguishable from zero. Fig.~\ref{fig:share_welfare} shows the same pattern: both fitted lines slope upward, but the blue (risk-only) line is steeper, whereas the lavender ($R$ \& $E$) line is flatter with a wide confidence interval. Thus, shifting risk to buyers tends to scale the pie, but layering enforcement burdens on buyers converts some of those gains into compliance costs, leaving average welfare statistically unchanged.

At the same time, magnitudes are economically meaningful at realistic movements in $share$. A move from 0 to 0.5 increases expected trades by $\approx 0.01$  (risk-only) and $\approx 0.035$ ($R$ \& $E$) per period; and volume by $\approx 0.53$ and $\approx 0.81$ per period, respectively---consistent with the visual ``upward tilt'' of binned points. The results imply that allocating some risk to buyers (without imposing enforcement costs) is Pareto-leaning in expectation: more trades, higher volume, and weakly higher total welfare. However, adding enforcement sharing further boosts throughput, but not welfare on average---suggesting that the extra compliance burden largely relabels gains rather than creating new surplus. Converting the throughput dividend into welfare would require complementary measures.

Therefore, increasing buyers’ liability share is shown to be an effective quantity lever---especially when coupled with enforcement---but only the risk-sharing (risk-only) version translates into statistically detectable welfare gains. The risk-and-enforcement version appears to reallocate rather than expand surplus: it brings more matches to fruition, yet leaves mean buyer, seller, and total surplus statistically flat once compliance costs are internalized.

\section{Conclusion}
This article addresses the fundamental dilemma of the AI data economy: the misalignment between regulatory tools and an opaque, highly dynamic market. By constructing a spatially explicit computational laboratory, we have attempted to move the governance debate from the realm of intuition to the light of controlled institutional comparison. Our findings offer both a substantive corrective to current doctrine and a methodological road-map for the future of legal empiricism. Substantively, our results serve as a economic vindication of the ``least cost avoider'' in the digital age. While the dominant regulatory paradigm focuses heavily on the seller (controller), our simulation reveals that this approach is not structurally efficient. Property-rule mechanisms that treat data as ``outside the law'' systematically externalize the social costs of personal harms. Conversely, we demonstrate that social welfare is maximized when liability is shifted to the downstream buyer. In a market defined by quality uncertainty and re-identification risk, the downstream user is the best actor capable of cheaply attenuating risk through post-acquisition safeguards.

This finding suggests that the ``two-sided reachability'' emerging in practice, like Business Associate provisions under HIPAA, is not merely an equitable expansion of the net, but the efficiency-maximizing path forward. Our model does not invent this rule, rather, it provides the missing doctrinal foundation for it. Much like Le Verrier’s calculations directed astronomers to Neptune, our computational policy lab points legal scholars toward a reality that practice has already quietly adopted. Sophisticated buyers are willing to assume risk to signal quality, and the law should formalize this private ordering rather than obstruct it. 

Methodologically, this study charts a ``third path'' for legal scholarship in the age of AI. For too long, AI's role in social science has been confined to two paths, knowledge synthesis or data processing. Instead, we demonstrate how to actively adopt and integrate new AI tools to forge a complete, novel empirical paradigm: translating micro-mechanisms observed in fieldwork into agent constraints, calibrating them with LLMs discrete choice experiments, and testing them in an ABM. This approach overcomes the traditional barriers of data scarcity and access to specific populations, allowing researchers to observe the ``unobservable'' elasticities of the data market.

Ultimately, the value of this computational laboratory lies in its epistemic humility. It does not aim to replace the legal theorist with a machine, but to provide the theorist with a telescope. By isolating the variable of institutional design and simulating its systemic effects, we confirm what standard priors had obscured, and practice often prefigures theory. For a field in transformation, this work suggests that the future of social science lies not in passively waiting for computer scientists to define the tools, but in actively forging a new empirical paradigm—one that uses computation to illuminate the hidden logic of our laws.

\footnotesize
\section*{Acknowledgements}
The authors appreciate Xin Dai for the helpful comments. This study was supported by the Open Research Grant ``The Restructuring of the Agenda of Legal Theory by Intelligent Technologies'' (Grant No. 8410103413) under the National Intelligent Social Governance Experimental Base of Donghu High-tech Zone and Peking University. 


\normalsize
\bibliography{main}

\end{document}